\documentclass[twocolumn,twocolappendix]{aastex63}

\usepackage{amsmath}
\usepackage{amssymb}

\usepackage{graphicx}
\begin{document}

\title{The Effect of Supernovae on the Turbulence and Dispersal of Molecular Clouds}

\author{Zu-Jia Lu}
\affiliation{Institut de Ci\`{e}ncies del Cosmos, Universitat de Barcelona, IEEC-UB, Mart\'{i} i Franqu\`{e}s 1, E08028 Barcelona, Spain; luzujia@icc.ub.edu}

\author{Veli-Matti Pelkonen}
\affiliation{Institut de Ci\`{e}ncies del Cosmos, Universitat de Barcelona, IEEC-UB, Mart\'{i} i Franqu\`{e}s 1, E08028 Barcelona, Spain}

\author{Paolo Padoan}
\affiliation{Institut de Ci\`{e}ncies del Cosmos, Universitat de Barcelona, IEEC-UB, Mart\'{i} i Franqu\`{e}s 1, E08028 Barcelona, Spain}
\affiliation{ICREA, Pg. Llu\'{i}s Companys 23, 08010 Barcelona, Spain}

\author{Liubin Pan}
\affil{School of Physics and Astronomy, Sun Yat-sen University, 2 Daxue Road, Zhuhai, Guangdong, 519082, China; panlb5@mail.sysu.edu.cn}

\author{Troels Haugb{\o}lle}
\affiliation{Niels Bohr Institute, University of Copenhagen, {\O}ster Voldgade 5-7, DK-1350 Copenhagen K, Denmark}

\author{{\AA}ke Nordlund}
\affiliation{Niels Bohr Institute, University of Copenhagen, {\O}ster Voldgade 5-7, DK-1350 Copenhagen K, Denmark}

\begin{abstract}

While the importance of supernova feedback in galaxies is well established, its role on the scale of molecular clouds is still debated. In this work, we focus on the impact of supernovae on individual clouds, using a high-resolution magneto-hydrodynamic simulation of a region of 250~pc where we resolve the formation of individual massive stars. The supernova feedback is implemented with \emph{real supernovae}, meaning supernovae that are the natural evolution of the resolved massive stars, so their position and timing are self-consistent. We select a large sample of molecular clouds from the simulation to investigate the supernova energy injection and the resulting properties of molecular clouds. We find that molecular clouds have a lifetime of a few dynamical times, less then half of them contract to the point of becoming gravitationally bound, and the dispersal time of bound clouds of order one dynamical time is a factor of two shorter than that of unbound clouds. We stress the importance of internal supernovae, that is massive stars that explode inside their parent cloud, in setting the cloud dispersal time, and their huge overdensity compared to models where the supernovae are randomly distributed. We also quantify the energy injection efficiency of supernovae as a function of supernova distance to the clouds. We conclude that intermittent driving by supernovae can maintain molecular cloud turbulence and may be the main process of cloud dispersal, and that the full role of supernovae in the evolution of molecular clouds cannot be fully accounted for without a self-consistent implementation of the supernova feedback. 

\end{abstract}


\section{Introduction}

Molecular clouds (MCs) are the link between galactic-scale processes and star formation. The origin and evolution of MCs is influenced by galactic dynamics, and the largest MC complexes may arise from galactic-scale processes. These include the disk differential rotation, the galactic tidal field, spiral-arm shocks, gravitational and Parker instabilities, and magneto-Jeans instability to name a few \citep[e.g.][]{Mouschovias+74,Elmegreen79,Blitz+shu80,Balbus88,Kim+Ostriker02,Shetty+Ostriker06,Dobbs2008,Jeffreson+18,Inoue+Yoshida19}. Because MCs are the sites of star-formation, their evolution is also driven by stellar feedbacks, primarily supernovae (SNe), winds and radiation from massive stars. Understanding the relative importance of all these galactic and internal energy sources in the evolution of MCs is a major problem whose solution may depend on the type of galactic environment \citep{Howard+17}.

As demonstrated by modern cosmological simulations of galaxy formation, SNe are the dominant stellar-feedback process on a galactic scale. Disk properties, galactic outflows, and the chemo-dynamical evolution of galaxies can not be reproduced without a proper implementation of SN-feedback \citep[e.g.][]{Kortgen+2016MNRAS,Hopkins+18,Smith2018MNRAS.478..302S,Keller+Kruijssen2020}. Thus, the effect of SNe on the interstellar medium (ISM) of a galaxy is arguably stronger than that of the galactic-scale dynamical processes mentioned above. More accurately, SNe may be viewed as the end product of those processes, because SNe are essential to the overall balance between gravity and pressure forces in galactic disks \citep[e.g.][]{Kim+Ostriker15,Sun+2020}. Given this recognized role of SNe in the ISM of galaxies, it would be surprising if MCs could be modelled in any realistic way while neglecting SNe. Regardless, most models of MCs do neglect SNe (particularly those where the ISM turbulence and its driving mechanism are also neglected), and some recent numerical studies of the interaction of SNe with MCs claim that the effect of SNe on the clouds is relatively mild \citep[e.g.][]{Walch+2015MNRAS,Geen+2016,Ibanez+MacLow2016ApJ...824...41I,Seifried+2018ApJ,Lucas+2020}. 

It is often stated that SNe are not important for the dispersal of a MC, because they occur too late, after the winds and ionizing radiation from their stellar precursors has already dispersed the cloud. However, the timescale of MC destruction depends on several factors (e.g.~cloud mass and virial parameter, density structure, star-formation history) and is not very short in general, particularly in the case of relatively massive clouds \citep[e.g.][]{Walch+12_dispersal,colin+2013,Dale+14,Howard+16,Dale+2017MNRAS,Kim+18_ionization,Haid+19}. Based on self-consistent models of star-forming clouds, \citet{Dale+2017MNRAS} found that only 20-60\% (depending on the virial parameter) of the mass of a cloud of $10^{4}$~M$_{\odot}$ is unbound by the ionization feedback after 3~Myr, and \citet{Kim+18_ionization} found that it takes $\sim 6.5$~Myr to destroy a cloud of $10^{5}$~M$_{\odot}$. In both cases, the total cloud destruction by the ionizing radiation would take about twice as long as the main-sequence lifetime of the most massive stars. Furthermore, SNe may be considered indirectly responsible for the dispersal of unbound MCs (possibly a significant fraction of the whole cloud population), as such transient clouds originate from large-scale compressions in the ISM turbulence \citep[e.g.][]{Larson81,deAvillez+Breitschwerdt05,Padoan18}, which is driven primarily by SNe. 

In this work, we study the effect of SNe on the evolution of MCs using a simulation that was developed in the spirit of isolating the role of SNe relative to stellar winds and radiation or galactic-scale processes \citep{Padoan+17sfr}. In other words, stellar winds, radiation and the galactic context is neglected to investigate if SNe alone can explain the formation and dispersal of MCs, their observational properties, and their star-formation rate. We have already shown in a number of publications that star-formation rates and MC properties from this simulation, and from its lower-resolution precursor, are consistent with the observations \citep{Padoan+SN1+2016ApJ,Pan+Padoan+SN2+2016ApJ,Padoan+SN3+2016ApJ,Padoan+17sfr}. The overall conclusion of those studies is that SNe are the most important feedback mechanism on MC scales, besides their already recognized role at galactic scale. Winds, radiation and galactic-scale processes may still play some role, but could not change the results to a large extent, as that would break the considerable agreement with the observations that we have already demonstrated with SNe alone. Because our study is based on a single simulation, it only addresses the case of a generic region of a galactic disk with a column density of $\sim 30$~M$_{\odot}$~pc$^{-2}$, a reasonable value for spiral arms of disk galaxies.  It is possible that with very different values of the gas column density, MCs could not be modeled without a self-consistent description of the galactic environment.

Here, we focus on the role of SNe in the dispersal of MCs, as this process has never been modelled in the context of a self-consistent simulation of a large ISM volume including single star formation. Large-scale simulations that could at least partly resolve the formation of MCs, did not resolve the formation of individual massive stars. Therefore, the timing and location of SNe and the location and dynamical state of nearby MCs could not be self-consistent. In our simulation, we resolve the formation of individual massive stars, and the position and timing of the SNe, as well as the formation of MCs, are modelled {\it ab initio}. Thanks to the large scale of the simulation (250~pc), and its long evolution time ($\sim 30$~Myr), we collect a large sample of hundreds of SNe and MCs spanning a realistic range of properties such as the local gas density at the site of the SN explosions, distance between SNe and MCs, SN frequency experienced by each cloud. In contrast, previous simulations that could resolve the internal dynamics of MCs and individual SN remnants were based on a few MCs, chosen arbitrarily for zoom-in re-runs of galactic fountain simulations, and run for only a few Myr, without following the star-formation process (and the emergence of SNe) self-consistently \citep[e.g.][]{Ibanez+MacLow+2017ApJ,Seifried+2018ApJ}.             

The structure of the paper is as follows. In \S~\ref{simulation} we outline the numerical simulation. In \S~\ref{mc_sample} we describe the selection of MCs in the simulation and the computation of MC parameters, including a discussion on the estimation of the virial parameter. Our main results on the MC evolution, the energy injection by SNe, the MC lifetime, the overdensity of internal SNe, and the kinetic energy decay are presented in \S~\ref{sn_feedback}. In \S~\ref{discussion} we discuss the derivation of the rms velocity of MCs, the role of SNe according to previous works, and the MC lifetime in relation to observational estimates. Our main conclusions are summarized in \S~\ref{conclusions}.

\section{Simulation} \label{simulation}

This work is based on a SN-driven MHD simulation of an ISM region of size $L_{\rm box}=250$ pc 
and total mass $M_{\rm box}=1.9\times 10^6$ $\rm M_{\odot}$. The 3D MHD equations are solved with the AMR code RAMSES \citep{Teyssier+2002A&A,Fromang+06,Teyssier07}, with periodic boundary conditions, a mean density $n_{\rm H,0}=5$ cm$^{-3}$ and a mean magnetic field $B_0=4.6$ $\mu$G. The simulation was presented in \citet{Padoan+17sfr} and details of the numerical setup can be found there. After self-gravity is included, the simulation is run for approximately 30 Myr, with a root grid of $512^3$ cells and six AMR levels, reaching a minimum cell size of $dx=0.0076$ pc (1568 AU). 250 million passively advected tracer particles, each representing a fluid element with a characteristic mass of approximately 0.008 $\rm M_{\odot}$, are included. The tracer particles record all the hydrodynamic variables and are tagged once they accrete onto a sink particle; they are crucial for the analysis of this work, because we use them to track the evolution of the molecular clouds we study.

The spatial resolution of the simulation is sufficient to resolve the formation of individual massive stars \citep{Padoan+2020} and thus to predict the time and location of SNe. To follow the collapse of prestellar cores, sink particles are created in cells where the gas density is larger than $10^6$ cm$^{-3}$, if i) the gravitational potential has a local minimum value, ii) the three-dimensional velocity divergence is negative, and iii) no other previously created sink particle is present within an exclusion radius $r_{\rm excl} =16 \, dx=0.12$~pc \citep[see][for details of the sink particle algorithm]{Haugboelle+2018}. Sink particles gradually accrete the gravitationally-bound surrounding gas within an accretion radius $r_{\rm accr} =4 \, {dx}=0.03$~pc and with an efficiency $\epsilon_{\rm out}=0.5$, to account for unresolved outflows. Therefore, only half of the infalling gas contributes to the growth of the sink-particle mass. When a sink particle of mass larger than 7.5 $\rm M_{\odot}$ has an age equal to the corresponding stellar lifetime for that mass \citep{Schaller+92}, a sphere of $10^{51}$ erg of thermal energy is injected at the location of the sink particle to simulate the SN explosion, as described in \citet{Padoan+SN1+2016ApJ}. The stellar mass is deposited on to the gas inside the sphere and the sink particle is removed. Because we resolve the formation of the massive stars that explode as SNe, the energy injection is fully consistent with the star-formation rate, as well as the spatial distribution of massive stars and of their parent clouds. For this reason, we have referred to this method as \emph{real SNe}. Being able to capture a realistic spatial distribution of SNe, consistent with the distribution and dynamical state of MCs, is a crucial innovation for the purpose of this work, because previous studies have shown that the effect of SNe on a MC is strongly dependent on the SN location with respect to the cloud \citep{Iffrig+2015A&A}. 

During the 30 Myr evolution $\sim 3000$ stars with mass $> 2.5\, \rm M_{\odot}$ and $\sim 800$ stars with mass $> 8\, \rm M_{\odot}$ formed. $304$ SNe exploded. These SNe inject energy into the surrounding ISM, whose random motion results in the formation and dispersal of a large number of clouds, where new stars are born. The star formation rate (SFR) per free-fall time in such clouds was studied in \citet{Padoan+17sfr}, where it was found to exhibit a very large scatter from cloud to cloud, with an average value of a few percent, consistent with the observations.

\section{MC Sample} \label{mc_sample}

\subsection{MC Selection}

MCs are selected from the simulation following the same procedure as in \citet{Padoan+17sfr}. We use a subsample of 83 of the clouds analyzed in that work. The clouds are defined as connected regions above a threshold density, $n_{\rm H,min}=200$ cm$^{-3}$, in the root-grid of the simulation ($512^3$ cells, or a spatial resolution of 0.49 pc). Having saved one snapshot every $0.03\, \rm Myr$, we have accumulated a total of approximately 1000 snapshots. In \citet{Padoan+17sfr}, MCs were selected from 10 snapshots at intervals of 1.5 Myr, with the first one starting 4 Myr after the inclusion of self-gravity in the simulation. The following conditions were also imposed: 1) the cloud mass is $M_{\rm cl} > 1000$ M$_{\odot}$, 2) the rms velocity is $\sigma_{\rm v}< 4\,\rm km\,\rm s^{-1}$ (to avoid MCs too strongly affected by recent nearby SNe, 3) the cloud does not disperse during the next 1.5 Myr, meaning that it is not doubling its effective size in that time interval. This selection resulted in a sample of 203 clouds with $n_{\rm H,min}=200$ cm$^{-3}$. 

Not all of these clouds are independent of each other, in the sense that a cloud selected at a given time may have a large fraction of its mass (tracer particles) in common with a cloud selected at an earlier or later time. Once a cloud is selected and the associated tracer particles identified, we follow its evolution through those initial tracer particles alone, and no more tracer particles are assigned to the cloud (though the tracer particle mass may decrease as a result of star formation). This is the reason why clouds selected at different times may be related to each other, but are never identical. In order to reduce the number of clouds that are continuation of previous ones, or at least to reduce the fraction of tracer particles that a cloud shares with other clouds, in this work we have reduced the cloud sample by selecting MCs from only four snapshots (instead of 10), at intervals of approximately 3 Myr, with the first snapshot at approximately 8 Myr from the inclusion of self-gravity in the simulation, bringing the cloud number down from 203 to 83. With this reduction in the cloud number, only 10 of the 83 clouds are non-unique, in the sense that they belong to a pair where each member contains more than 50\% of the tracers of the other member.

Although the clouds are selected at the root-grid resolution, the cloud properties are computed using all the tracer particles identified within each cloud. Because the tracers record all the hydrodynamical variables interpolated at their position, and due to the very large number of tracers in high density regions, MC properties are derived with the hydrodynamical variables sampled at the highest local spatial resolution of the AMR grid, up to the highest resolution of 0.0076 pc in the densest regions.

\subsection{Computation of MC Parameters}

The basic properties of the MCs are measured at the time when the MCs are first identified (except for $\alpha_{\rm vir,min}$, as explained below), such as the observational virial parameter, $\alpha_{\rm vir,obs}$, the total mass, $M_{\rm cl}$, the radius, $R_{\rm cl}$, and the one-dimensional velocity dispersion, $\sigma_{\rm v}$. For reference, the cloud properties are listed in Table \ref{tab_1_all} in Appendix \ref{app_table}. Here, in Table \ref{table_mc_properties}, we give the same properties only for two representative clouds, C10 and C31, which we have chosen as examples of a low-mass cloud that is readily dispersed by a single SN, and a high-mass cloud that forms a stellar cluster of $\sim 10^4\,\rm M_{\odot}$ and requires several local SNe to be fully dispersed. We will focus on the same two clouds in later figures as well.   

\begin{table}
\centering
\caption{Basic properties of two representative clouds in the simulation, C10 and C31, measured at the time when the clouds are identified. From left to right: cloud number, mass, observational virial parameter, star formation rate, radius and one-dimensional velocity dispersion. The same quantities are listed for all clouds in Table \ref{tab_1_all} in Appendix \ref{app_table}.}
\begin{tabular}{lccccc}
\hline
\hline
$\rm Cloud$ & $M_{\rm cl}$ & $\alpha_{\rm vir,obs}$ & $SFR$ & $R_{\rm cl}$ & $\sigma_{\rm v}$ \\
\, & $\rm [M_{\odot}]$ & \, & $\rm [Myr^{-1}]$ & $\rm[pc]$ & $\rm [km\,\rm s^{-1}]$ \\
\hline
C10 & 3.25 $ \times 10^3 $ & 0.9 & 0.02 & 5.44 & 1.18 \\
C31 & 5.22 $ \times 10^4 $ & 3.24 & 0.02 & 25.46 & 4.14 \\
\hline
\hline
\end{tabular}
\label{table_mc_properties}
\end{table}

The virial parameter \citep{Bertoldi+McKee1992ApJ...395..140B} is defined as the ratio of turbulent to gravitational energy, 
\begin{equation}
\alpha_{\rm vir} \equiv \frac{2 E_{\rm k}}{E_{\rm g}} \, ,
\label{eq_alpha_vir}
\end{equation}
and is used to gauge whether a cloud is bound or unbound. A bound cloud is characterized by $\alpha_{\rm vir} \lesssim 2$, while $\alpha_{\rm vir} > 2$ implies an unbound cloud. To determine the virial parameter, we compute the internal kinetic energy of the clouds, $E_{\rm k}$, and their gravitational energy, $E_{\rm g}$, using the velocities and positions of the tracer particles in the cloud\footnote{Tracer particles that have accreted onto sink particles are not included in the computation of the cloud $E_{\rm g}$, because the potential energy of individual stars, which would be very large, is irrelevant to the cloud evolution.},
\begin{equation}
E_{\rm k} = \frac{1}{2}\,m \sum_{n=1}^{N} u_{n}^2 \, ,
\label{eq_Ek}
\end{equation}
\begin{equation}
E_{\rm g} = \frac{1}{2}\,G\, m^2 \sum_{j=1}^{N} \sum_{k=1}^{N} \frac{1 }{r_{jk}} \, ,
\label{eq_Eg}
\end{equation}
where $N$ is the total number of tracer particles in the cloud, $m$ is the mass associated to a tracer particle, $u_{n}$ is the modulus of the velocity of the $n$-th particle, and $r_{jk}$ is the distance between the $j$-th and the $k$-th particles. This expression for $E_{\rm g}$ assumes that the cloud is isolated. 

The virial parameter is often used to characterize observed MCs, based on its simplified expression for a uniform, spherical cloud:
\begin{equation}
\alpha_{\rm vir,obs} \equiv \frac{5 \sigma^2_{\rm v} R_{\rm cl}}{G M_{\rm cl}} = \frac{40} {3\pi^2}\left(\frac{t_{\rm ff}}{t_{\rm dyn}}\right)^2  \, ,
\label{eq_alpha_vir_obs}
\end{equation}
where $R_{\rm cl}$ is the cloud radius, $M_{\rm cl}$ the cloud mass, $\sigma_{\rm v}$ the one-dimensional velocity dispersion and $G$ the gravitational constant. In the second equality, $t_{\rm dyn}$ is the dynamical time, defined as
\begin{equation}
t_{\rm dyn} \equiv R_{\rm cl}/\sigma_{\rm v,3D} \, ,
\label{eq_tdyn}
\end{equation}
and $t_{\rm ff}$ the free-fall time,
\begin{equation}
t_{\rm ff}=\left(\frac{3\pi}{32G\rho_0}\right)^{1/2} \, ,
\end{equation}
where $\rho_0$ is the cloud mean density. Because the simplified expression is widely used in the literature, we compute $\alpha_{\rm vir,obs}$ for the clouds as well. For the cloud velocity dispersion we adopt the one-dimensional rms velocity of the tracer particles, so the velocity is weighted by mass (it is weighted by emission in the observations):
\begin{equation}
\sigma_{\rm v} \equiv  \left[{\frac{1}{3N}}\,\sum_ {i=1}^{3} \, \sum_{n=1}^{N} ( u_{i,n} - \bar{u}_i )^2  \right ]^{1/2} \, ,
\label{eq_v}
\end{equation}
where $\bar{u}_i \equiv  \sum_{n=1}^{N} u_{i,n} /N$ are the components of the mean tracer particle velocity (the velocity of the center of mass), $u_{i,n}$ is the $i$-th velocity component of the $n$-th tracer particle, and $N$ is the total number of tracer particles in the cloud. For the cloud radius, $R_{\rm cl}$, we adopt the rms of the particle positions,
\begin{equation}
R_{\rm cl} \equiv  \left[{\frac{1}{N}}\,\sum_ {i=1}^{3} \, \sum_{n=1}^{N} ( x_{i,n} - \bar{x}_i )^2  \right ]^{1/2} \, ,
\label{eq_cloud_radius}
\end{equation}
where $\bar{x}_i \equiv \sum_{n=1}^{N} x_{i,n} /N$ are the components of the mean 3D particle position and $N$ is the total number of tracer particles in the cloud.

\subsection{Estimation of the Virial Parameter} \label{sec_alpha_vir}

\begin{figure*}
\centering
\includegraphics[width=0.48\textwidth]{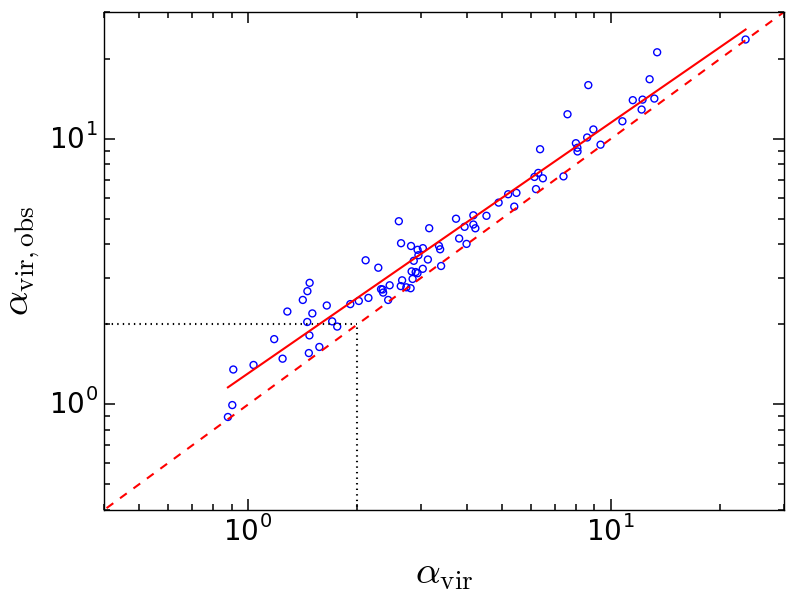}
\includegraphics[width=0.48\textwidth]{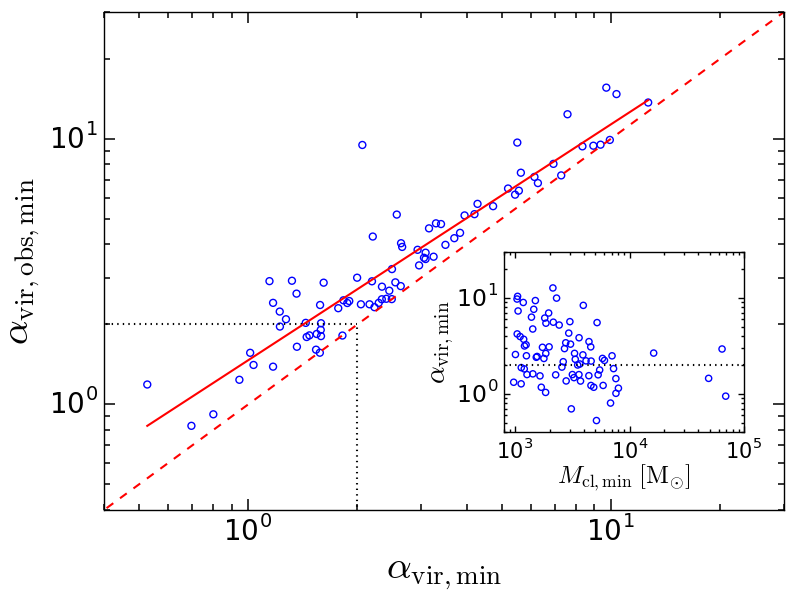}
\caption{Left panel: Virial parameter computed with the expression for a uniform sphere in Equation~(\ref{eq_alpha_vir_obs}), $\alpha_{\rm vir,obs}$, versus the virial parameter computed from the energy ratio as in Equation~(\ref{eq_alpha_vir}), $\alpha_{\rm vir}$, for all the clouds in the sample, evaluated at the time of their selection in the simulation. We calculate the mean ratio: $\langle\alpha_{\rm vir,obs}/\alpha_{\rm vir}\rangle=1.24 \pm 0.23$. The solid line is a least-squares fit giving $\alpha_{\rm vir,obs} = (1.31 \pm 0.05) \; \alpha_{\rm vir}^{(0.94 \pm 0.02)}$. The dotted lines isolate the clouds that appear to be bound based on both definitions of the virial parameter, and the dashed line corresponds to $\alpha_{\rm vir,obs} = \alpha_{\rm vir}$. Right panel: Same as the left panel, but for $\alpha_{\rm vir,obs}$ and $\alpha_{\rm vir}$ measured at the time when $\alpha_{\rm vir}$ reaches its minimum value, giving $\langle\alpha_{\rm vir,obs,min} / \alpha_{\rm vir,min}\rangle=1.37 \pm 0.47$. The solid line is a least-squares fit to the data points giving $\alpha_{\rm vir,obs,min} = (1.47 \pm 0.07) \; \alpha_{\rm vir,min}^{(0.89 \pm 0.04)}$. The relation between $\alpha_{\rm vir,min}$ and cloud mass $M_{\rm cl,min}$ measured at the same time are shown in the inset. The horizontal dotted line corresponds to $\alpha_{\rm vir,min}=2$.}
\label{fig_virial_versus_virial}
\end{figure*}

\citet{Padoan+SN1+2016ApJ} showed that, in the absence of self-gravity, the choices of cloud rms velocity and radius described above yield values of $\alpha_{\rm vir,obs}$ that are on average only 20\% larger than the virial parameter from the energy ratio, $\langle\alpha_{\rm vir,obs}/\alpha_{\rm vir}\rangle=1.20$ (see Figure 17 in \citet{Padoan+SN1+2016ApJ}). The $\alpha_{\rm vir,obs}/\alpha_{\rm vir}$ ratio was found to be essentially constant over three orders of magnitude in virial parameter, with a very small scatter. This suggests the presence of some underlying universality in the fragmentation properties of MCs due to their supersonic turbulence. 

The nature of the fragmentation was found to persist also in the presence of self-gravity (see Figure 18 in \citet{Padoan+SN1+2016ApJ}), except in a fraction of clouds with low virial parameters. These clouds contained collapsing clumps, resulting in a strong decrease in the energy ratio. Because the clouds were not undergoing a global collapse, $\alpha_{\rm vir,obs}$ did not drop with the inclusion of self-gravity, leading to a strong increase in $\alpha_{\rm vir,obs}/\alpha_{\rm vir}$. However, the simulation in \citet{Padoan+SN1+2016ApJ} did not include sink particles (the SN feedback was modeled with SNe generated randomly in space and time). Therefore, the $\alpha_{\rm vir,obs}/\alpha_{\rm vir}$ ratio was certainly overestimated, because a lot of very dense collapsed gas, which should have been continuously accreted to stars, was added to the potential energy of the clouds. Thus, the $\alpha_{\rm vir,obs}/\alpha_{\rm vir}$ ratio under self-gravity should be addressed again using the current simulation that include both gravity and sink particles at much higher resolution.

The left panel of Figure~\ref{fig_virial_versus_virial} shows a scatter plot of $\alpha_{\rm vir,obs}$ versus $\alpha_{\rm vir}$ computed at the time of cloud selection, as in \citet{Padoan+SN1+2016ApJ}. The mean ratio, $\langle\alpha_{\rm vir,obs}/\alpha_{\rm vir}\rangle=1.24 \pm 0.23$, is consistent, inside the error bars, with our previous result for clouds evolved without self-gravity, despite the fact that clouds in this sample have been identified $\sim 10-19$~Myr after the inclusion of self-gravity in the simulation, and many of them are actively forming stars. A power-law least-squares fit gives $\alpha_{\rm vir,obs} = (1.31 \pm 0.05) \; \alpha_{\rm vir}^{(0.94 \pm 0.02)}$. We don't find a significant number of clouds with $\alpha_{\rm vir,obs}/\alpha_{\rm vir} \gg 1$, unlike in \citet{Padoan+SN1+2016ApJ}, because in this simulation the collapsing gas is accreted onto the sink particles. The tight correlation between $\alpha_{\rm vir,obs}$ and $\alpha_{\rm vir}$ could be due to the existence of a global cloud density profile with an exponent of approximately $-2$ \citep[see][]{Bertoldi+McKee1992ApJ...395..140B}.
However, we have verified that the density profiles show large variations from cloud to cloud, and are often not fit well by power laws, due to the complex cloud structure.
Instead, the correlation is most likely a result of the universal fractal sub-structure in MCs
\citep[e.g.][]{Elmegreen+Falgarone96,Chappell+Scalo2001,Kritsuk+2007}. As it extends to high values of $\alpha_{\rm vir}$, this correlation is strong evidence of the dominant role of supersonic turbulence relative to gravity in the dynamics of the clouds. 

The cloud kinetic energy evolution is characterized by discrete peaks that decay rapidly and do not represent the global state of a cloud, because the excess energy corresponding to the peaks is not fully distributed to the cloud initially (see the discussion of individual energy peaks in \S~\ref{sec_energy_injection}). To better characterize the dynamical state of the clouds, instead of using the values of $\alpha_{\rm vir}$ and $t_{\rm dyn}$ measured at the time of the cloud identification, we adopt the minimum value of the virial parameter, $\alpha_{\rm vir,min}$, and the corresponding value of the dynamical time, $t_{\rm dyn,\alpha}$ (measured at the time of the minimum of $\alpha_{\rm vir}$). These values represent what the dynamical state of a cloud would be in the absence of the effect of discrete SN events. The comparison of $\alpha_{\rm vir,min}$ with $\alpha_{\rm vir,obs,min}$, the value of $\alpha_{\rm vir,obs}$ computed at the time of the minimum value of $\alpha_{\rm vir}$, is shown in the right panel of Figure~\ref{fig_virial_versus_virial}. The mean value of this virial-parameter ratio is $\langle\alpha_{\rm vir,obs,min} / \alpha_{\rm vir,min}\rangle=1.37 \pm 0.47$. The best fit to the scatter plot is only slightly shallower than that in the left panel: $\alpha_{\rm vir,obs,min} = (1.47 \pm 0.07) \; \alpha_{\rm vir,min}^{(0.89 \pm 0.04)}$. Based on this minimum value of the virial parameter, the sample contains 32 bound clouds and 51 unbound ones. 

Despite the relatively large fraction of bound clouds, it is important to stress that the bound state of a cloud is a rather transient feature. As further discussed in \S~\ref{MC_lifetime}, the clouds undergo an initial contraction phase (we will refer to its duration as the formation time) until a minimum radius is reached, followed by an expansion phase (we will refer to its duration as the dispersal time). The minimum value of the virial parameter is reached in the neighborhood of the minimum radius. Not long before or after that time, the virial parameter is larger than 2 also for the clouds that have $\alpha_{\rm vir,min}\le 2$, so MCs are bound only for a limited time (typically a few Myr). An important consequence is that the initial contraction (formation) phase is not driven by the cloud self-gravity ($\alpha_{\rm vir}> 2$) for much of that phase, meaning that the clouds must be assembled primarily by random compressive motions from the large-scale turbulence, rather than by a gravitational collapse. Furthermore, clouds with virial parameter values larger than two may be in either of the two phases (formation or dispersal). Additional probes of cloud evolution, besides the virial parameter, could be used to assess the evolutionary state of a cloud (e.g.~the star-formation efficiency, signatures of SN remnants, kinematical evidence for large-scale compressions).       

\section{Effect of SN Feedback on MCs} \label{sn_feedback}

In this section we present the main results from analysing the effect of SN feedback on MCs. We first present the time evolution of basic MC properties and then study in more detail how energy is injected and decays in individual MCs. We also discuss the difficulty of estimating the global effect of the SN feedback on the velocity dispersion of MCs based on average quantities such as the SN rate and different expressions for the energy decay rate.

\subsection{Cloud Evolution} \label{mc_evolution}

\begin{figure*}
\centering
\includegraphics[width=0.98\textwidth]{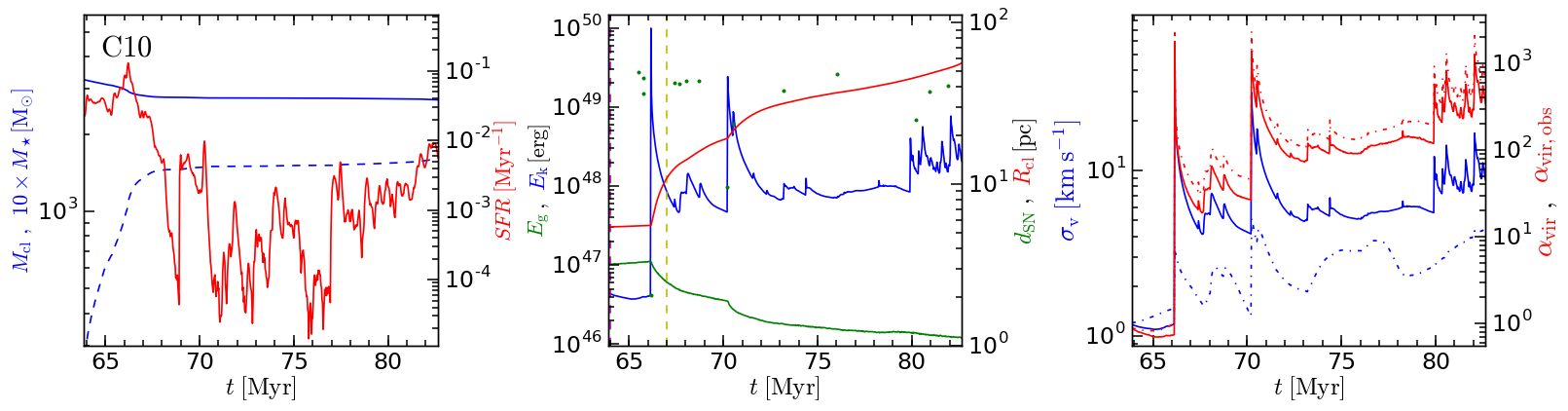}\\
\includegraphics[width=0.98\textwidth]{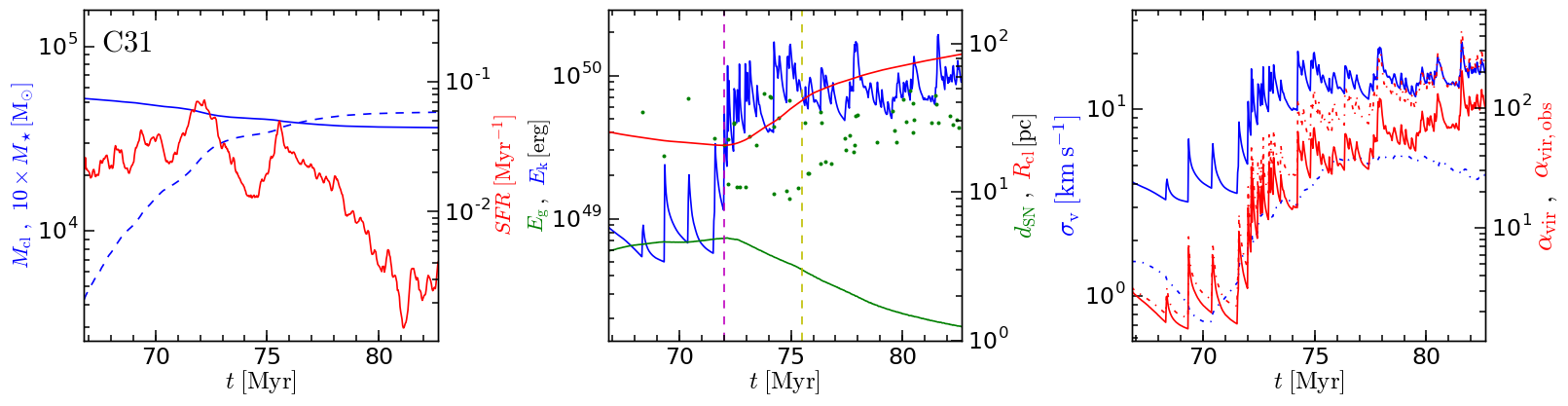}
\caption{Time evolution of properties of two clouds: C10 (upper panels) and C31 (lower panels). Left panel: Time evolution of cloud mass, $M_{\rm cl}$ (blue solid line), total star mass multiplied by a factor of 10, $M_{\rm \star}$ (blue dashed line) and star formation rate, $SFR$ (red solid line, right y axis). Middle panel: Time evolution of total kinetic energy, $E_{\rm k}$ (blue line), total gravitational potential energy, $E_{\rm g}$ (green line), cloud radius, $R_{\rm cl}$ (red line, right y axis). The explosion times of all the SNe with distance $d_{\rm SN} \leq 50\,\rm pc$ from the mass center of the cloud are marked by the green filled circles that also give the distance of the SNe from the cloud center (right y axis). Dashed magenta and yellow lines indicate the times when the cloud reaches the minimum radius and twice the minimum radius. Notice that past the yellow line, the cloud is considered to be dispersed. Right panel: Time evolution of the one-dimensional velocity dispersion, $\sigma _{\rm v}$ (blue solid line), and average velocity (blue dotted line) of all tracers in the cloud, virial parameter calculated from the energy ratio $\alpha_{\rm vir}$ (red solid line, right y axis), and from the uniform-sphere expression, $\alpha_{\rm vir,obs}$ (red dotted line, right y axis). The corresponding plots for all 83 clouds are publicly available online (URL: http://www.erda.dk/vgrid/ramses/).}
\label{fig_cloud_properties}
\end{figure*}

To follow the evolution of the clouds under the effect of SNe, we compute several quantities in all available snapshots, and plot them in a series of three-panel figures. Here we show the plots for our chosen representative clouds, C10 and C31, in Figure \ref{fig_cloud_properties}. The corresponding plots for all 83 clouds are publicly available online (URL: http://www.erda.dk/vgrid/ramses/). 

The left panel of Figure \ref{fig_cloud_properties} shows the time evolution of the cloud mass $M_{\rm cl}$ (blue line) and of the star-formation rate (red line). The total mass of the cloud is the sum of all the tracer particles in that snapshot (excluding those that have been accreted onto stars). Because we follow the evolution of only the tracer particles selected at the initial snapshot when a cloud is identified, the cloud mass cannot increase, but it decreases due to the formation of stars. The star-formation rate, $SFR$ is defined as the inverse of the gas consumption time, that is the variation of the cloud mass over an interval of time divided by that time interval, and normalized to the cloud mass \citep[e.g.][]{Inoue00,Clark08,Padoan+17sfr}:
\begin{equation}
SFR(t) = \frac{[M_{\rm cl} (t_{j}) - M_{\rm cl} (t_{j+1})]/M_{\rm cl}(t_{j})} {t_{j+1}-t_{j}} \, ,
\label{eq_sfe}
\end{equation}
where $M_{\rm cl}(t)$ is the mass of tracers in the cloud that have not been accreted onto sink particles at the time $t$, $t_{j}$ is the time corresponding to the $j$-th snapshot in which the cloud is identified. The MC mass loss, $M_{\rm cl} (t_{j}) - M_{\rm cl} (t_{j+1})$, between the time interval $\Delta t = [t_{j+1}-t_{j}]$ is the mass that is accreted onto stars (sink particles in the simulation). Our two representative cloud have initial masses of $\sim 3.25 \times 10^{3} \,\rm M_{\odot}$ and $\sim 5.22 \times 10^{4} \,\rm M_{\odot}$ and $SFR$ corresponding to a gas-consumption time that reaches a minimum value of order 10 Myr, but is typically significantly longer than that and may increase by orders of magnitude when the cloud start to expand.

The middle panel of Figure \ref{fig_cloud_properties} shows the time evolution of the internal kinetic energy of the cloud, $E_{\rm k}$ (blue line), the gravitational potential energy, $E_{\rm g}$ (green line), the radius of the cloud, $R_{\rm cl}$ (red line), and all the SNe exploded within 50 pc distance from the mass center of the cloud (green filled circles). The right y axis gives both the cloud radius and the distance of the SNe from the cloud center, $d_{\rm SN}$. The magenta, vertical, dashed line marks the time when the radius reaches its minimum value, $t_{\rm R,min}$, while the yellow, vertical, dashed line marks the time of cloud dispersal, defined as the time when the cloud radius has increased to twice its minimal value (see \S~\ref{MC_lifetime} for a discussion of the time evolution of the cloud radius).  

As shown in the middle panels of Figure \ref{fig_cloud_properties}, the two clouds C10 and C31 are initially bound, although that is true for the cloud C31 only in the time intervals when the energy injected by the first SNe has decayed enough. 
While the low-mass C10 is quickly dispersed by the first close SN, which explodes inside the MC at only 2 pc from the cloud center of mas, the cloud C31 slightly contracts during the first $\sim 5\,\rm Myr$, despite some SNe outside of the cloud, and requires a number of internal SNe to be fully dispersed. Although the SFR is reduced by approximately a factor of 10 after the clouds have doubled their radius, SF may still continue (presumably in the densest clumps) once the clouds are unbound.

The right panel of Figure \ref{fig_cloud_properties} shows the one-dimensional velocity dispersion, $\sigma_{\rm v}$ (blue solid line), and the average velocity of the tracer particles (blue dotted line). It also shows the time evolution of the cloud virial parameter, $\alpha_{\rm vir}$, computed in two different ways, either from the formula for a uniform sphere, $\alpha_{\rm vir,obs}$ (red dotted line), or from the energy ratio $\alpha_{\rm vir}$ (red solid line). The two expressions yield values of $\alpha_{\rm vir,obs}$ and $\alpha_{\rm vir}$ that are sometimes similar, but may also differ significantly, particularly after the dispersal of the cloud C31. 

\subsection{Supernova Energy Injection} \label{sec_energy_injection}

\begin{figure}
\centering
\begin{tabular}{cc}
C10 & C31 \\
\includegraphics[width=0.22\textwidth]{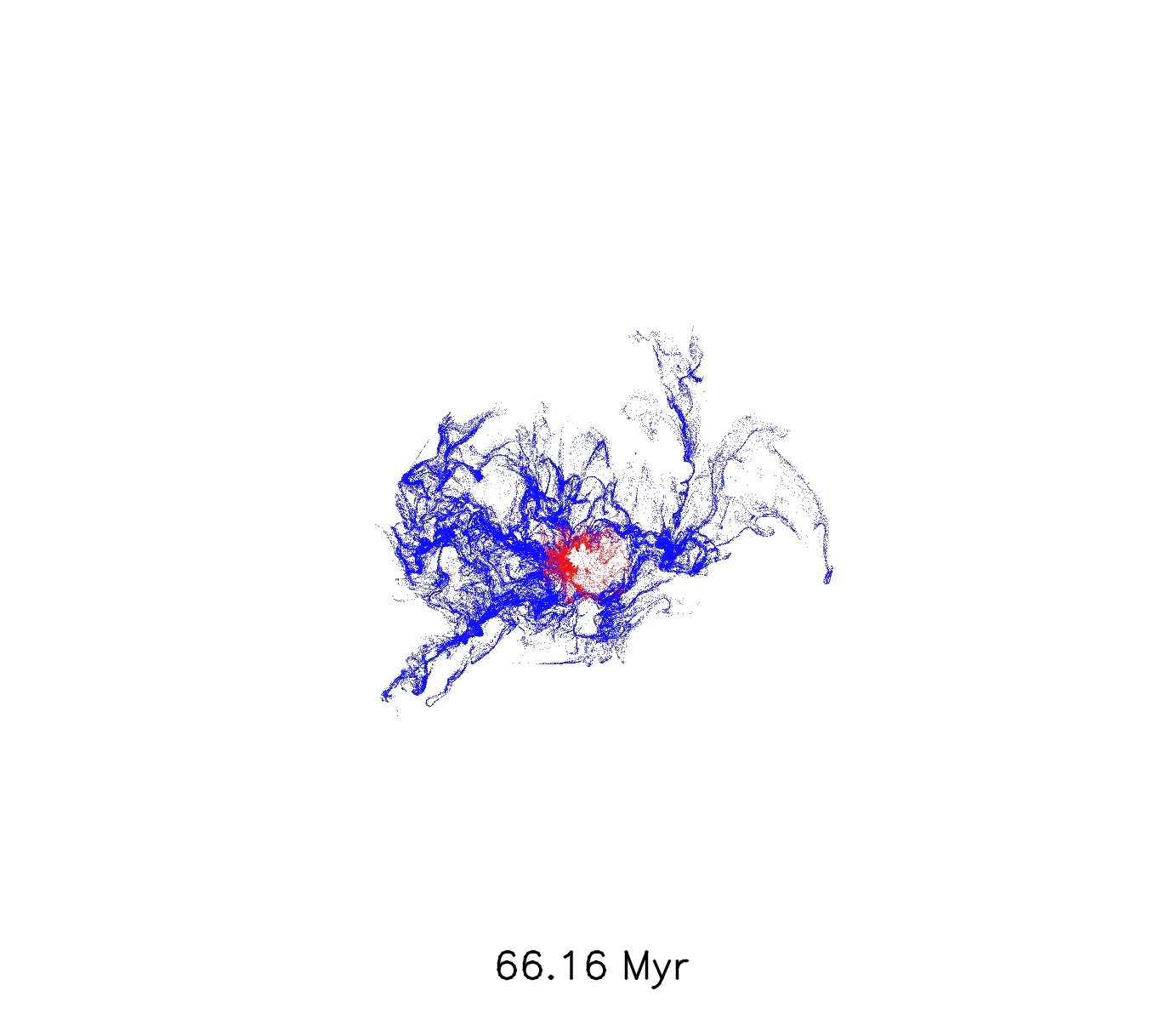}
&\includegraphics[width=0.22\textwidth]{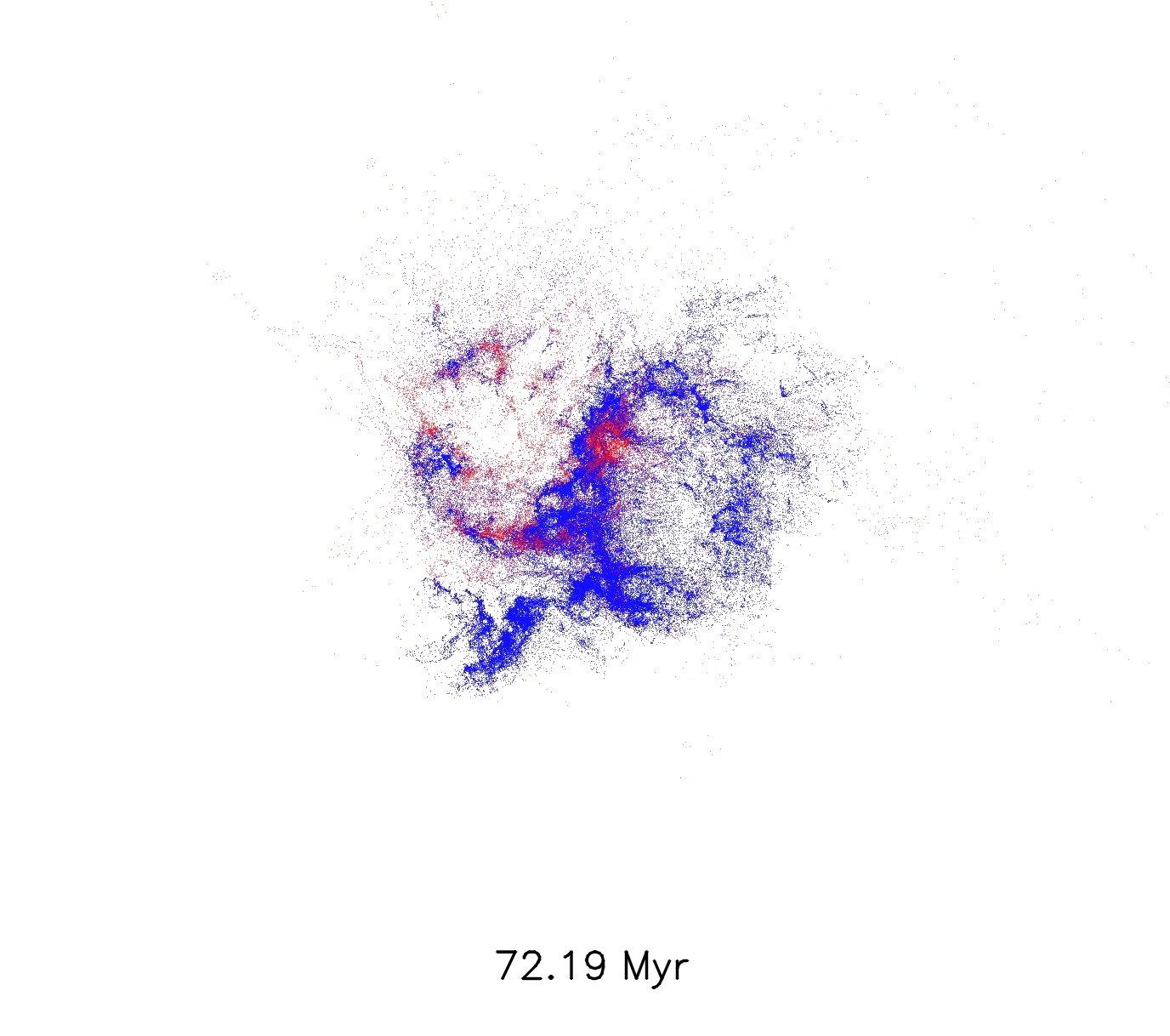}\\
\includegraphics[width=0.22\textwidth]{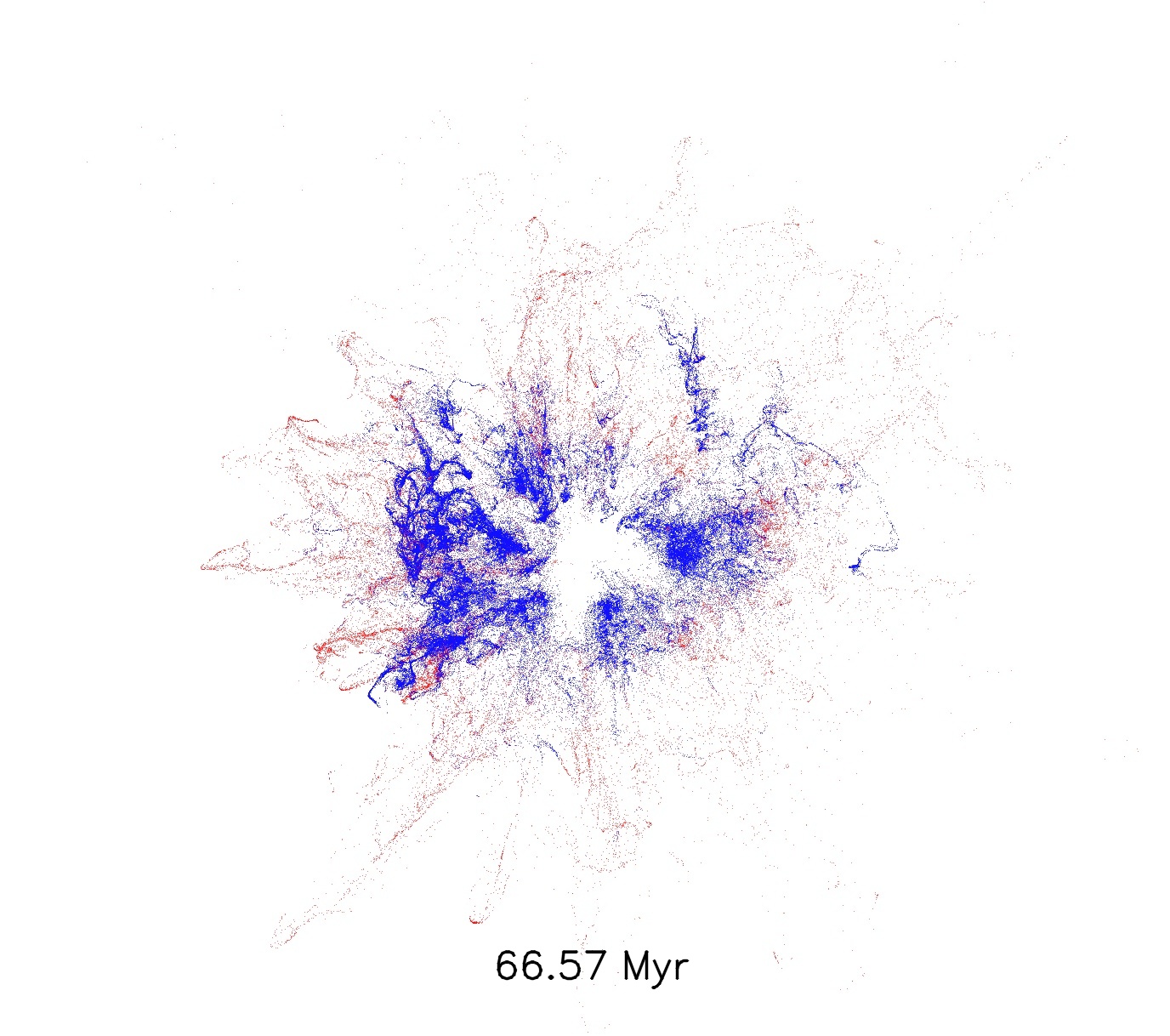}
&\includegraphics[width=0.22\textwidth]{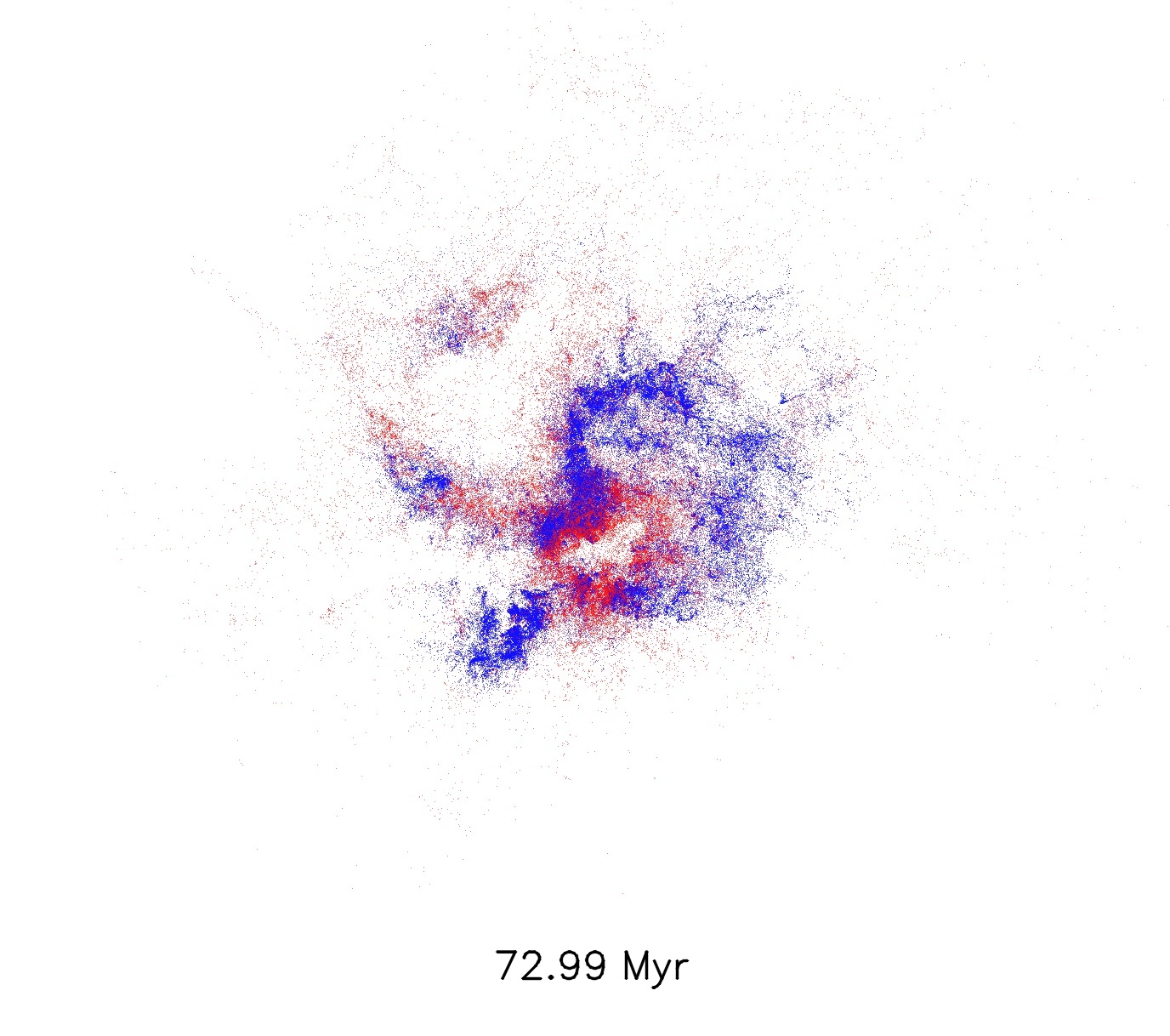}\\
\includegraphics[width=0.22\textwidth]{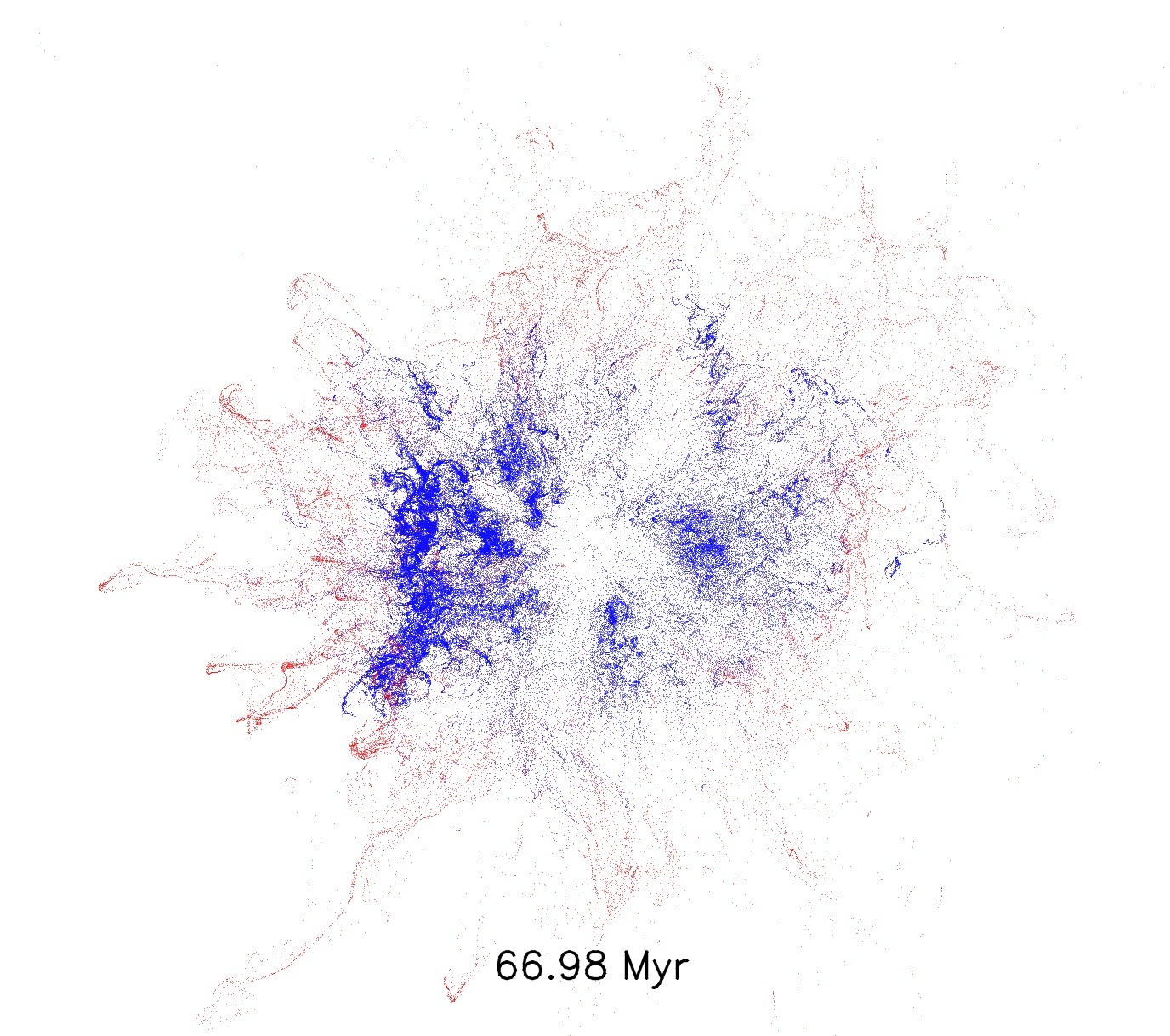}
&\includegraphics[width=0.22\textwidth]{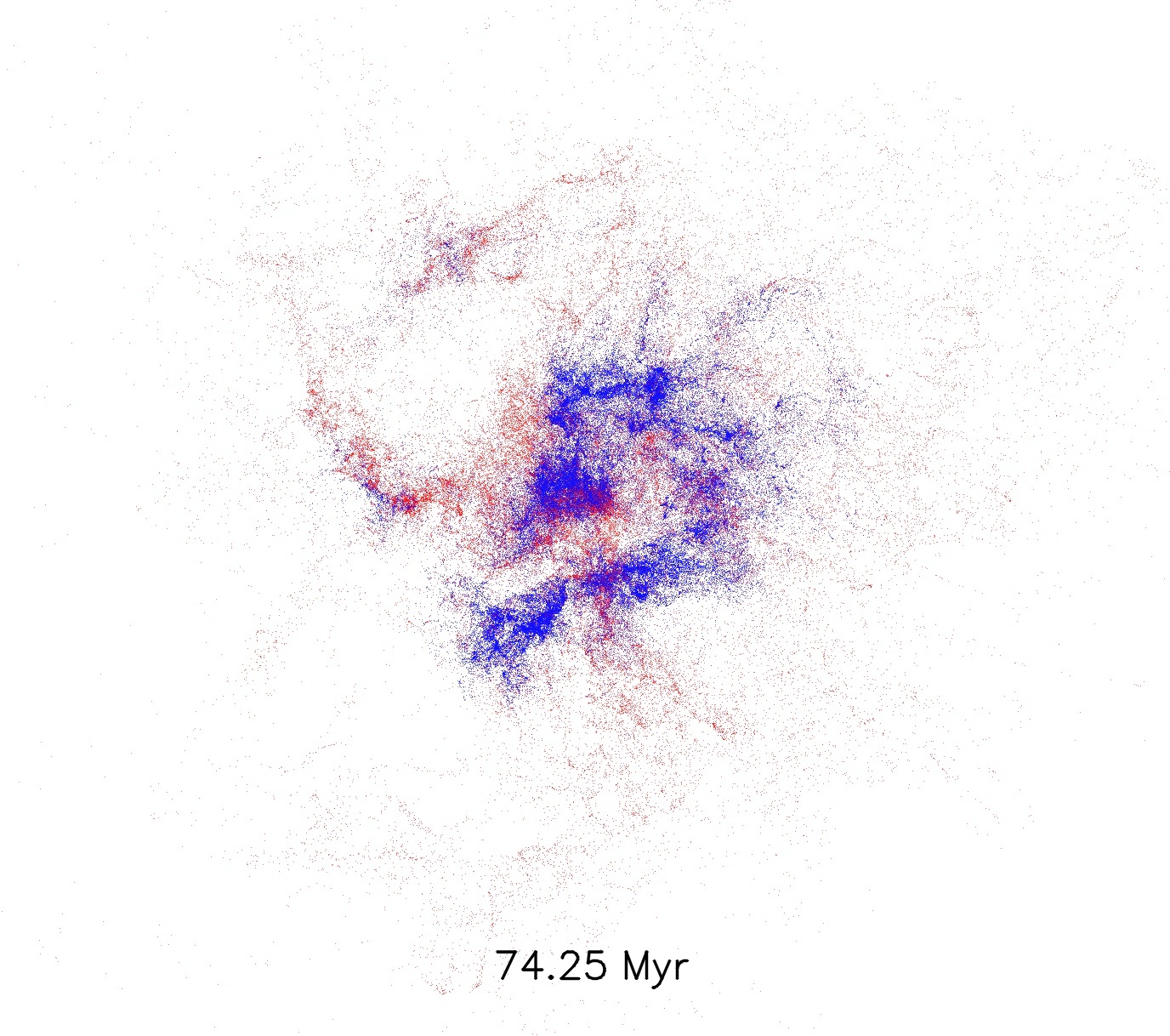}\\
\includegraphics[width=0.22\textwidth]{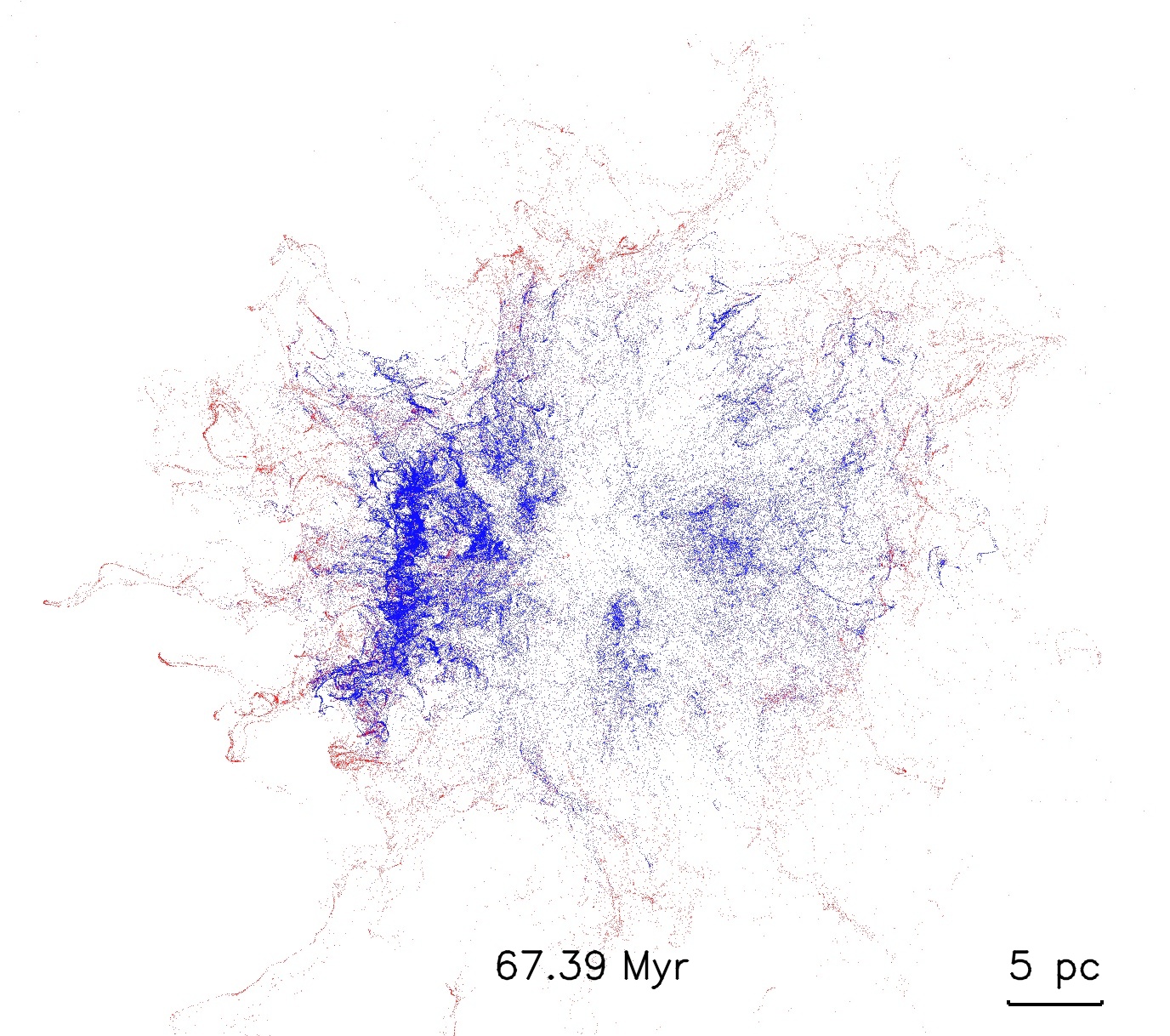}
&\includegraphics[width=0.22\textwidth]{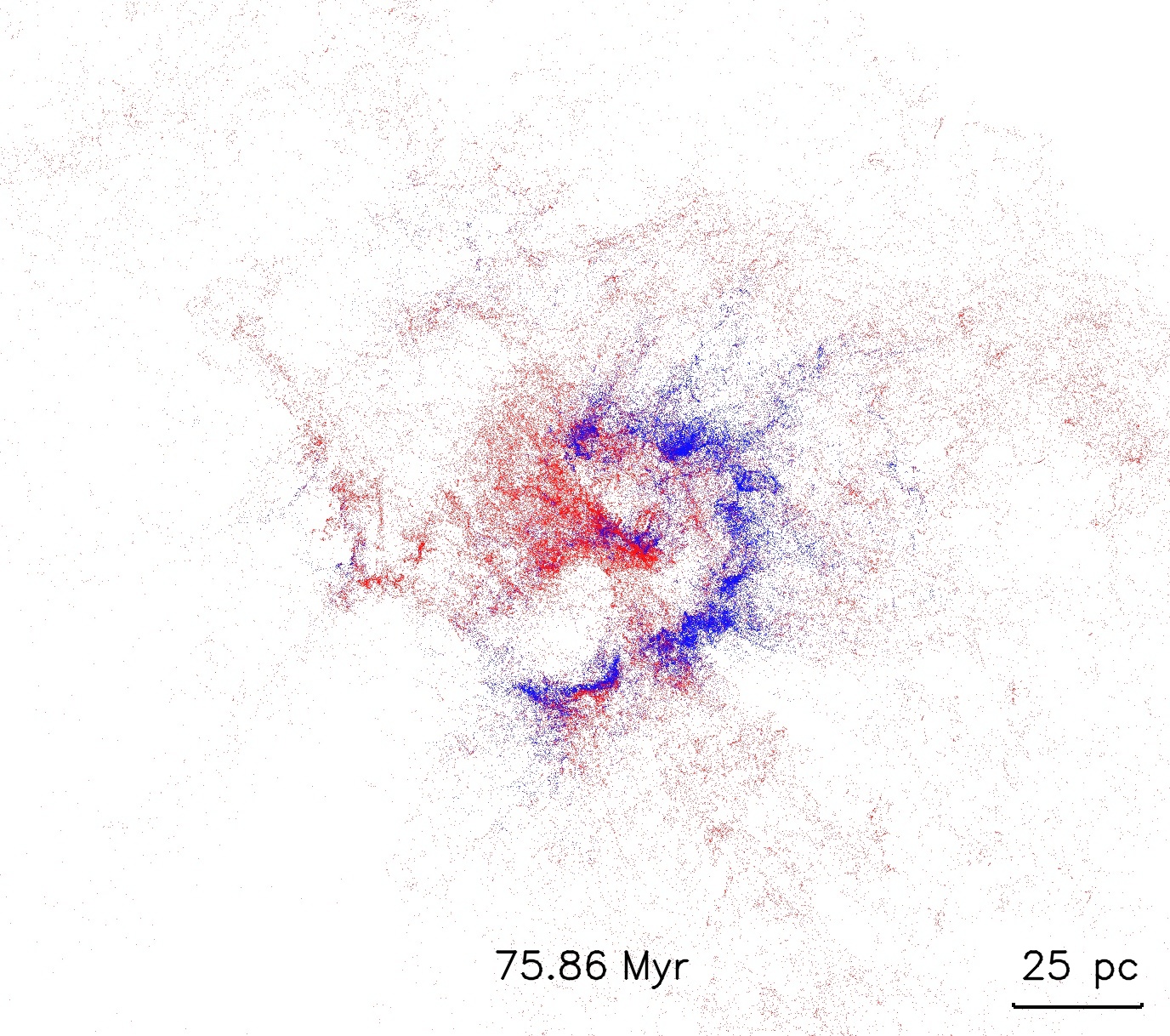}\\
\end{tabular}
\caption{Time evolution of the tracer particle distribution for the clouds C10 (left) and C31 (right). For cloud C10, the first time frame corresponds to the high energy peak caused by a SN near the cloud center that is responsible for the cloud dispersal ($t_{\rm SN}=66.16 \,\rm Myr$, $d_{\rm SN}=2.06 \,\rm pc$, $R_{\rm cl}=5.54 \,\rm pc$), while the other three time frames are equally spaced until the time before the next SN. For cloud C31, the time frames correspond to the energy peaks due to four consecutive internal SNe. All tracer particles are initially blue, and they turn red once they are accelerated to a velocity $> 10\,\rm km\,\rm s^{-1}$. Even if they were to slow down later on, their color remains red, to allow to visually track the mass fraction of the cloud that has been affected by SNe. The full movies for these clouds and others can be found online (\url{http://www.erda.dk/vgrid/ramses/}).}
\label{fig_mc_image_v}
\end{figure}

\begin{figure*}
\centering
\includegraphics[width=0.48\textwidth]{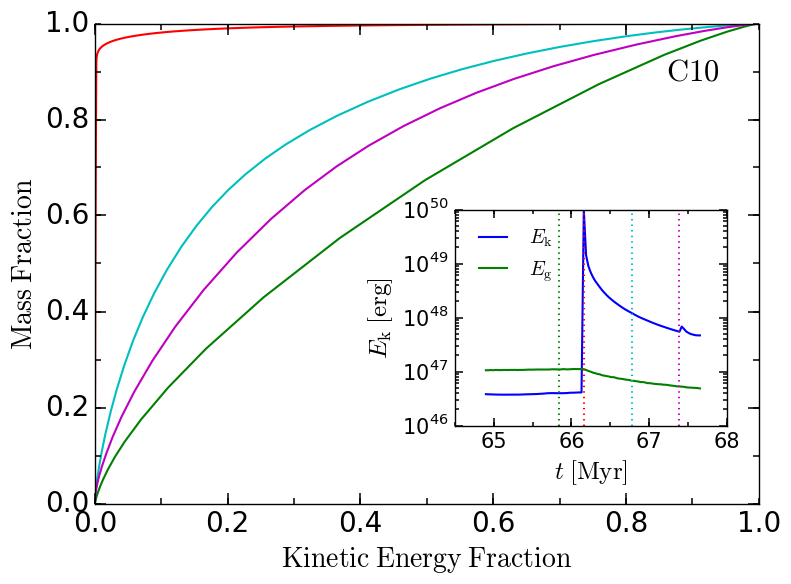}
\includegraphics[width=0.48\textwidth]{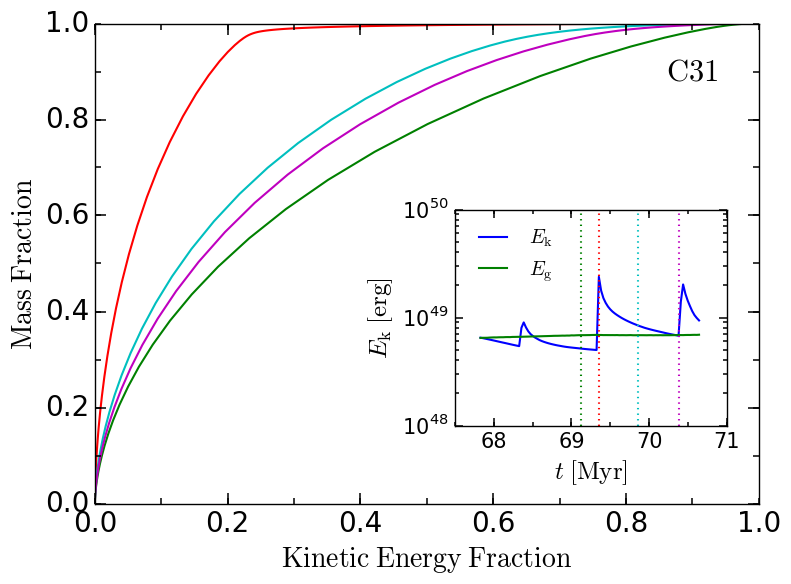}\\
\includegraphics[width=0.48\textwidth]{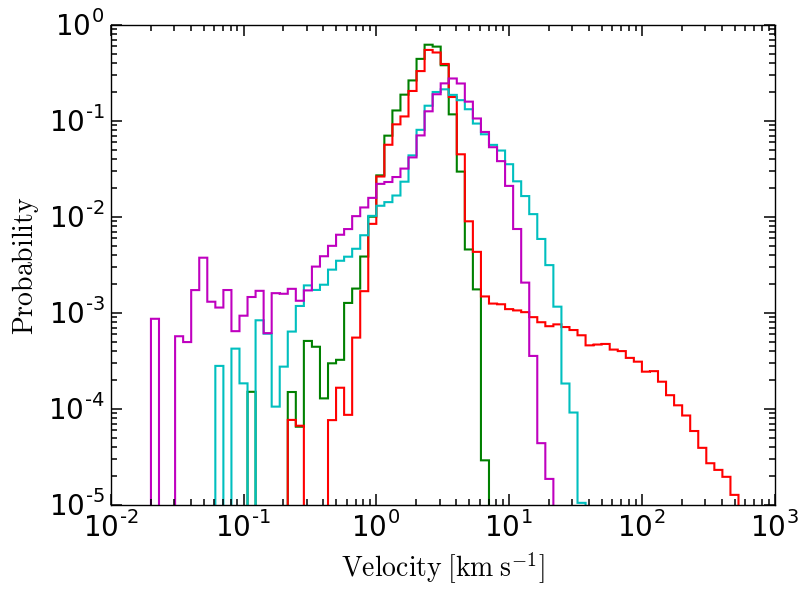}
\includegraphics[width=0.48\textwidth]{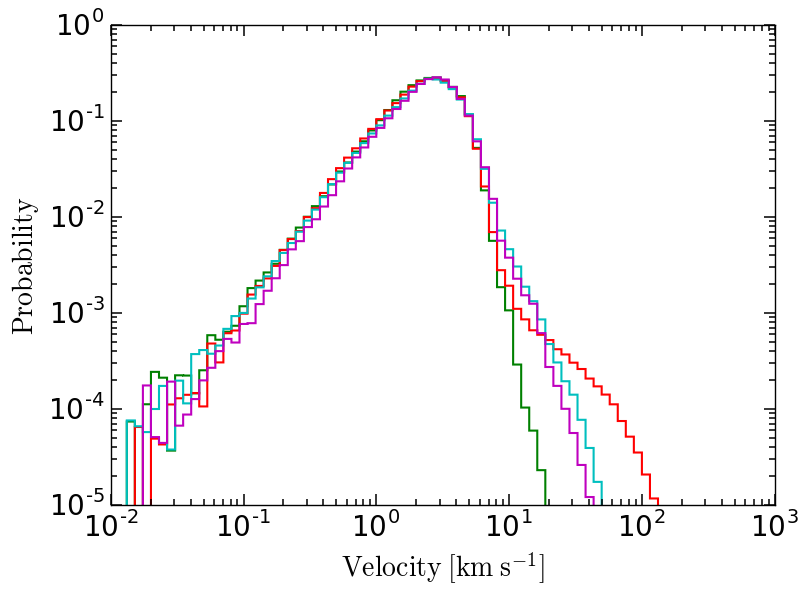}
\caption{Upper panels: Cumulative density function (CDF) of kinetic energy of the clouds C10 (left) and C31 (right), expressed as mass fraction versus kinetic energy fraction. The four lines correspond to four different times, marked by the vertical dashed lines in the insets. We focus on one specific peak in each cloud, as examples, and show the CDF just before the energy peak (green), and the energy peak (red), and at two times after the peak (cyan and magenta). 
Lower panels: PDF of tracer-particle velocity, relative to the center-of-mass velocity before the energy peak. Each PDF corresponds to one of the times as in the upper panels, with the same color coding.}
\label{fig_v_cdf}
\end{figure*}

Figure \ref{fig_mc_image_v} shows the time evolution of the spatial distribution of tracer particles in the two representative clouds, C10 (left) and C31 (right). The corresponding movies for these clouds and others can be found online\footnote{\url{http://www.erda.dk/vgrid/ramses/}}. The tracer particles are initially all blue. We change their color to red if they are accelerated and their velocity (relative to the cloud center of mass) becomes $> 10\,\rm km\,\rm s^{-1}$, which is usually the case when a SN remnant hits a cloud. In order to visually keep track of what fraction of the cloud mass is accelerated by SNe at least once, red tracers remain red even if their velocity were to decrease again to a value $< 10\,\rm km\,\rm s^{-1}$, which is often the case (e.g. 96\% and 38\% of red tracers in the fourth time frame of clouds C10 and C31 respectively). In the left column, the time frames of cloud C10 are a series between the first significant kinetic energy peak caused by a SN internal to the cloud ($t_{\rm SN}=66.16 \,\rm Myr$, $d_{\rm SN}=2.06$ $\rm pc$, $R_{\rm cl}=5.54$ $\rm pc$) and the minimum kinetic energy before the following SN peak. The time interval between frames is $\sim 0.4\,\rm Myr$. In the right column, the time frames for cloud C31 correspond to four internal SNe ($t_{\rm SN}=72.16$, $72.95$, $74.23$, $75.82 \;\rm Myr$, $d_{\rm SN}=11.39$, $10.77$, $9.79$, $13.42 \;\rm pc$, $R_{\rm cl}=20.70$, $22.45$, $29.75$, $44.73 \;\rm pc$) that are responsible for the cloud dispersal. 

\begin{figure*}
\centering
\includegraphics[width=0.49\textwidth]{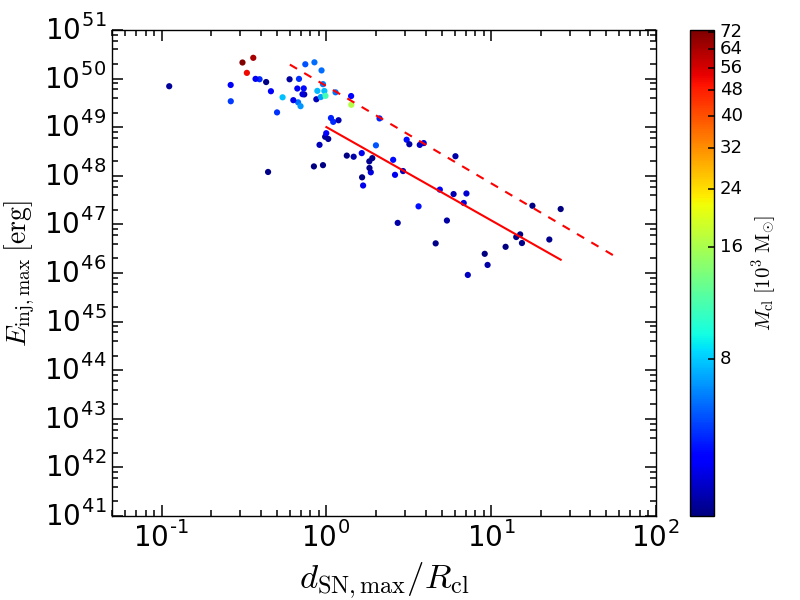}
\includegraphics[width=0.49\textwidth]{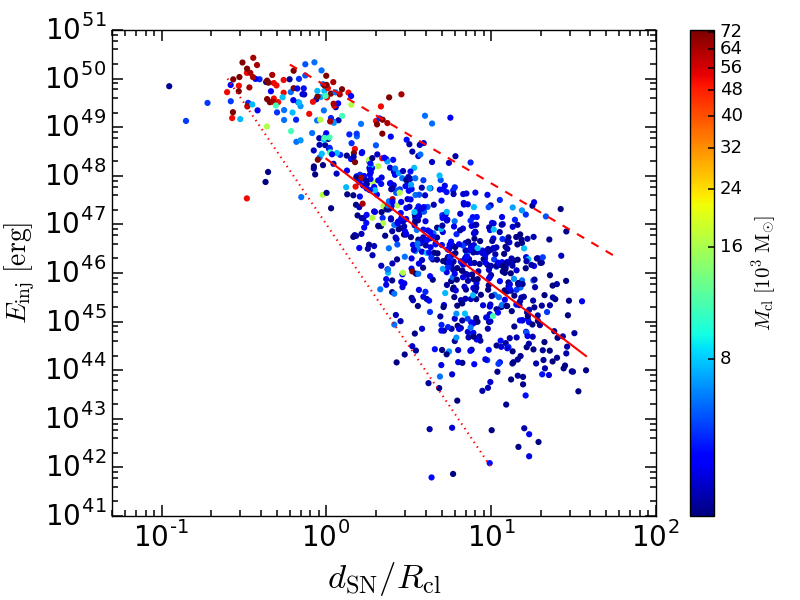}
\caption{Left panel: Maximum kinetic energy, $E_{\rm inj,max}$, injected by a single SN in each cloud versus the distance of that SN to the cloud center of mass, $d_{\rm SN,max}$, in units of the cloud radius, $R_{\rm cl}$.   
The colors correspond to the cloud mass. The red solid line is a least-square-fit to all points with $d_{\rm SN}/R_{\rm cl}>1$, and the red dashed line is the prediction for the Sedov-Taylor stage of the remnant. Right panel: Same as left panel, but included all energy peaks (may be several per cloud). }
\label{fig_kinetic_inj_vs_d_sn}
\end{figure*}

To further analyze how the energy from a SN is injected in a cloud and redistributed in the form of turbulent kinetic energy, we focus on a single peak of kinetic energy in each of the two representative clouds, and compute the cumulative distribution function (CDF) of kinetic energy and the velocity normalized probability distribution function (PDF) at four  different times: before the energy peak, at the energy peak, and during the decay from the peak. These quantities are shown in Figure \ref{fig_v_cdf}. 

The CDFs (top panels of Figure \ref{fig_v_cdf}) are plots of the cloud mass fraction versus the kinetic energy fraction, to show if the energy injected by the SN is affecting only a small part of the cloud or is distributed to most of the cloud. One can see that, before the energy peak (green line), the kinetic energy is distributed rather uniformly across the clouds, with 60\% of the energy contained in approximately 80\% of the cloud mass. On the other hand, at the energy peak (red line) 80\% of the kinetic energy is contained in only 1-2\% of the cloud mass, showing that only a small portion of the cloud has been accelerated due to the interaction with the SN. After approximately 0.5 Myr (cyan line), the energy is again distributed much more uniformly, and it is even closer to the original distribution after 1 Myr. 

The velocity PDFs in the lower panels of Figure \ref{fig_v_cdf} show that, at the time of the peak, the kinetic energy is dominated by a tail of high-velocity tracers (up to several hundreds km\,s$^{-1}$ in the case of the cloud C10), which is then rapidly reabsorbed into a smoother PDF shape at later times. In the case of a low-mass cloud and a SN very close to the cloud center (the peak in cloud C10), once the high-velocity tail is gone, the whole PDF has shifted to larger velocities compared to the PDF prior to the energy peak, showing that, after the initial energy decay, part of the energy injected by the SN has been retained and distributed through the whole cloud, resulting in a significant increase in the cloud turbulent energy. In the case of a further-away SN affecting a much more massive cloud (the peak in cloud C31), the velocity PDF is not shifted significantly to higher velocity after the velocity tail has disappeared, consistent with the fact that the total kinetic energy has decayed again to a value close to that prior to the SN, and the cloud is gravitationally bound again. Only a sequence of internal SNe can fully disperse such a massive cloud, which occurs at later times for the cloud C31.

To characterize the effect of SNe on all the MCs in the sample, we first consider the most-significant energy peak in each cloud prior to the time when the cloud is dispersed, which usually corresponds to the closest SN to the cloud center. We compute the energy injected by the corresponding SN, $E_{\rm inj}$, as the difference between the peak energy and the energy just before the peak (in practice the minimum value of the kinetic energy closest to and preceding the peak, because the kinetic energy is usually decaying between peaks). The most-significant peak (prior to the cloud dispersal) is simply the one with the largest injected energy, $E_{\rm inj,max}$. 
The value of $E_{\rm inj,max}$ for each cloud is plotted versus the distance of the corresponding SN, $d_{\rm SN,max}$, expressed in units of the cloud radius, $R_{\rm cl}$, in the left panel of Figure \ref{fig_kinetic_inj_vs_d_sn}.
The right panel shows the kinetic energy injection measured at all energy peaks prior to the dispersal of each cloud. 

The total kinetic energy in a supernova remnant (SNR) is conserved during the adiabatic phase is conserved \citep[e.g.][]{Shu92}, and based on an integration of the Sedov-Taylor solution, 20-30\% of the SN energy is expected to be in the kinetic form \citep{Spitzer68,Chevalier74,Walch+2015MNRAS}.
Considering the cross section, $\pi$, of an MC with the SNR, an estimate of the total amount of kinetic energy available to be injected to the cloud is given by  
\begin{equation}
E_{\rm inj,ST} = 0.28\,E_{\rm SN}\, (2 \,d_{\rm SN}/R_{\rm cl})^{-2} \, ,
\label{eq_e_st}
\end{equation}
where $E_{\rm SN}=10^{51}\,\rm erg$, and the coefficient, 0.28, is adopted from \cite{Chevalier74}. A seen in Figure \ref{fig_kinetic_inj_vs_d_sn}, this equation, shown as the red dashed line, overestimates the measured kinetic energy jumps in the MCs. This is perhaps expected since Equation~(\ref{eq_e_st}) represents the total kinetic energy available for injection, and the actual amount energy that an MC may receive from an SN impact may depend on the details of the interaction between the cloud and the remnant.  For example, the density of cloud may play a role in the interaction as the radiative cooling depends on the density. The fraction of kinetic energy that may be transferred to the cloud may also depend on the geometry of the cloud. A bow shock is expected to form when the SN shock encounters the cloud, and a significant fraction of kinetic energy in the SN shock may simply pass by the cloud through the oblique shock.  
 
The dashed line in the left panel of Figure \ref{fig_kinetic_inj_vs_d_sn} is the least-square fit, 
\begin{equation}
E_{\rm inj,max} = 0.04\,E_{\rm SN}\, (2 \,d_{\rm SN,max}/R_{\rm cl})^{-1.92}\, ,
\label{eq_e_inj_max}
\end{equation}
to the data points for the strongest SN peaks of all the clouds.  Equation (\ref{eq_e_inj_max}) for $E_{\rm inj,max}$  is smaller by a factor of 7 than Equation (\ref{eq_e_st}), meaning that, on average, only $\sim 15\%$ of the kinetic energy available from an energetic SN shock manages to enter an MC. 

\begin{table*}
\centering
\caption{List of the properties of two representative MCs and the distance of the most significant SN of the two MCs. From left to right: cloud number, cloud mass, virial parameter, SN rate within $25\,\rm pc$, SN rate within $50\,\rm pc$, total energy injection, the ratio of total energy injection to initial potential energy, distance of the most significant SN, and the energy injection by the most significant SN. The same quantities are listed for all clouds in Table \ref{tab_2_all} of Appendix \ref{app_table}.}
\begin{tabular}{lccccccccc}
\hline
\hline
$\rm Cloud$ & $M_{\rm cl} \,\rm [M_{\odot}]$ & $\alpha_{\rm vir,min}$ & $f_{\rm SN,25} \,\rm [Myr^{-1}]$ & $f_{\rm SN,50} \,\rm [Myr^{-1}]$ & $E_{\rm inj,tot} \,\rm [erg]$ & $E_{\rm inj,tot}/E_{\rm g}$ & $d_{\rm SN,max} \,\rm [pc]$ &  $E_{\rm inj,max} \,\rm [erg]$ \\
\hline
C10 & 3.25 $ \times 10^3 $ & 0.70 & 0.32 & 1.28 & 9.87 $ \times 10^{49} $ & 9.67 $ \times 10^2 $ & 2.06 & 9.87 $ \times 10^{49} $ \\
C31 & 5.22 $ \times 10^4 $ & 1.45 & 1.61 & 2.76 & 9.66 $ \times 10^{50} $ & 1.63 $ \times 10^2 $ & 9.79 & 1.31 $ \times 10^{50} $ \\
\hline
\hline
\end{tabular}\\
\label{table_energy_injection}
\end{table*}

The kinetic energy jumps of all SNe within the cloud lifetime are well below the Sedov-Taylor energy (see right panel of Figure \ref{fig_kinetic_inj_vs_d_sn}). The least-square fit gives  
\begin{equation}
E_{\rm inj} = 0.01\,E_{\rm SN}\, (2 \,d_{\rm SN}/R_{\rm cl})^{-2.58} \, ,
\end{equation}
that is a factor of 28 below the Sedov-Taylor energy at $d_{\rm SN}=R_{\rm cl}$, and a factor $\sim 100$ below for $d_{\rm SN}=10 \, R_{\rm cl}$. The range of cloud distances is approximately the same as for 
the most significant energy peaks, showing that the SN distance from the cloud center is not the only parameter controlling the efficiency of the kinetic energy injection, $E_{\rm inj} /E_{\rm SN}$. 
The lower envelope of the data points in the right panel roughly behaves as $\propto (d_{\rm SN}/R_{\rm cl})^{-5}$, as indicated by the dotted line. The $d_{\rm SN}^{-5}$ dependence 
may be explained by the late phase of an SNR evolution where the momentum is approximately conserved. In this phase, the velocity of the shock, $v_{\rm sh}$, decreases with time as $t^{-3/4}$ 
or with the remnant radius, $R_{\rm snr}$, as  $R_{\rm snr}^{-3}$.  Therefore, the total kinetic energy in the entire SN shell goes like, $\propto \frac{4 \pi}{3} \, R_{\rm snr}^{3} \, v_{\rm sh}^2 \propto R_{\rm snr}^{-3}$. 
Considering the cross section of an MC cloud with the remnant, we expect that the kinetic energy jump due to an SNR in the momentum conservation phase scales as $R_{\rm snr}^{-5}$, or equivalently as  $d_{\rm SN}^{-5}$. 
Evidently, due to the density, the geometry  of MCs, as well as the evolution stage of the SNR,  the fraction of  explosion energy of an SN  that  
is converted to kinetic energy in an MC can vary a lot, even at the same distance to the cloud center. In other words, the energy injection is strongly dependent on the details of the SNR and the physical and geometric conditions  of the MC, as pointed out in previous works.

\begin{figure}
\centering
\includegraphics[width=0.48\textwidth]{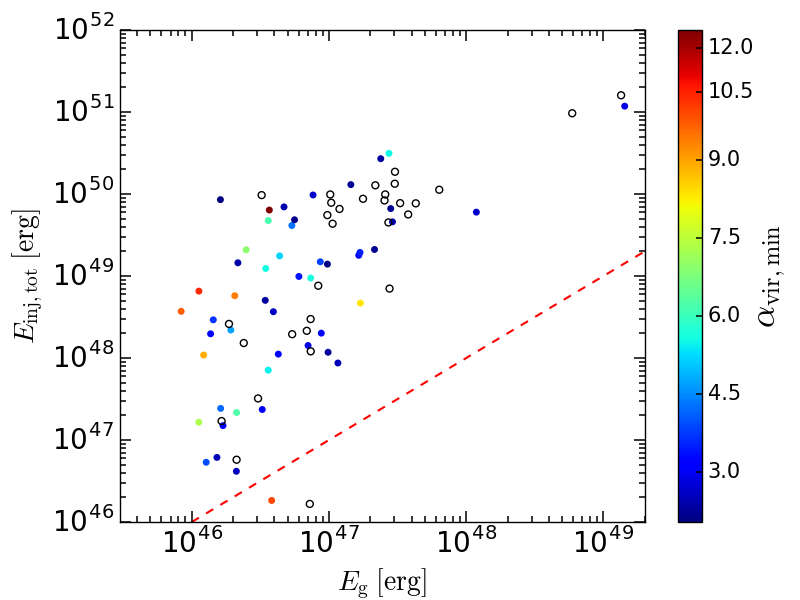}
\caption{Total kinetic energy injected prior to a cloud dispersal versus initial gravitational energy. The colors correspond to the virial parameter of the clouds $\alpha_{\rm vir}$, with black empty circles denoting bound MCs, with virial parameter $\alpha_{\rm vir,min} \leq 2$. The dashed line corresponds to $E_{\rm inj,tot}=E_{\rm g}$.}
\label{fig_kinetic_inj_vs_Eg}
\end{figure}

Despite the large energy losses prior to the interaction of most remnants with the MCs, and the initially rapid decay of the highest energy peaks, initially bound MCs are still dispersed by SNe, thanks for the large number of SNe during a cloud lifetime. Figure \ref{fig_kinetic_inj_vs_Eg} shows that the total kinetic energy injected in a cloud prior to its dispersal is typically $\sim 100$ times larger than its initial gravitational energy. Thus, even if the energy efficiency of the feedback were only 1\%, it would still be enough to unbound the clouds. More than half of the clouds are unbound, and would disperse anyway, but even unbound clouds are injected, during their lifetime, a total amount of kinetic energy well in excess of their initial potential energy.

The quantities discussed above, as well as SN frequencies, are listed in Table~\ref{table_energy_injection} for our two representative clouds, C10 and C31, while the full version of this table with values for all clouds are given in Table~\ref{tab_2_all} of Appendix~\ref{app_table}. 

\subsection{MC Lifetime} \label{MC_lifetime}

We define the cloud lifetime based on the time evolution of the cloud radius, as in \citet{Padoan+SN1+2016ApJ}. The cloud lifetime is assumed to have started when its radius was twice its minimum value, and to end when the cloud has expanded to double again its minimum radius. While all other data-analysis in this work has been carried out from the time of cloud selection onward, in the case of the cloud evolution we must follow the tracer particle positions both backward and forward in time from the cloud selection time. 

\begin{figure}
\centering
\includegraphics[width=0.47\textwidth]{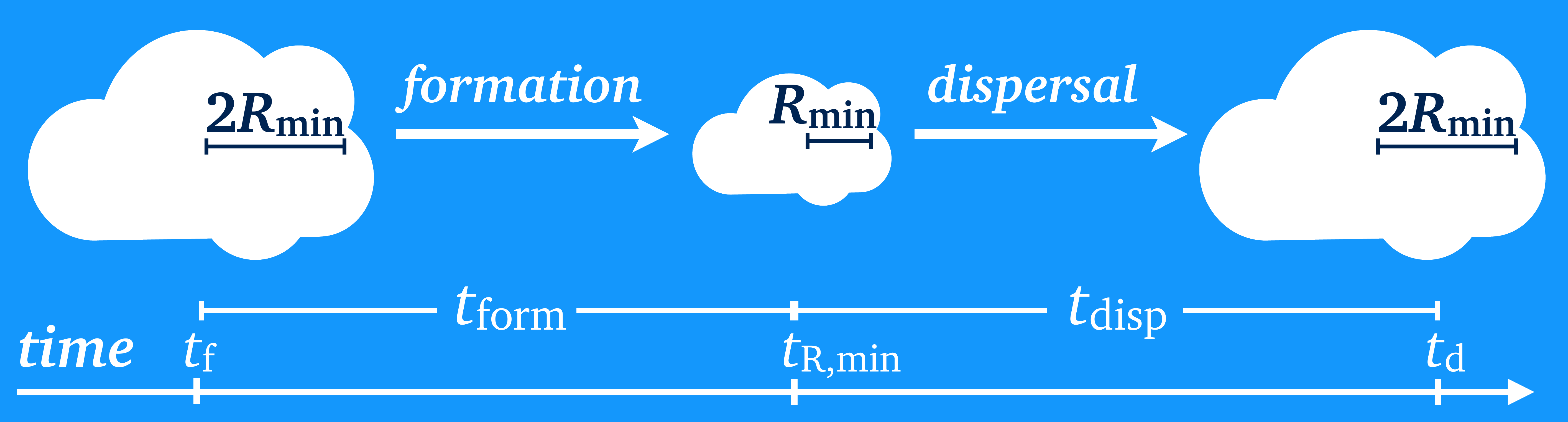}
\caption{Sketch of the definition for the cloud formation and dispersal time.} 
\label{fig_cloud_evol}
\end{figure}

\begin{table}
\centering
\caption{List of timescales for two representative MCs. From left to right: cloud number, cloud formation time, cloud dispersal time, cloud lifetime, cloud delay time and the cloud dynamical time. The same quantities are listed for all clouds in Table \ref{tab_lifetime_all} in Appendix \ref{app_table}.}
\begin{tabular}{lccccc}
\hline
\hline
$\,\rm Cloud$ & $t_{\rm form}$ & $t_{\rm disp}$ & $t_{\rm life}$ & $t_{\rm delay}$ & $t_{\rm dyn,\alpha}\,$ \\
\, & $\rm [Myr]$ & $\rm [Myr]$ & $\rm [Myr]$ & $\rm [Myr]$ & $\rm [Myr]\,$ \\
\hline
\,C10 & 7.41 & 3.02 & 10.43 & 0.85 & 4.88\, \\
\,C31 & 16.55 & 3.52 & 20.07 & 1.26 & 6.79\, \\
\hline
\hline
\end{tabular}
\label{tab_lifetime}
\end{table}

We first find the time, $t_{\rm R,min}$, when the cloud radius has its minimum value, $R_{\rm min}$. This time is typically less than two Myr from the time of the snapshot where the cloud was identified. We then search for the times when the cloud radius is $2\,R_{\rm min}$ before and after $t_{\rm R,min}$, and refer to them as $t_{\rm f}$ and $t_{\rm d}$ respectively. The radius increases nearly monotonically going both backward and forward in time from $t_{\rm R,min}$, so these two times are well defined.
We then define the cloud formation time as the time interval during which the cloud contracts, starting from a radius equal to twice its minimum value,
\begin{equation}
t_{\rm form} = t_{\rm R,min} - t_{\rm f} \, ,
\end{equation}
and the cloud dispersal time as the time interval during which the cloud expands up to twice the minimum radius,
\begin{equation}
t_{\rm disp} = t_{\rm d} - t_{\rm R,min} \, .
\end{equation}
Finally, we define the cloud lifetime as the sum of the contraction and expansion phases:
\begin{equation}
t_{\rm life} = t_{\rm d} - t_{\rm f} = t_{\rm form} + t_{\rm disp} \, .
\end{equation}
The definition of the three times and the corresponding three time intervals is illustrated by the sketch in Figure~\ref{fig_cloud_evol}, and the values of the time intervals are given in Table~\ref{tab_lifetime} for our two reference clouds (see Table \ref{tab_lifetime_all} in Appendix \ref{app_table} for the values in the other clouds). The table also includes the values of $t_{\rm dyn,\alpha}$ and the time interval, $t_{\rm delay}$, between the most significant kinetic energy peak, $t_{\rm p,max}$, corresponding to $E_{\rm inj,max}$ defined in \S~\ref{sec_energy_injection}, and the time of dispersal, $t_{\rm d}$:
\begin{equation}
t_{\rm delay} = t_{\rm d} - t_{\rm p,max} \, .
\end{equation}

For 28 of the 83 clouds (particularly among those from the first two selection times) the radius has not doubled from its minimum value backward in time before the first available snapshot of the simulation (approximately 8 and 11~Myr for the first two selection times), so we cannot determine $t_{\rm f}$ and we can only set a lower limit to the formation time, $t_{\rm form}$, and to the lifetime, $t_{\rm life}$. We also have two clouds for which the radius has not expanded beyond twice its minimum value before the last available snapshot, so $t_{\rm d}$ could not be determined and we can also measure a lower limit for $t_{\rm disp}$. Thus, we have a total of 53 clouds with values of $t_{\rm life}$, and 30 clouds with only lower limits for $t_{\rm life}$.  

\begin{figure}
\centering
\includegraphics[width=0.48\textwidth]{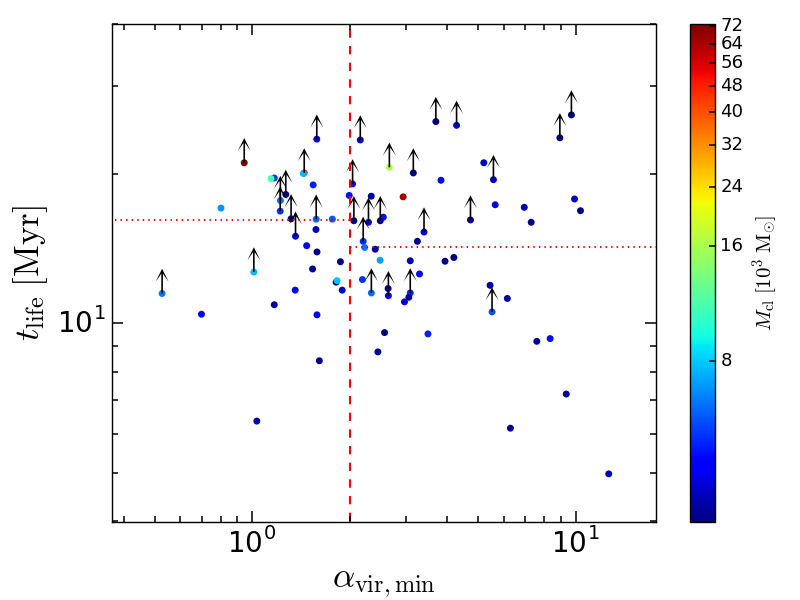}\\
\includegraphics[width=0.48\textwidth]{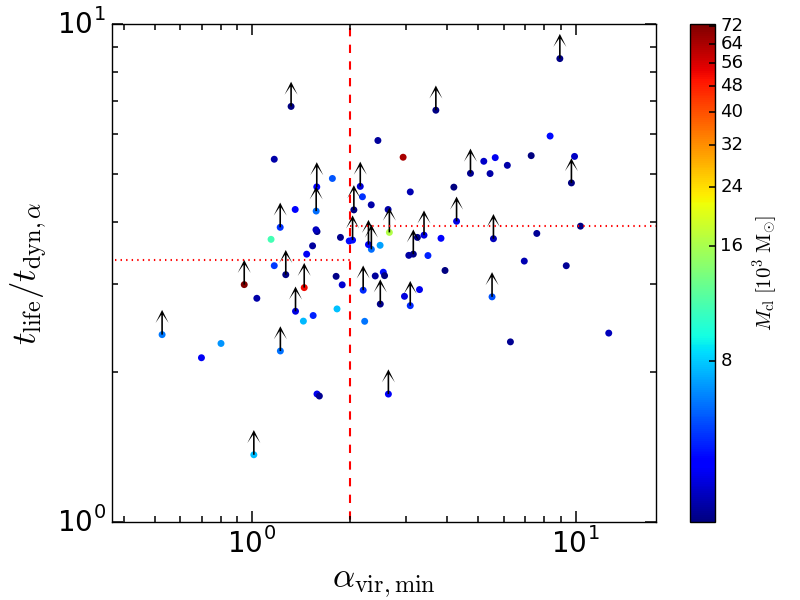}
\caption{Upper panel: Dependence of the cloud lifetime, $t_{\rm life}$, on the minimum virial parameter, $\alpha_{\rm vir,min}$, with colors indicating the cloud mass, $M_{\rm cl}$. The arrows mark the lower limits (clouds where $t_{\rm f}$ or $t_{\rm d}$ could not be measured), and the dotted horizontal lines are the median values for bound (to the left of the vertical dashed line) and unbound (to the right of the vertical dashed line) clouds based on the last two selection times, 16.1~Myr and 14.2~Myr, respectively. Lower panel: The same as the upper panels, but with $t_{\rm life}$ in units of $t_{\rm dyn,\alpha}$. The median values from the last two selection times are 3.36 and 3.93 for bound and unbound clouds, respectively.}
\label{fig_lifetime}
\end{figure}

\begin{figure}
\centering
\includegraphics[width=0.48\textwidth]{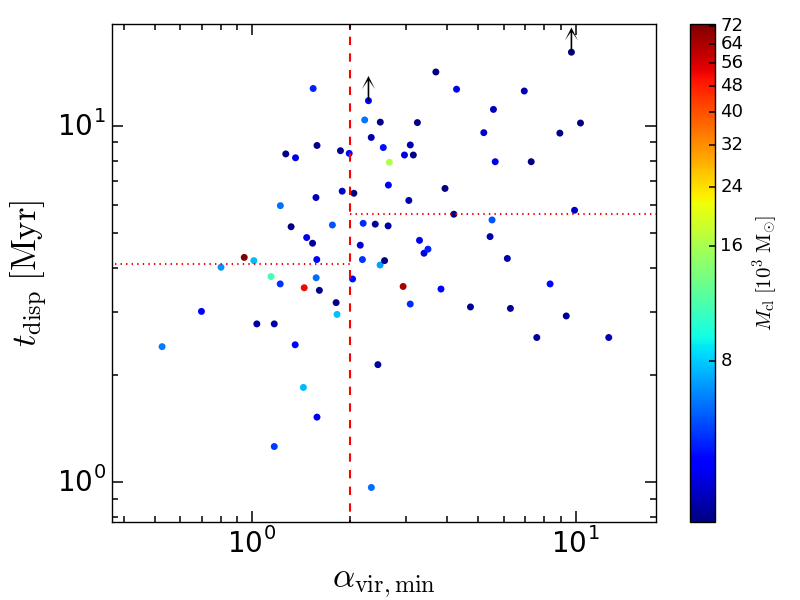}\\
\includegraphics[width=0.48\textwidth]{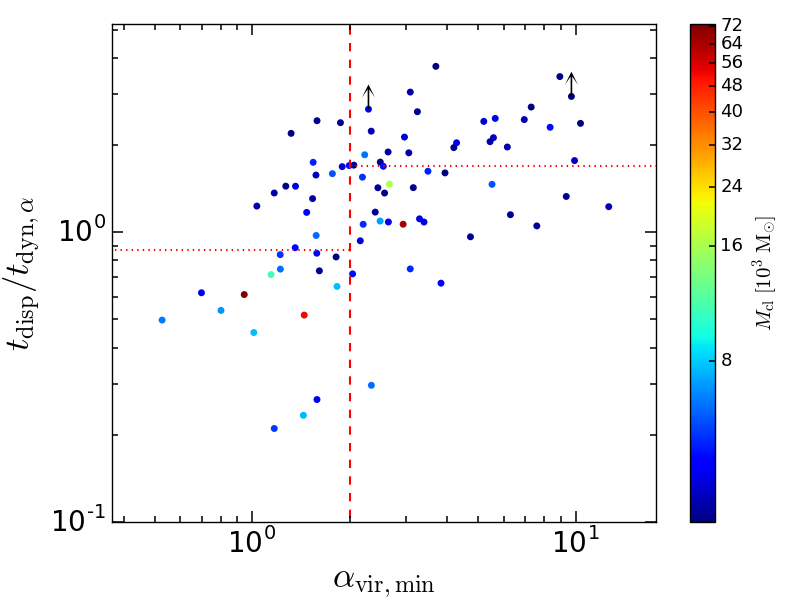}
\caption{The same as in Figure~\ref{fig_lifetime}, but for the dispersal time, $t_{\rm disp}$. Here the medians are evaluated using all four selection times, because only two clouds are missing a determination of $t_{\rm d}$. The median values are 4.1~Myr and 5.7~Myr for bound and unbound clouds, respectively. The corresponding median values of $t_{\rm disp} /t_{\rm dyn,\alpha}$ are 0.87 and 1.69.}
\label{fig_dispersion_time}
\end{figure}

\begin{figure}
\centering
\includegraphics[width=0.48\textwidth]{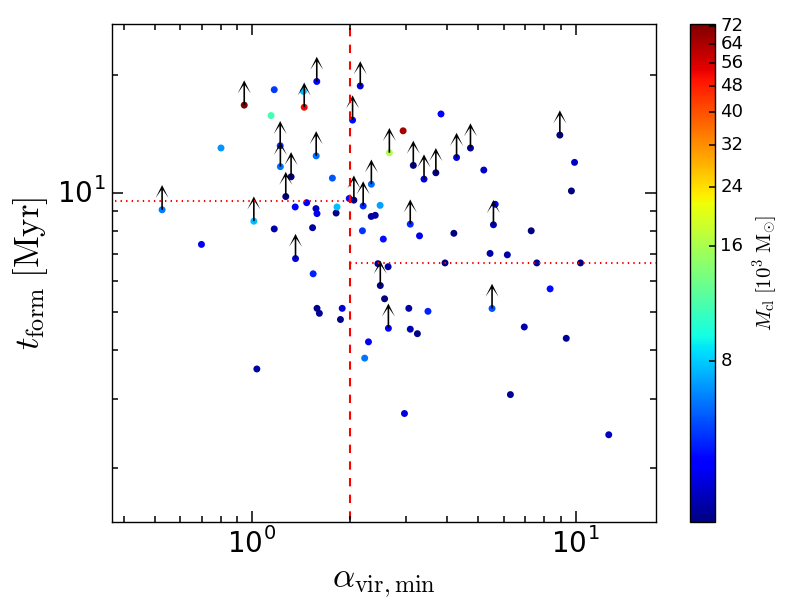}\\
\includegraphics[width=0.48\textwidth]{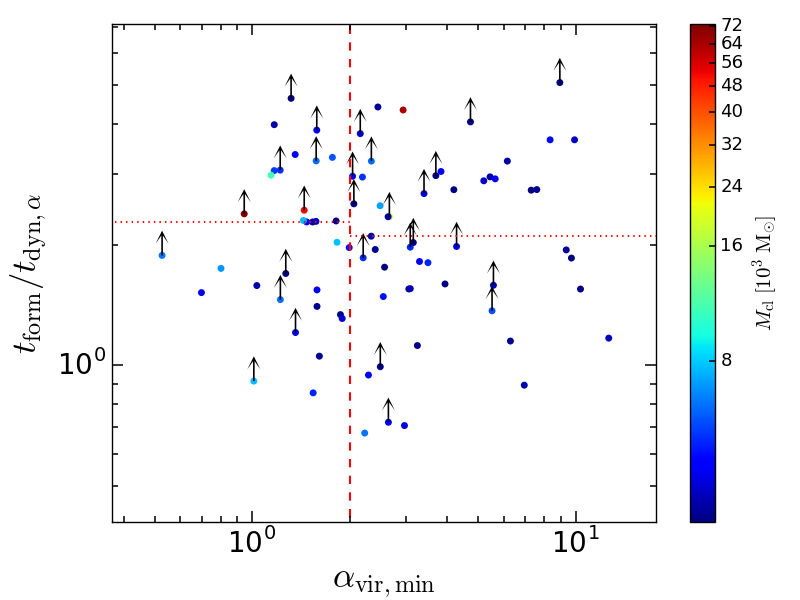}
\caption{The same as in Figure~\ref{fig_lifetime}, but for the formation time, $t_{\rm form}$. The medians, evaluated using only the last two selection times, are 9.6~Myr and 6.7~Myr for bound and unbound clouds, respectively. The corresponding median values of $t_{\rm form} /t_{\rm dyn,\alpha}$ are 2.28 and 2.1.}
\label{fig_formation_time}
\end{figure}

\begin{figure*}
\centering
\includegraphics[width=0.48\textwidth]{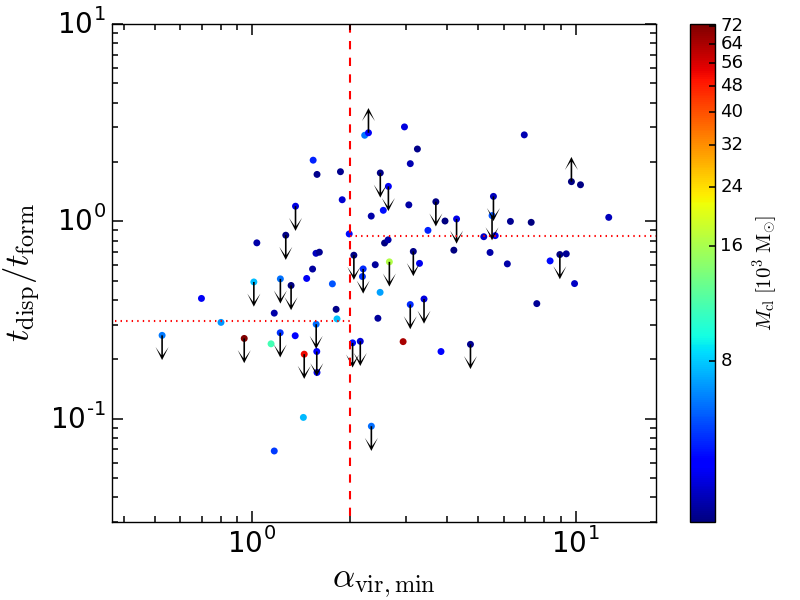}
\includegraphics[width=0.48\textwidth]{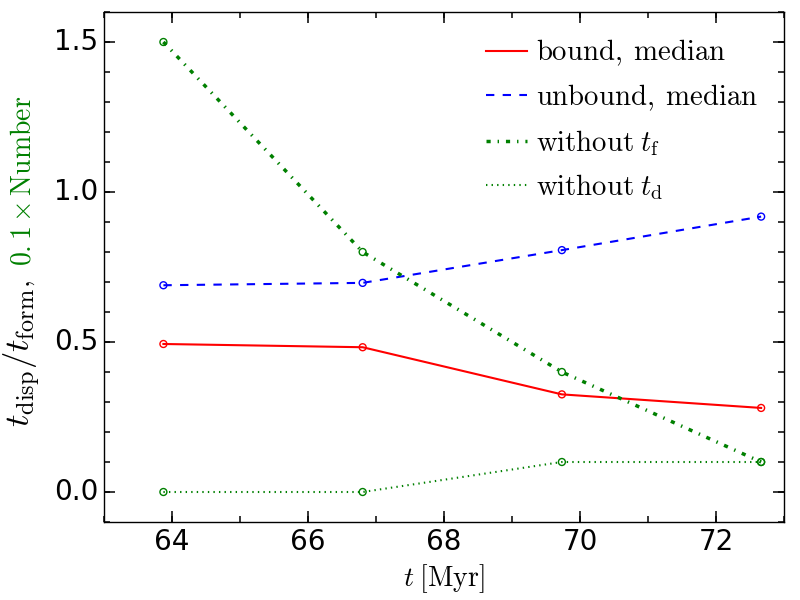}
\caption{Left panel: Ratio of cloud dispersal to formation time, $t_{\rm disp}/t_{\rm form}$, versus minimum virial parameter, $\alpha_{ \rm vir,min}$.
The dotted horizontal lines are the median values for bound (to the left of the vertical dashed line) and unbound (to the right of the vertical dashed line) clouds based on the last two selection times, 0.31 and 0.85, respectively. Right panel: The median values of $t_{\rm disp}/t_{\rm form}$ for bound (solid line) and unbound (dashed line) clouds at the four cloud selection times. The numbers of clouds without a determination of $t_{\rm f}$ of $t_{\rm d}$, multiplied by 0.1, are shown by the dash-dotted and dotted lines, respectively.}
\label{fig_ratio_t_disp_to_t_form}
\end{figure*}

\begin{figure}
\centering
\includegraphics[width=0.48\textwidth]{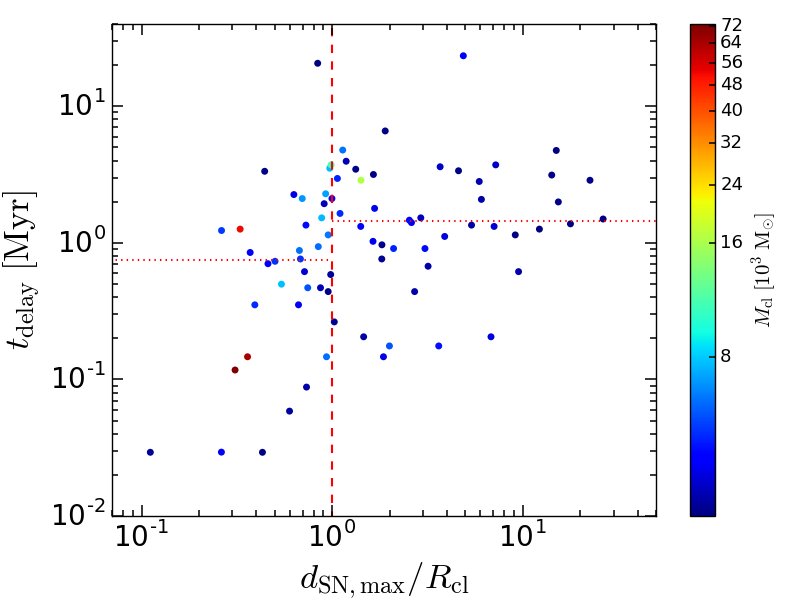}\\
\includegraphics[width=0.48\textwidth]{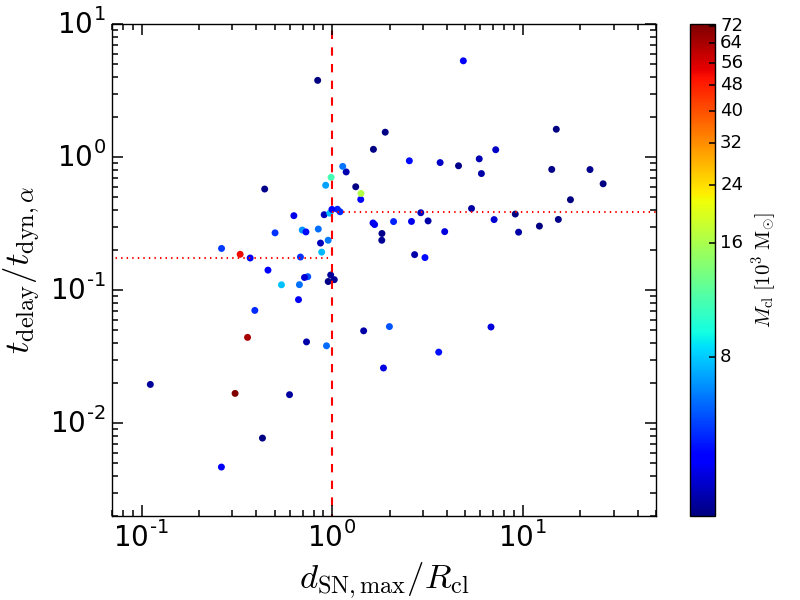}
\caption{Upper panel: Delay time between the most significant kinetic-energy peak and the cloud dispersal time, $t_{\rm delay}$, versus the distance of the corresponding SN to the cloud center, in units of the cloud radius. The dotted horizontal lines are the median values of $t_{\rm delay}$ for internal ($d_{\rm SN,max}/R_{\rm cl}<1$) and external ($d_{\rm SN,max}/R_{\rm cl}>1$) SNe, 0.75~Myr and 1.44~Myr, respectively (the median of $t_{\rm delay}$ for all the clouds is 1.25~Myr). Lower panel: The same as the upper panel, but with $t_{\rm delay}$ in units of the dynamical time, $t_{\rm dyn,\alpha}$. The median values of $t_{\rm delay}/t_{\rm dyn,\alpha}$ are 0.18 and 0.38 for internal and external SNe (the median of $t_{\rm delay}/t_{\rm dyn,\alpha}$ for all the clouds is 0.29).} 
\label{fig_t_delay_vs_d_sn}
\end{figure}

The lifetime of all clouds in the sample is shown in Figure~\ref{fig_lifetime} in units of Myr in the upper panel, and in units of the dynamical time, $t_{\rm dyn,\alpha}$, in the lower panel (see \S~\ref{sec_alpha_vir} for the justification of adopting the corresponding value of the dynamical time measured at the time of minimum virial parameter and the minimum virial parameter). The scatter plots show the dependence of $t_{\rm life}$ and $t_{\rm life}/t_{\rm dyn,\alpha}$ on $\alpha_{\rm vir,min}$. The lifetime spans a wide range of values between approximately 5 and 20~Myr, corresponding to approximately to 2 to 7 dynamical times. 
There are only 53 clouds for which we have measured the value of $t_{\rm life}$. To estimate the median values, we include lower limits, but consider only the last two selection times, for which only a small fraction of the cloud lifetimes are upper limits. With that choice, the median values of $t_{\rm life}$ are 16.1 and 14.2~Myr for bound and unbound clouds, respectively, while the median for all clouds (in the final two selection times) is 15.4~Myr. In units of the cloud dynamical times, the median computed in the same way are 3.36 and 3.93 for bound and unbound clouds, respectively, and 3.42 from all clouds. 

The dispersal times for all the clouds in the sample are shown in Figure~\ref{fig_dispersion_time} in units of Myr in the upper panel, and in units of the dynamical time, $t_{\rm dyn,\alpha}$, in the lower panel. The scatter plots show the dependence of $t_{\rm disp}$ and $t_{\rm disp}/t_{\rm dyn,\alpha}$ on $\alpha_{\rm vir,min}$. The dispersal time spans a wide range of values above and below the cloud dynamical time, between approximately 1 and 20~Myr. The figure also shows a trend of increasing $t_{\rm disp}$ (or $t_{\rm disp}/t_{\rm dyn,\alpha}$) with increasing $\alpha_{\rm vir,min}$.     
The median values of $t_{\rm disp}$ are 4.1 and 5.7~Myr for bound and unbound clouds respectively (dotted horizontal lines in the upper panel of Figure~\ref{fig_dispersion_time}), while the median values of $t_{\rm disp}/t_{\rm dyn,\alpha}$ are 0.87 and 1.69, also for bound and unbound clouds, respectively (dotted horizontal lines in the lower panel of Figure~\ref{fig_dispersion_time}). 

Figure~\ref{fig_formation_time} shows the same as Figure~\ref{fig_dispersion_time}, but for the formation time, $t_{\rm form}$. The values of $t_{\rm form}$ are clearly larger than those of $t_{\rm disp}$, with a slight trend of decreasing $t_{\rm form}$ with increasing $\alpha_{\rm vir,min}$ (opposite to the trend for $t_{\rm disp}$). The median values of $t_{\rm form}$ are 9.6~Myr for bound clouds, and 6.7~Myr for unbound ones. This slight dependence on the virial parameter is expected because the virial parameter tends to decrease with increasing cloud mass, and the dynamical time increases with cloud mass, hence should decrease with increasing virial parameter. Because the clouds are assembled by the large-scale turbulence and tend to follow a velocity-size relation consistent with the observed Larson relation, we expect the formation time to scale with the dynamical time and so to decrease with increasing virial parameter. As the formation time should scale with the dynamical time, the ratio of formation time and dynamical time should not depend on the virial parameter, which is in fact confirmed by the lower panel of Figure~\ref{fig_formation_time}. The median values of $t_{\rm form}/t_{\rm dyn,\alpha}$ for bound and unbound clouds are nearly identical, 2.28 and 2.1, respectively.

The ratio between dispersal and formation times, $t_{\rm disp}/t_{\rm form}$, is an important quantity to gauge the effect of the SN feedback on the cloud evolution. \citet{Padoan+SN1+2016ApJ} found that formation and dispersal times of the clouds in their simulation were on average the same, both of the order of two dynamical times. As mentioned above, in that simulation the SN feedback was implemented with randomly-generated SNe. In the current simulation, when star formation is included and the timing and position of SNe (relative to the MCs) are accounted for self-consistently, we find that the $t_{\rm form}\sim 2 \, t_{\rm dyn,\alpha}$, as in \citet{Padoan+SN1+2016ApJ}, while the dispersal time is significantly shorter, particularly for bound clouds where the median value is $t_{\rm disp}\sim 0.8 \, t_{\rm dyn,\alpha}$. This is an important result, because it shows the crucial role of locally-formed SNe in the dispersal of their parent MC. Because the star-formation rate is strongly correlated with the virial parameter \citep[e.g.][]{Krumholz+McKee05sfr,Padoan+Nordlund11sfr,Federrath+Klessen12,Padoan+12sfr,Padoan+17sfr}, locally-formed SNe are more likely to play a role in clouds with lower virial parameter (hence large star-formation rate). Thus, the dispersal time is shorter in bound clouds than in unbound ones. This result also stresses the importance of simulating the SN feedback in a self-consistent way (which we refer to as {\it real SNe}), which has been achieved in our simulation for the first time. 

The $t_{\rm disp}/t_{\rm form}$ is plotted in the left panel of Figure~\ref{fig_ratio_t_disp_to_t_form}. The horizontal dotted lines mark the median values computed from the last two selection times (where only a small fraction of the ratios are lower or upper limits). The median values are 0.31 and 0.85 for bound and unbound clouds respectively. The right panel of Figure~\ref{fig_ratio_t_disp_to_t_form} shows the variation of these median ratios for the four selection times. The dash-dotted and dotted lines show the decrease with time in the number of clouds with only lower limits for $t_{\rm form}$ and $t_{\rm disp}$ respectively. At the final selection time, we only have one cloud without a determination of $t_{\rm f}$ and one more cloud without a determination of $t_{\rm d}$, so the median values derived there are nearly unbiased, but the sample size from a single selection time is small. One can see that the $t_{\rm disp}/t_{\rm form}$ ratio for bound clouds is almost unchanged between the third and the fourth selection times, so the median value we obtain from the last two selection times should be a good approximation of the unbiased value. For unbound clouds, the ratio increases at a constant rate between the second and fourth selection times, reaching a value of 0.9, so the unbiased median may be very close to 1, as in the case of randomly-generated SNe \citep{Padoan+SN1+2016ApJ}.

The importance of locally-formed SNe is also stressed by the significant correlation between the time of the most significant kinetic energy peak, $t_{\rm p,max}$, and the dispersal time, $t_{\rm d}$, in each cloud. The interval between the two times is measured by $t_{\rm delay}$, which is plotted in Figure~\ref{fig_t_delay_vs_d_sn} as a function of the distance, $d_{\rm SN,max}$, of the SN responsible for the most significant peak from the cloud center, in units of the cloud radius, $R_{\rm cl}$. The delay time is usually very short (in some case even less than one percent of a dynamical time) when the SN occurs within one cloud radius, with a median value of 0.75~Myr, or $0.18\, t_{\rm dyn,\alpha}$. 

\subsection{Overdensity of Internal SNe} \label{sec_overdensity}

As discussed above, locally-formed SNe are responsible for the short dispersal time of MCs with the highest star-formation rate. They are also part of the reason why SNe appear to have a more important role in our simulation than in previous studies, where SNe were generated randomly or clustered, but never with self-consistent timing and positions relative to the MCs. Thus, it is important to quantify the local overdensity of SNe relative to the case of a random distribution. To do so, we define as {\it internal} SNe those that occur inside a MC, $d_{\rm SN}\le R_{\rm cl}$, where $d_{\rm SN}$ is the SN distance to the cloud center and $R_{\rm cl}$ is the cloud radius, after the time of cloud selection and before the cloud is dispersed. As we will show below, the chance that an internal SN was not born in the cloud is very small, so we can safely assume that the majority of the internal SNe comes from stars that were formed locally. We aim at computing the distribution of overdensities of internal SNe relative to the case of SNe randomly generated with a uniform distribution (in which case a SN would be internal by chance, not because it is formed in that cloud).

Although all SNe come from stars that were formed in MCs, predicting this overdensity without a realistic simulation would be far from trivial, because the overdensity depends on the MC lifetimes relative to the stellar lifetimes, on the stellar mass distribution, on the star-formation history of each cloud, and on the rate of spatial diffusion of massive stars from their formation sites. For example, the median lifetime of the clouds is $t_{\rm life} \sim 15.4$~Myr, while the lifetime of the lowest-mass stars to explode as SNe is $\sim 40$~Myr. Therefore, a significant fraction of the SNe from stars formed in the MCs cannot be classified as internal to their parent clouds, as these have already been dispersed when those SNe occur. 

\begin{figure*}
\centering
\includegraphics[width=0.48\textwidth]{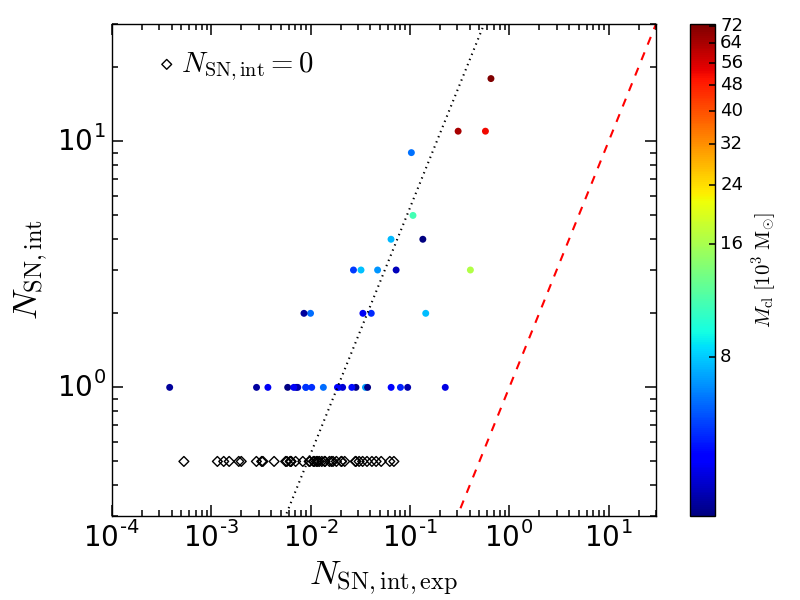}
\includegraphics[width=0.48\textwidth]{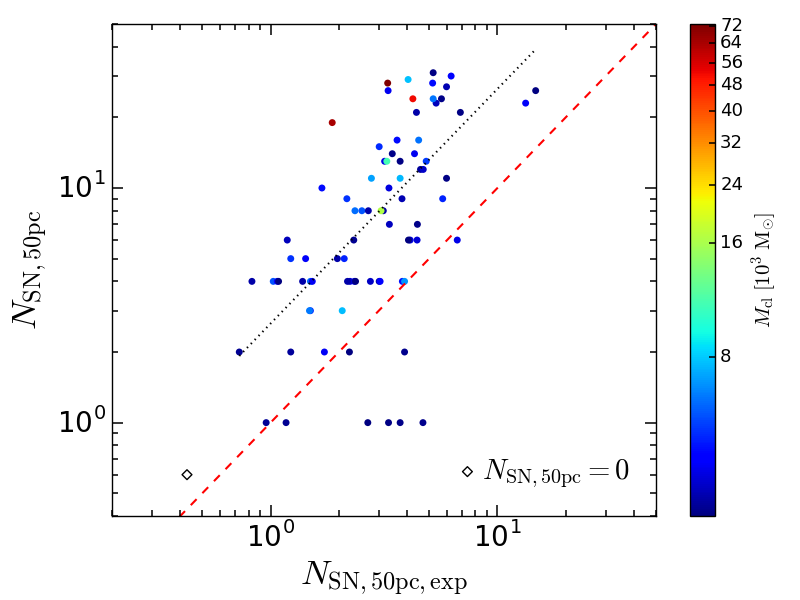}
\caption{Left panel: Number of SNe exploding within the MCs versus expected number of internal SNe assuming a uniform spatial distribution of SNe. There are 39 MCs with internal SNe and 44 without (black diamond symbol). The red dashed line corresponds to $N_{\rm SN,int}=N_{\rm SN,int,exp}$. The black dotted line shows the median ratio, $med(N_{\rm SN,int}/N_{\rm SN,int,exp})=53.7$. Right panel: The same as the left panel, but for SNe within 50 pc from the cloud center. The median ratio is 2.6.}
\label{fig_sn_int_num}
\end{figure*}

\begin{figure}
\centering
\includegraphics[width=0.48\textwidth]{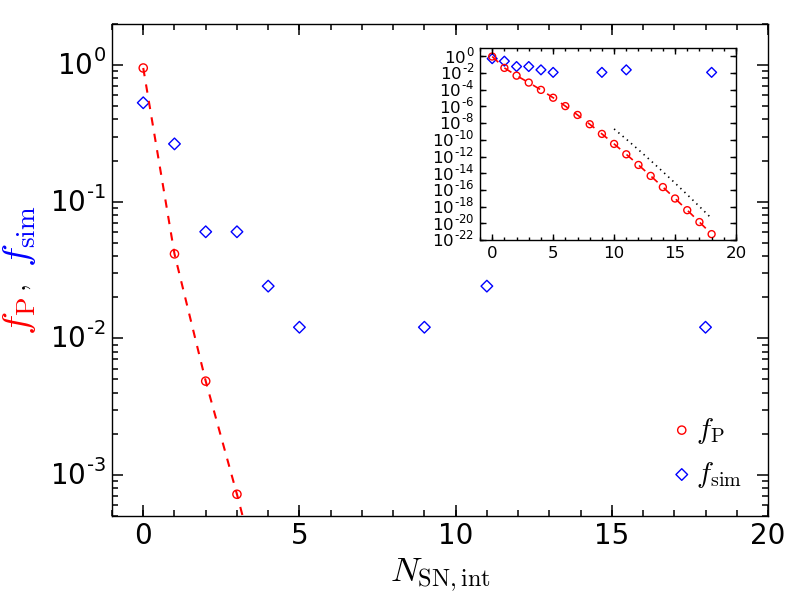}
\caption{The expected fraction of MCs with $k$ internal SNe from the case of a spatially uniform SN distribution ($f_{\rm P}$, red circles), and the fraction from the simulation ($f_{\rm sim}$, blue diamonds), as a function of the number of internal SNe. The difference in the MC fractions is striking even at one internal SN, and they diverge even more at higher $k$.}
\label{fig_sn_fraction}
\end{figure}

Figure~\ref{fig_sn_int_num} shows the number of internal SNe found in each of the MCs, versus the expected number if the SNe were uniformly distributed, both within an MC radius (left panel) and within 50~pc from an MC (right panel). The expected number of internal SNe in a cloud $i$, $N_{\rm SN,int,exp,i}$, is calculated from the following formula: 
\begin{equation}
N_{\rm SN,int,exp,i} = \frac{N_{\rm SN,s1,end}}{(t_{\rm end}-t_{\rm s1})} \times \frac{V_{\rm mean,i}}{V_{\rm box}} \times (t_{\rm d,i}-t_{\rm s,i}) \, ,
\end{equation}
where $N_{\rm SN,s1,end}$ is the number of SNe in the simulation from the first time we select MCs ($t_{\rm s1} \sim 63.8$~Myr) to the end of the simulation ($t_{\rm end}$), $V_{\rm mean,i}$ is the mean MC volume from its selection time ($t_{\rm s,i}$) to its dispersal time ($t_{\rm d,i}$), and $V_{\rm box}$ is the volume of the simulation box. For the expected SNe within 50~pc, we simply replace the mean MC volume by the volume of a 50~pc radius sphere. About half of the sample, 44 MCs, are without any internal SNe, and are represented by open symbols at $N_{\rm SN,int}=0.5$ (left panel), to compare with their expected values from the uniform distribution. As the expected number of SNe is 0.1 or below for these MCs, we would not expect to find any internal SNe. For the other half of the sample, 39 MCs, we have a clear overdensity of internal SNe, with a median value of 53.7 (see the dotted line in Figure~\ref{fig_sn_int_num}, left panel). Thus, in order for a uniform SN distribution to explain the number of internal SNe we find, the SN rate would have to be 53.7 times higher. Even when going out to 50~pc radius (right panel), we can still see a clear overdensity, even though the median value has reduced to 2.6 (see the dotted line in Figure~\ref{fig_sn_int_num}, right panel).

To further appreciate the meaning of the internal SNe overdensity shown in Figure~\ref{fig_sn_int_num} (left panel), we quantify the probability of its occurrence relative to the case of a uniform SN distribution. If the SNe were uniformly distributed, the probability of an MC $i$ of having $k$ internal SNe is given by the Poisson distribution, with the mean value, $\lambda$, equal to $N_{\rm SN,int,exp,i}$:
\begin{equation}
P_{\rm i}(k) = \frac{\lambda^{k} e^{-\lambda}}{k!} \, .
\end{equation}
If we assume that MCs are independent of each other, then the expected fraction of MCs in the sample that have $k$ internal SNe, $f_{\rm P}$, can be calculated as a mean of the individual MCs' Poisson probability distributions:
\begin{equation}
f_{\rm P}(k) = \sum_{i=1}^{N_{\rm C}} P_{\rm i}(k) / N_{\rm C} \, ,
\end{equation}
where $N_{\rm C}=83$ is the number of MCs in the sample. We calculate $f_{\rm P}$ for $k$ between 0 and 18, the range of values of internal SNe in the MCs.

Figure \ref{fig_sn_fraction} shows the expected fraction of MCs with $k$ internal SNe from the uniform SN distribution, $f_{\rm P}$, and the fraction in the sample, $f_{\rm sim}$, where $f_{\rm sim}(k) = N_{\rm C}(k)/N_{\rm C}$ and $N_{\rm C}(k)$ is the number of MCs with $k$ internal SNe. The overdensity is clear even at $k = 0$, as the uniform distribution would predict that 95\% of the sample should have no internal SNe at all, whereas our result is 53\%. The difference becomes even more dramatic at the higher values of $k$, as the prediction is just 5\%, or 4 MCs, having just one SN, rather than 39 MCs with one or more SNe. At $k=18$, the probability of so many internal SNe arising from the sample of MC sizes and lifetimes by uniform distribution of SNe is by a factor of $10^{-20}$ lower. Even if we assume that all of the MCs would have been as large as our largest MC (see the black dotted line in Figure~\ref{fig_sn_fraction} inset), this factor is still about $10^{-18}$. This shows that a uniform distribution of SNe is inadequate to capture the true distribution of internal SNe, which are the most important ones for the cloud dispersal, and hence underestimates the effect that the SNe have on their parent MCs.

We identify the SNe that originate from the massive stars born within the MCs by crossmatching the accreted tracer particles of the MCs to the progenitor sink particles of the SNe. We find that out of 273 SNe that explode after 63.8~Myr, the majority, 205, are born within the MCs of the sample. Out of those 205 SNe, 155 explode within 50~pc from the center of any of the MCs, and 55 within any MC's radius (these SN numbers are computed in both cases including only SNe occurring before the dispersal time of each MC, $t_d$). In the context of the parent MCs, 89 SNe explode during the parent MC's lifetime, 82 of them within 50~pc from the center of their parent MC, and 50 within their parent MC's radius. In other words, of a total 56 internal SNe (regardless of their birthplace), 50 explode within their parent MC, showing the importance of self-consistent modelling of star formation and SN feedback. Since we do not update the number of tracers after the selection of an MC, the other 5 SNe that are born in the MCs, but explode within an MC other than their parent one, may be due to the merging of the parent MC to the host MC after the host MC's selection time. These results are compiled in Table~\ref{table_internal_SNe}.

\begin{table}
\centering
\caption{The numbers of SNe born anywhere and in the MCs of the sample, exploding within the whole box (irrespective of any MC), within 50~pc of a MC or within a MC. The numbers in parenthesis are with regards to their parent MC specifically.  See the text for details.}
\begin{tabular}{lccc}
\hline
\hline
\; \; \; Explosion within & Whole box & 50~pc & MC\\
Birth & & & \\
\hline
Anywhere & 273 & 186 & 56 \\
In the MCs & 205 (89) & 155 (82) & 55 (50)\\
\hline
\hline
\end{tabular}
\label{table_internal_SNe}
\end{table}

\subsection{Decay of Kinetic Energy in MCs} \label{sec_decay}

\begin{figure*}
\centering
\includegraphics[width=0.48\textwidth]{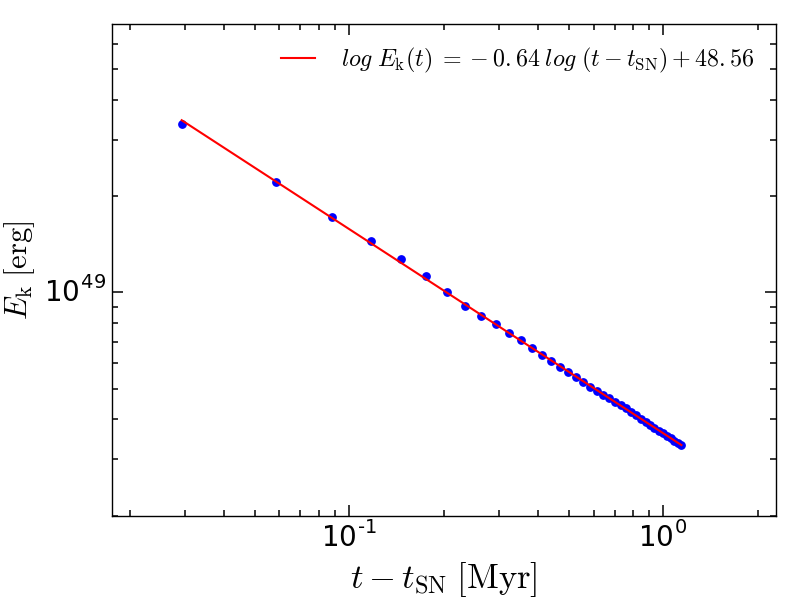}
\includegraphics[width=0.48\textwidth]{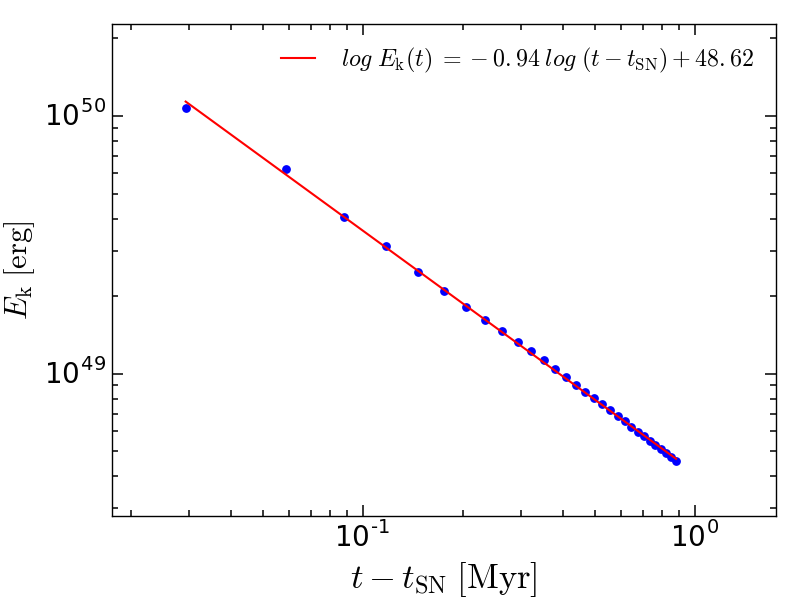}\\
\includegraphics[width=0.48\textwidth]{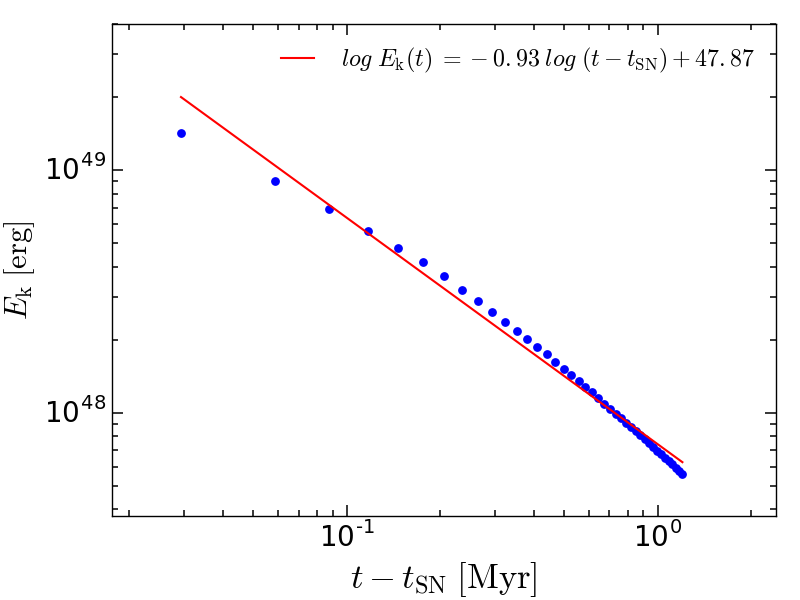}
\includegraphics[width=0.48\textwidth]{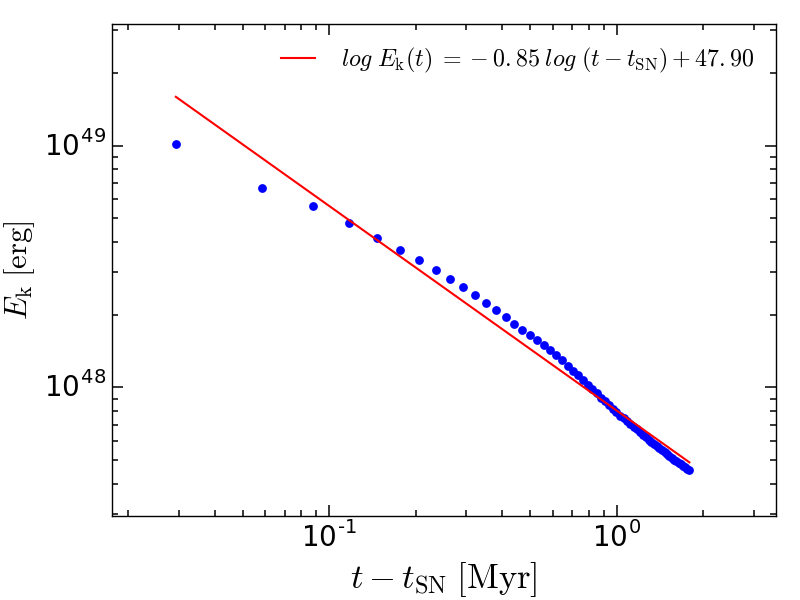}
\caption{Four examples of the evolution of kinetic energy in MCs right after an impact of a strong SN (the peak kinetic energy $>10$ the kinetic energy before the peak) at $t_{\rm SN}$. The blue filled circles are the kinetic energies after the peak, in the energy decay phase, and the red solid line is the power-law function fit.}
\label{fig_powerlaw_fit}
\end{figure*}

\begin{figure}
\centering
\includegraphics[width=0.48\textwidth]{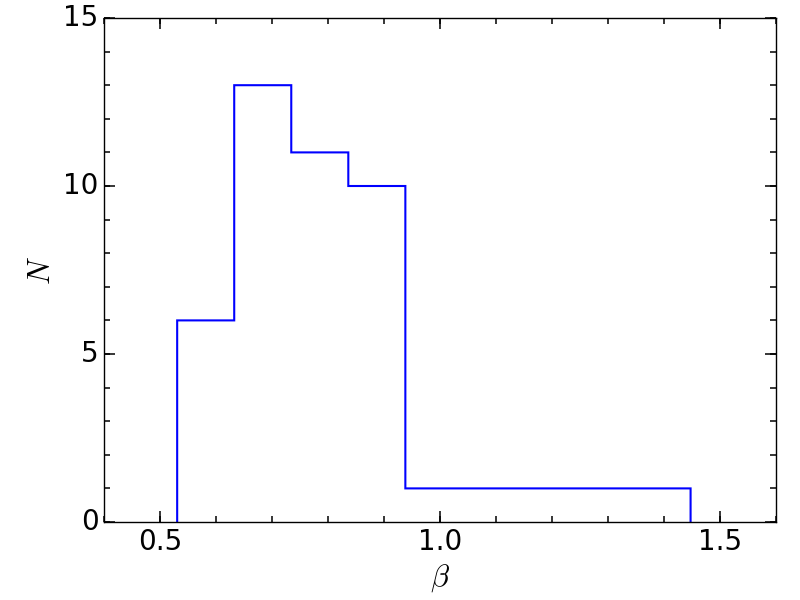}
\caption{Distribution of the power-law decay exponent, $\beta$.}
\label{fig_powerlaw_exponent}
\end{figure}

Kinetic energy in MCs is subject to viscous dissipation, and it decays in between two successive SN impacts.   
Figure \ref{fig_powerlaw_fit} shows a few examples of the evolution of kinetic energy in MCs right after an energy jump by a factor of $>10$ due to a strong SN impact.  We find that a strong kinetic energy jump is typically followed by a power-law decay. The red solid lines in Figure \ref{fig_powerlaw_fit} are the power-law fits, $E_{\rm k} \propto (t -t_{\rm SN})^{-\beta}$, to the data points, where $t_{\rm SN}$ is the time of the SN impact. The decay exponent, $\beta$, measured from least-square fitting, lies in the range from 0.53 to 1.45, as shown in Figure \ref{fig_powerlaw_exponent}, with a median value of $\beta \simeq 0.8$. 

The power-law decay is reminiscent of kinetic energy decay in unforced, statistically homogeneous and isotropic turbulence. Since the prediction of a $t^{-10/7}$ decay law by \citet{Kolmogorov41} for incompressible hydrodynamical turbulence, the kinetic energy decay has been extensively investigated and has been extended to the study of MHD and compressible turbulence. Theoretical models for turbulent energy decay are usually based on a power-law energy spectrum at small wave numbers, whose shape is assumed to remain invariant \citep[a hypothesis termed ``permanency of large eddies'', e.g.][]{Frisch95}, and on a turbulent energy cascade (or equivalently the Kolmogorov self-similarity hypothesis) in the inertial range. A power-law decay develops, as the cascade transfers kinetic energy from energy-containing large scales to the small scales in the dissipative range where the viscosity converts it into heat. 
It follows from the cascade picture that the timescale for turbulent energy decay is determined by the large-eddy turnover time, the dynamical time of the turbulent flow. Numerous theoretical, experimental and numerical studies have been devoted to predicting or measuring the decay exponent in turbulent flows. The exponent ranges  
from $-2/3$ to $-2$, depending on the initial energy spectrum at small wave numbers, the turbulent cascade model/mechanism, and whether the flow is magnetized, etc. \citet{MacLow+1998PhRvL} conducted numerical simulations of kinetic energy decay in supersonic HD and MHD turbulence, and they found that the decay exponent is in between $0.85$ and $1.1$ \citep{MacLow+1999ApJ}. Note, however, that the decay exponent measured by \citet{MacLow+1999ApJ} may not be universal, because the setup of the initial condition in their simulations corresponds to a particular choice for the initial energy spectrum, i.e., $\propto k^2$, at small $k$. 

Although the power-law decay and the measured decay exponent in the MCs after a strong SN impact are consistent with kinetic energy decay in homogeneous and isotropic turbulence, we stress that the actual decay mechanism may not be simply attributed to turbulent decay. A significant difference of the SN-driven turbulence in MCs from idealized simulations with artificial initial and boundary conditions is that the velocity field set up by SN impact is neither homogeneous nor isotropic. Energy injection by a SN is through the localized effect of shocks. As shown in Figure \ref{fig_v_cdf}, right after the SN impact, the velocity PDF in an MC typically has an extended component at the right tail, corresponding to the effect of an SN shock(s), 
suggesting that the majority of kinetic energy supply is contained in only a small mass fraction of the cloud.  Since the statistical homogeneity and isotropy are needed for the derivation of the power-law decay based on the Kolmogorov similarity hypothesis, it is not clear whether kinetic energy decay in an anisotropic turbulent flow is power law in general. 

Kinetic energy decay following an SN impact in an MC does not require an energy cascade from large to small scales. As kinetic energy is injected in the form of SN shocks, the dissipation starts immediately once the cloud is impacted. This is another difference from turbulent decay in isotropic flows with a large-scale initial field, where the cascade process that bridges energy-containing scales and the dissipative scales plays a key role in the prediction of a power-low decay.
Since most kinetic energy from an SN shock is injected ``locally" and the viscous dissipation starts immediately, the energy decay timescale right after SN impact is found to be much shorter than the dynamical time of the cloud. Also note that energy injection by SN shocks is unlikely to set up a power-law energy spectrum at large scales, which is typically assumed in theoretical models for power-law turbulent decay.  
   
Here we give a simple argument to explain the power-law energy decay in an MC based on the kinetic energy loss of a SN shock propagating into the cloud. For simplicity, we will assume that the total momentum is conserved as the SN shock sweeps into an MC. This assumption is valid for SNRs that have already reached the momentum conservation phase before hitting the MC. For SNRs that are still in the adiabatic, Taylor-Sedov phase, the assumption of momentum conservation is valid only if the higher radiative efficiency due to the higher density in the MC is able to sufficiently cool the postshock region. If a SN shock moves into a cloud at a velocity of $v$, it will sweep up a total mass $\propto v t$ in time $t$. The conservation of momentum then implies that $v_{\rm sh}^2 t $ is constant, which predicts that kinetic energy, $\propto v_{\rm sh}^2$, decays as $ t^{-1}$. A $t^{-1}$ decay is consistent with the measured values of $\beta$ shown in Figure \ref{fig_powerlaw_exponent}. On average, the measured $\beta$ is $\simeq 0.8$, suggesting that the energy decay is typically slower than the predicted $t^{-1}$ behavior. This is perhaps because the SNRs that cause a large kinetic energy jump are still in the adiabatic phase when reaching the MC, and the assumption of momentum conservation is inaccurate. If not efficiently cooled, the pressure in the postshock gas would push the shock, and the pdV work would make the decay of kinetic energy slower. 

\begin{figure}
\centering
\includegraphics[width=0.48\textwidth]{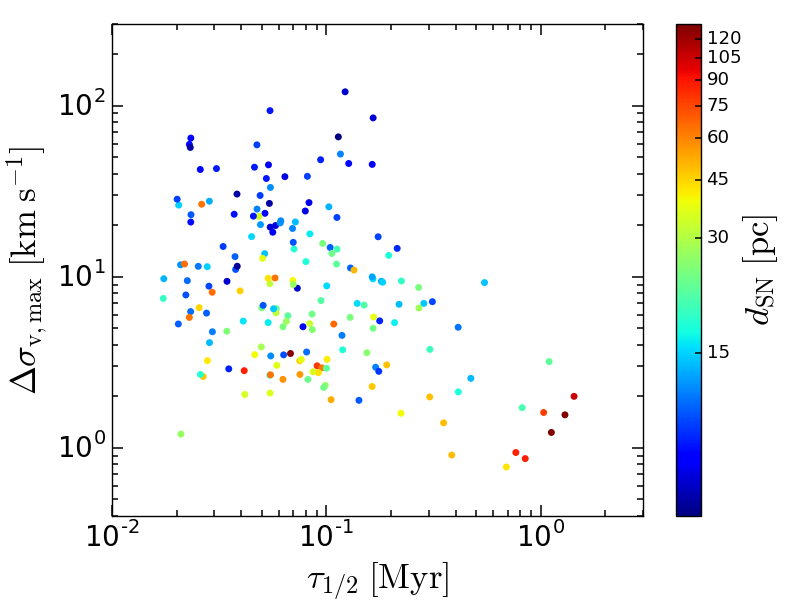}
\caption{Relation between the decay time $\tau_{\rm 1/2}$ and the maximum increase $\Delta \sigma_{\rm v,max}$ of the additional turbulent velocity component caused by SNe. The color bar is the distances of SNe, $d_{\rm SN}$, to the mass center of MCs.}
\label{fig_t_one_half}
\end{figure}

We quantify how fast the kinetic energy decays right after an SN impact by measuring the timescale at which the velocity dispersion in the MC decreases. 
Following \citet{Seifried+2018ApJ}, we first define the velocity dispersion jump, $\Delta  \sigma_{\rm v}$, 
driven by a SN at time $t_{\rm SN}$ as,  
\begin{equation}
\Delta \sigma_{\rm v} =  \sigma_{\rm v} (t_{\rm SN+}) - \sigma_{\rm v}(t_{\rm SN-}) \, ,
\label{eq_sigma_max}
\end{equation}
where $t_{\rm SN+}$ and $t_{\rm SN-}$ are the times right before and right after the MC is impacted by 
the SNR, respectively. 
As in \citet{Seifried+2018ApJ}, we then define a half-life decay time, $\tau_{\rm 1/2}$, as the time 
during which $\sigma_{\rm v} (t) - \sigma_{\rm v}(t_{ \rm SN-})$ drops from $\Delta \sigma_{\rm v} $ to $0.5 \, \Delta \sigma_{\rm v}$, i.e.,
\begin{equation}
\sigma_{\rm v} (t_{\rm SN} + \tau_{\rm 1/2}) - \sigma_{\rm v}(t_{\rm  SN-}) = 0.5 \, \Delta \sigma_{\rm v} \, .
\label{}
\end{equation}
Essentially, $\tau_{\rm 1/2}$ represents the time needed for the MC to lose half of its gain of velocity dispersion from an SN. 

\begin{figure}
\centering
\includegraphics[width=0.48\textwidth]{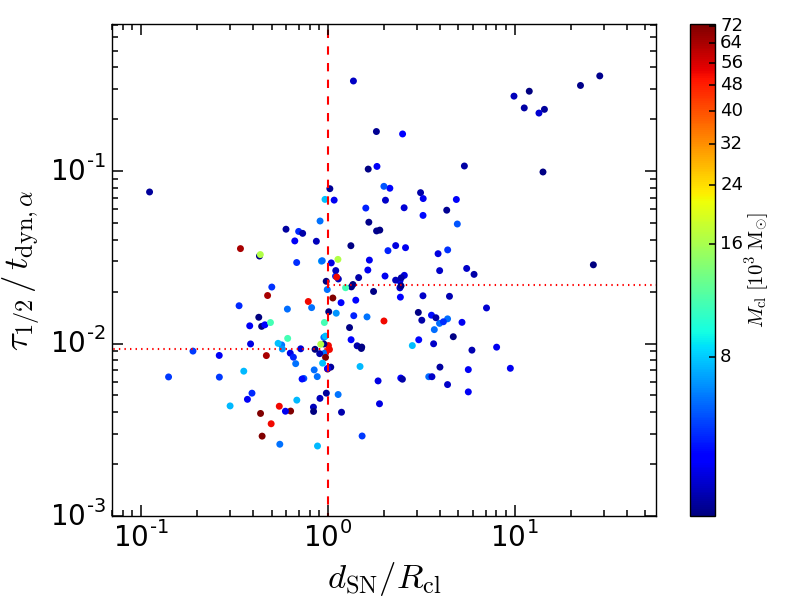}
\caption{Decay time $\tau_{\rm 1/2}$, in units of the dynamical time $t_{\rm dyn,\alpha}$, versus SN distance to cloud center $d_{\rm SN}$, in units of the cloud radius $R_{\rm cl}$, colored with cloud mass $M_{\rm cl}$. The vertical dashed line corresponds to $d_{\rm SN}/R_{\rm cl}=1$. The median values of $\tau_{\rm 1/2}/t_{\rm dyn,\alpha}$ are 0.01 and 0.02 for internal ($d_{\rm SN}/R_{\rm cl} \leq 1$) and external ($d_{\rm SN}/R_{\rm cl}>1$) SNe. The median of $\tau_{\rm 1/2}/t_{\rm dyn,\alpha}$ for all the clouds is 0.015. }
\label{fig_t_one_half_vs_d_sn}
\end{figure}

\begin{figure*}
\centering
\includegraphics[width=0.48\textwidth]{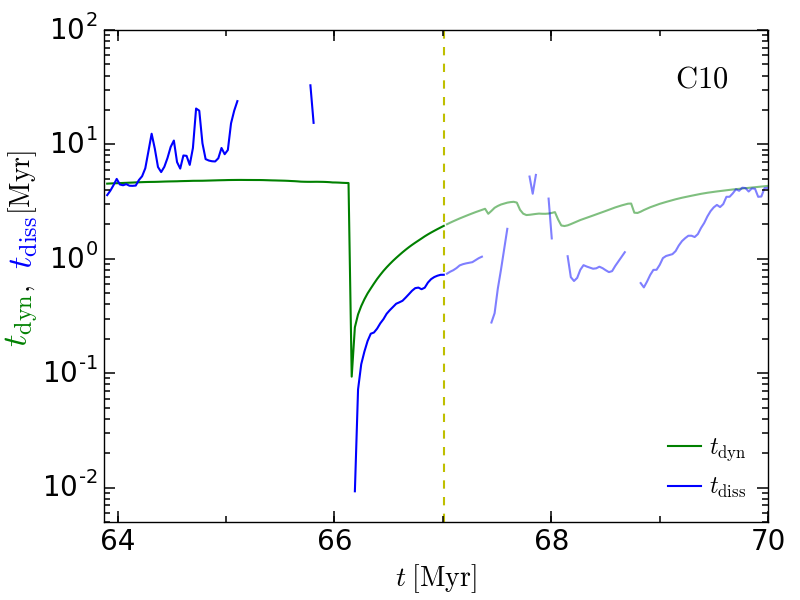}
\includegraphics[width=0.48\textwidth]{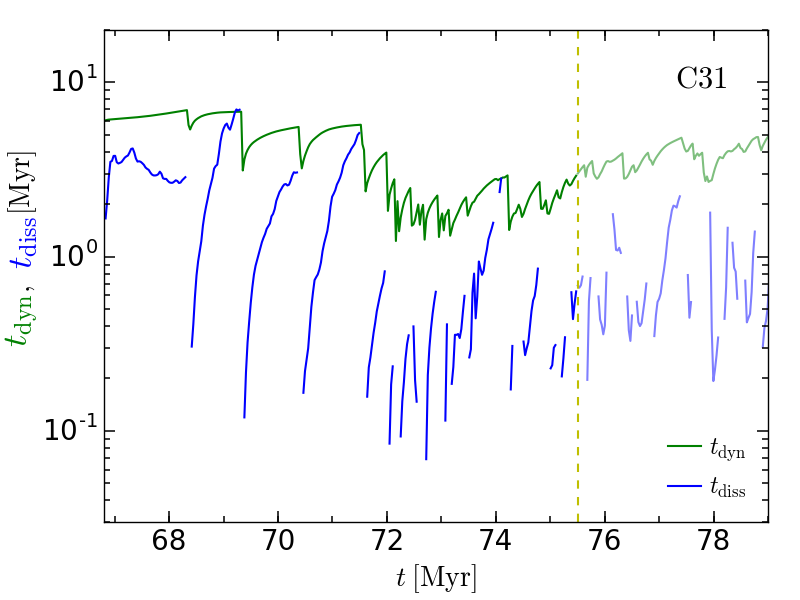}
\caption{The evolution of the instantaneous energy dissipation time of the cloud (blue line) and its instantaneous dynamical time (green line). The yellow vertical dashed line corresponds to the time of cloud dispersal ($t_{\rm d}$).}
\label{fig_dyn_evol}
\end{figure*}

The half-life decay time, $\tau_{\rm 1/2}$, is defined for each SN impact or each kinetic energy peak during the evolution of an MC. By following the energy evolution, we measured $\tau_{\rm 1/2}$ for the kinetic energy peaks satisfying the following criteria:
1) The kinetic energy experiences jumps by at least a factor of 2; 2) following the kinetic energy peak, there are more than 3 snapshots before the impact of the next SN; 3) The time of the kinetic energy peak occurs prior to the dispersal of the cloud. The first two criteria are motivated by the need to have a well enough defined peak and decay evolution in order to measure the decay time, while the third condition is used because we are not interested in following the evolution of the clouds after their dispersal (the tracer particles we follow do not really define a cloud anymore at that point). 
For some peaks, the dissipation of kinetic energy does not last long enough to lose half of the velocity dispersion gain, $\Delta \sigma_{\rm v}$, from an SN, because the interruption by another SN or by mass accretion may cause the kinetic energy to increase again. We ignore these peaks since $\tau_{\rm 1/2}$ cannot be quantified. In Figure \ref{fig_t_one_half}, we show a scatter plot of the velocity dispersion jump, $\Delta \sigma_{\rm v}$, versus the half-life decay time, $\tau_{\rm 1/2}$. Typically $\tau_{\rm 1/2}$ is $\sim 100-600\,\rm kyr$ in \citet{Seifried+2018ApJ}. We have more MCs in the sample and our result is consistent with Figure 5 in their paper. As shown in Figure \ref{fig_t_one_half}, the $\tau_{\rm 1/2}$ in our simulation range between $\sim 20$ and $2000\,\rm kyr$. Larger values of maximum gain in velocity dispersion $\Delta \sigma_{\rm v,max}$ due to the close distance SNe. It implies that the close SNe have more energy injection while dissipate fast resulting shorter $\tau_{\rm 1/2}$.
The figure shows the decay time of 184 energy peaks.
The colors give the distance, $d_{\rm SN}$, of the SN responsible for the energy peak, which we identify as the closest SN to the cloud within 5 snapshots ($\sim 0.15\,\rm Myr$) prior to the kinetic energy peak. Figure \ref{fig_t_one_half} shows that $\tau_{\rm 1/2}$ ranges from $\sim 20$ to $2000\,\rm kyr$, which is much smaller than the dynamical time, $t_{\rm dyn,\alpha}$ $\sim 4$ Myr, of the clouds. This is also seen in Figure \ref{fig_t_one_half_vs_d_sn}, which plots the ratio $\tau_{\rm 1/2}/t_{\rm dyn,\alpha}$ versus the distance of the SN. As discussed earlier, the fast energy decay right after the SN impact is because the energy injection occurs in the form of shocks, which are small-scale structures that dissipate immediately.  

Figure \ref{fig_dyn_evol} compares the instantaneous energy dissipation time (blue lines) with the instantaneous dynamical time (green lines) for the two representative clouds C10 and C31. The instantaneous dissipation time, $t_{\rm diss} (t)$, is defined as 
\begin{equation}
t_{\rm diss} \equiv \frac{E_{\rm k}(t)}{\left | d E_{\rm k}(t)/d t  \right |} \, ,
\label{eq_t_disp}
\end{equation}
while the dynamical time, $t_{\rm dyn}$, is calculated from Equation (\ref{eq_tdyn}), i.e., $t_{\rm dyn} \equiv R_{\rm cl}/\sigma_{\rm v,3D}$, where $\sigma_{\rm v,3D}$ is the measured three-dimensional rms velocity. The blue lines only show the dissipation timescale in time intervals where the kinetic energy decays, leaving out the phases in which the energy increases either due to SN impact or some weak energy sources entering the cloud from the boundary, such as mass accretion. 

Right after the SN impact, the instantaneous dissipation timescale is much smaller (by a factor of $ \ge 10$) than the dynamical time of cloud, corresponding to the immediate, local dissipation of the SN shock propagating into the cloud. Unlike idealized simulations of turbulent energy decay, the SN energy is injected locally, the majority of the kinetic energy supply initially affecting only a small faction of the cloud.   
At later times, the injected kinetic energy is more mixed into the entire cloud as the SN shock sweeps into the clouds and affect larger scales.  
A  kinetic energy transfer toward larger scales  may also occur through a so-called back scatter effect\footnote{The back scatter effect is essentially nonlocal energy transfer triad interactions in a turbulent flow.  Two wave vectors with large amplitudes may form a triangle with a short third side corresponding to a small wave number, and such triad interactions may cause kinetic energy transfer toward large scales.},
and the turbulence becomes more developed and more isotropic. Only at this stage does the instantaneous dissipation time, $t_{\rm diss}$, become comparable to the dynamical time of the cloud, suggesting that the kinetic energy is actually lost through turbulent decay. Prior to the arrival of the next SN impact, the instantaneous dissipation time scale sometimes exceeds the dynamical time, perhaps due to some weak energy sources from the cloud boundary. 

\section{Discussion} \label{discussion}

\subsection{Can SN Energy Injection Maintain the Velocity Dispersion in MCs?}

\begin{figure*}
\centering
\includegraphics[width=0.48\textwidth]{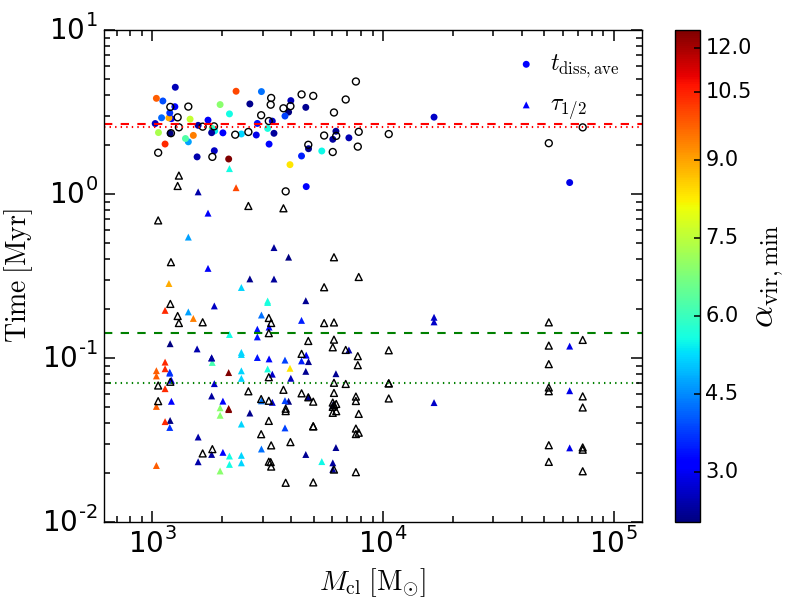}
\includegraphics[width=0.48\textwidth]{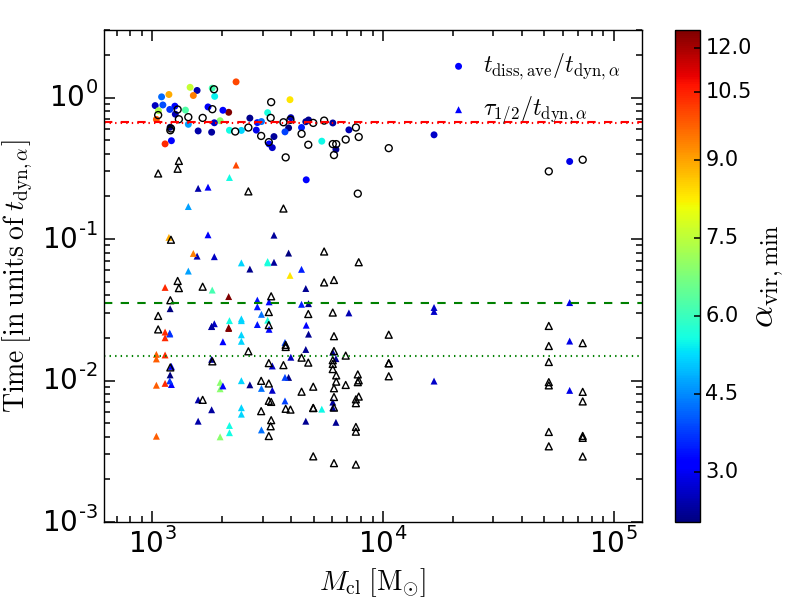}
\caption{Left panel: The average dissipation time $t_{\rm diss,ave}$ (circle symbol) and the half-life decay time $\tau_{\rm 1/2}$ (triangle symbol) versus the cloud mass, colored with virial parameter. The opened and filled symbols are denoted as bound and unbound clouds respectively. 
The mean (horizontal red dashed line) and median (horizontal red dotted line) values of the $t_{\rm diss,ave}$ are 2.68 Myr and 2.55 Myr, respectively. The mean (horizontal green dashed line) and median (horizontal green dotted line) values of the $\tau_{\rm 1/2}$ are 0.14 Myr and 0.07 Myr, respectively. Right panel: The same as the left panel, but with $t_{\rm diss,ave}$ and $\tau_{\rm 1/2}$ in units of the dynamical time $t_{\rm dyn,\alpha}$. The mean and median values for 
$t_{\rm diss,ave}/t_{\rm dyn,\alpha}$ are 0.67 and 0.66, respectively. The mean and median values for 
$\tau_{\rm 1/2}/t_{\rm dyn,\alpha}$ are 0.035 and 0.015, respectively. The mean and median values of $t_{\rm dyn,\alpha}$ are 4.29 Myr and 4.05 Myr, respectively.}
\label{fig_t_diss}
\end{figure*}

Our main motivation for estimating the timescale of kinetic energy decay is to examine whether and how the energy injection by SNe may maintain the observed velocity dispersion in MCs. Our simulation already provides ample evidence that the SNe impact does provide sufficient energy source for random motions in MCs, considering that most of the clouds in the simulation are dispersed by either a strong SN impact or by the cumulative impacts of a number of SNe. Nevertheless, an estimate of the amplitude of the velocity dispersion resulting from the balance of the SN energy injection with the viscous dissipation would help better understand the physical mechanism of SN driving in MCs. This type of estimate has been attempted in previous works \citep[e.g.][]{Seifried+2018ApJ}, and we find it useful to illustrate its large uncertainties that may potentially lead to opposite conclusions.  

Most previous estimates based on this idea simply adopt an average SN rate (assuming uniformly distributed SNe) and an assumed SN efficiency, i.e., the percentage of the explosion energy of each SN that contributes to the turbulent motions in MCs, and use an energy dissipation rate expected from turbulent decay. Such calculations clearly oversimplify the complicated physical processes of SN-driven turbulence in MCs. For example, we found that the energy received by an MC depends on various factors, e.g., the number of SN impacts, the distance and specific position of the SN with respect to the cloud, as well as the evolution phase of the SNR. The kinetic energy evolution between successive SN impacts is also very complicated, not simply determined by turbulent decay, as discussed above. As shown in Figure \ref{fig_dyn_evol}, the instantaneous timescale for kinetic energy decay exhibits strong temporal variations. 

The strong temporal variation of the instantaneous decay timescale makes it difficult, if not impossible, to accurately model the decay of kinetic energy in between SN impacts. Due to the fast decay right after the SN impact, using $\tau_{1/2}$ as the typical decay timescale would overestimate the overall dissipation rate between two successive SN events. An alternative choice is to characterize the overall dissipation by averaging the instantaneous dissipation timescale over the lifetime of each cloud. We refer to this average dissipation timescale as $t_{\rm diss, ave}$. The instantaneous dissipation time, $t_{\rm diss}$, is not available at times when kinetic energy increases, and for those times we use the instantaneous dynamical time, $t_{\rm dyn}$, for the calculation of $t_{\rm diss, ave}$. Furthermore, the instantaneous dissipation timescale (blue line in Figure \ref{fig_dyn_evol}) is replaced by the instantaneous dynamical time (green line in Figure \ref{fig_dyn_evol}) when $t_{\rm diss} > t_{\rm dyn}$. This choice is based on the fact that the kinetic energy decay timescale cannot be larger than the dynamical time of the turbulence, and  $t_{\rm diss} > t_{\rm dyn}$ is likely due to some energy sources through the cloud boundary (which can be neglected in comparison to the SN energy injection). Filled circles in Figure \ref{fig_t_diss} show the average dissipation timescale, $t_{\rm diss, ave}$, for all the clouds. It is much larger than $\tau_{1/2}$, by a factor of $\simeq 30$. We point out that $t_{\rm diss, ave}$ may give too much weight to the periods where the kinetic energy is unaffected by SN explosions. Therefore, it should be taken as an upper limit for the overall decay timescale. We will take $\tau_{1/2}$ and $t_{\rm diss, ave}$ as the lower and upper limits for the energy decay timescale in our estimate of the expected velocity dispersion in the MCs. The median values of $\tau_{1/2}$ and $t_{\rm diss, ave}$ are 0.07 Myr and 2.55 Myr, respectively (see left panel in Figure \ref{fig_t_diss}).

\begin{figure*}
\centering
\includegraphics[width=0.48\textwidth]{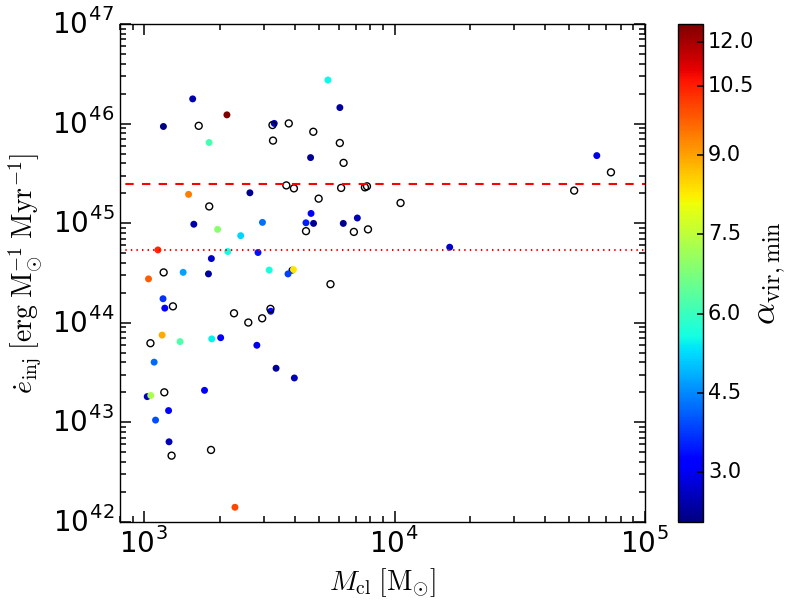}
\includegraphics[width=0.48\textwidth]{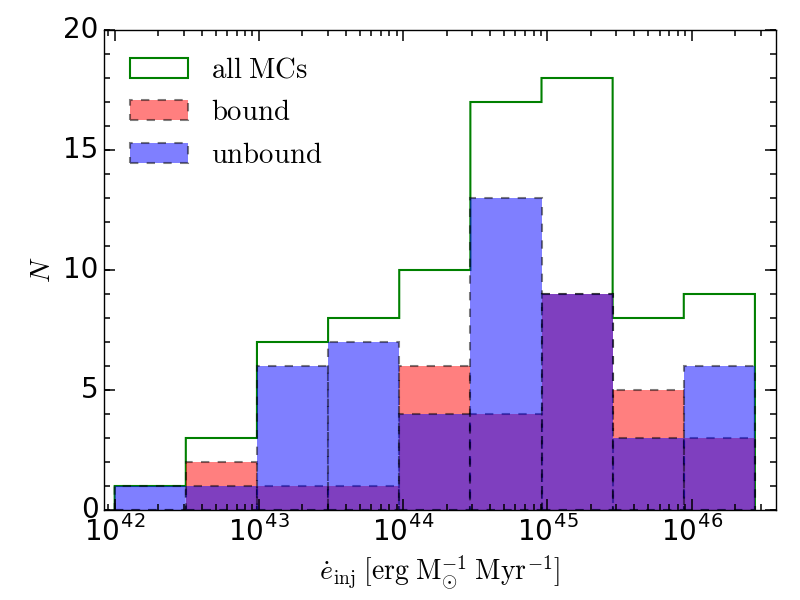}
\caption{Left panel: The relation between the cloud mass and the specific kinetic energy injection rate (see Eq.~\ref{eq_dote}). The mean (red horizontal dashed line) and median (red horizontal dot line) values of the specific kinetic energy injection rate are $2.48 \times 10^{45}$ erg $M_\odot^{-1}$ Myr$^{-1}$ and $5.39 \times 10^{44}$ erg $M_\odot^{-1}$ Myr$^{-1}$, respectively. Right panel: The distribution of the specific kinetic energy injection rate in $t_{\rm s,d}$. Red and blue histograms are the subsets of bound and unbound MCs, respectively.}
\label{fig_ki_inj_rate}
\end{figure*}

We compute the SN energy injection rate per cloud mass:
\begin{equation}
\dot e_{\rm inj} = E_{\rm inj, tot}  / (t_{\rm s,d} \, M_{\rm cl }) \, ,
\label{eq_dote}
\end{equation}
where $E_{\rm inj, tot}$ is the cumulative energy injection throughout the period from the selection of the cloud to its dispersal, $t_{\rm s,d}$, and $M_{\rm cl}$ is the cloud mass at the time of the selection. The expected velocity dispersion of the cloud is then calculated from the energy balance $\dot e_{\rm inj}  \simeq  \frac{1}{2} \, \sigma_{\rm v}^2/\tau_{\rm dec}$, 
i.e., $\sigma_{\rm v} = [ 2 \, (\dot e_{\rm inj} \, \tau_{\rm dec})]^{1/2}$, where $\tau_{\rm dec}$ is the characteristic overall timescale of kinetic energy decay. Instead of examining the energy budget in each cloud, here we conduct a ``global" estimate using the median values over all the clouds in the simulation. Due to the crude approximations made in the calculation, it is not meaningful to conduct a cloud-by-cloud analysis. The left panel of Figure \ref{fig_ki_inj_rate} shows that $\dot e_{\rm inj} \simeq  5.39 \times 10^{44} $ erg $M_\odot^{-1}$ Myr$^{-1}$ on average. If we set $\tau_{\rm dec} \simeq \tau_{1/2}$ and adopt  $\tau_{1/2} = 0.07$ Myr, we find a lower limit of 1.94~$\rm km \;\rm s^{-1}$ for the velocity dispersion. On the other hand, using the median value, 2.55 Myr, of $t_{\rm diss, ave}$ for the overall decay timescale, $\tau_{\rm dec}$, yields an estimate of the upper limit of $\sigma_{\rm v}$ that is about 11.72~$\rm km \;\rm s^{-1}$. Clearly, the estimated lower and upper limits of $\sigma_{\rm v}$ are consistent with the observed velocity dispersions in MCs, confirming that the SN injection is a sufficient energy source for turbulent motions in MCs.

Although our back-of-the-envelope calculation gives a reasonable estimate for the velocity dispersion, it is perhaps better viewed as a demonstration for the complexity of the problem of SN-driven turbulence in MCs rather than as a successful model. The complicated physical processes involved can hardly be well understood without the help of numerical simulations. Only with simulation data was it possible for us to carry out a ``global estimate", and yet we could only give upper and lower limits for the typical velocity dispersion due to the peculiar behavior of kinetic energy decay in between SN impacts. 
Previous calculations using assumed SN efficiency and the turbulent decay timescale cannot capture the rich physics in the problem and should be viewed at best as a zeroth-order, toy model of the energy budget for SN-driven turbulence in MCs. Results from such calculations are highly uncertain, since the SN efficiency is difficult to estimate, and the overall energy dissipation in between SN events is not simply controlled by turbulent decay. 
We stress that our calculation is much more physical and accurate than all the previous ``empirical" estimates in the sense that the timing of the SN impact, the location of the SN explosion, the SN frequency or the the number of SN impacts, as well  as the physical conditions of 
the MCs are realistic in our simulation, which aimed to self-consistently capture star formation and SN driving in MCs.

\subsection{The Importance of SN Feedback}

High-resolution galactic-fountain simulations established that SNe can explain the general ISM properties, and in particular the HI velocity dispersion \citep{deAvillez+Breitschwerdt05,Joung+MacLow+2006ApJ,Joung+2009,Hill+12}. It would seem natural that the turbulence in the HI gas would cascade down to smaller scales to generate the turbulence in MCs, especially if MCs are viewed as random density fluctuations in the turbulent ISM. However, those simulations did not have the spatial resolution to investigate if SNe could also reproduce the turbulence within MCs. Based on a more recent galactic-fountain simulation, \citet{Ibanez+MacLow2016ApJ...824...41I} concluded that SNe cannot drive the turbulence in MCs, but the spatial resolution of their uniform grid around the midplane was the same as in \citet{Joung+MacLow+2006ApJ}, 1.9~pc, and their maximum resolution of 0.95~pc was achieved with one AMR level of Jeans refinement, which is unlikely to improve the resolution of the turbulence. It is well-established that, in numerical simulations, the turbulence is not resolved at scales $\lesssim 10\,\Delta x$, where $\Delta x$ is the mesh size. Thus, in \citet{Ibanez+MacLow2016ApJ...824...41I}, velocity differences at scales $\lesssim 20$~pc are underestimated, and the turbulence inside clouds of size $\sim 10$~pc is not resolved. In \citet{Padoan+SN1+2016ApJ} we found that randomly generated SNe could explain the velocity-size relation of MCs \citep[see also][for the synthetic observations]{Padoan+SN3+2016ApJ}, thanks to a spatial resolution of 0.25~pc, nearly 10 times larger (MCs were uniformly covered at that maximum resolution).  

In order to study the interaction between SNe and MCs in greater details than in galactic-fountain simulations, recent studies have 
simulated either a single cloud in isolation \citep[e.g.][]{Walch+2015MNRAS,Iffrig+2015A&A}, or clouds selected from larger-scale simulations \citep[e.g.][]{Kortgen+2016MNRAS,Ibanez+MacLow+2017ApJ,Seifried+2018ApJ}.

\citet{Walch+2015MNRAS} performed high-resolution SPH simulations to investigate the combined effect of SNe, HII regions and density structure. They placed a single SN explosion in the center of a spherical, static MC ($R_{\rm cl} = 16 \,\rm pc$, $M_{\rm cl} = 10^5 \,\rm M_{\odot}$, $\alpha_{\rm vir} = 0$) without magnetic fields. They found that while ionization enhances the momentum transfer of the SN by 50\%, a single SN explosion was unable to disperse such a massive cloud. Their cloud was initially static without turbulence, so this is not surprising. Our most massive cloud was about half of theirs with an initial virial parameter of the order of unity, and it took several internal SNe exploding in close succession to destroy the cloud.

\citet{Iffrig+2015A&A} performed a series of numerical simulations to study the momentum injection into a more realistic, turbulent MC ($R_{\rm cl} = 6.8 \,\rm pc$, $M_{\rm cl} = 1.5 \times 10^4 \,\rm M_{\odot}$, $\alpha_{\rm vir} = 2$) by a single SN explosion at different positions: inside, at the edge, or outside of the MC. They concluded that SNe exploding inside MCs remove a significant fraction of the cloud mass, consistent with our results, while external SNe remove mainly lower density gas. Thus, it is essential that the location of the SNe with respect to the dense cloud are realistic and self-consistent in the simulation, something our simulation has achieved for the first time with a large statistical sample.

\citet{Kortgen+2016MNRAS} performed MHD simulations of cylindrical colliding flows without initial turbulence ($R_{\rm cl} = 64 \,\rm pc$, $M_{\rm cl} = 0.9 \times 10^5 \,\rm M_{\odot}$). They concluded that SN feedback alone fails to disrupt entire, gravitationally bound MCs, but is able to disperse small-sized $\sim10 \,\rm pc$ regions on timescales of less than $1 \,\rm Myr$. However, in their simulations, the SNe explode in clouds that are extremely bound, having energy rations between 1\% and 10\%. These energy ratios are unrealistic compared to observed clouds, resulting from the lack of realistic level of turbulence in their initial conditions. Despite this, their SNe were able to disperse up to 50\% of mass in some cases. Furthermore, the {\it ad hoc} geometry of their colliding flows yields a thin layer whose fragmentation results in bound clouds. The thickness of the layer is $\sim 10$~pc or less, so it is to be expected that when SN remnants reach a size of $\sim 10$~pc in their simulation most of the hot gas will escape through the low-density medium.  

\cite{Ibanez+MacLow+2017ApJ} used zoom-in techniques to select three MCs from a SN-driven galactic fountain MHD simulation. The MCs are characterized by $L_{\rm box} = 100 \,\rm pc$, $M_{\rm cl} = 3.6, \, 3.2, \, 7.5 \times 10^3 \,\rm M_{\odot}$, and $\alpha_{\rm vir} = 0.4, \, 0.45, \, 0.3$. The closest SN distance is $\sim 45\,\rm pc$, while most of them are over $50\,\rm pc$ away. They concluded that MCs are bound, always in a state of gravitational contraction, and SNe fail to disrupt them. The lack of effect from the distant SNe is not surprising and consistent with our findings. However, the complete absence of nearby SNe is fundamentally inconsistent with bound, star-forming clouds, such as those found in our more realistic implementation of SN feedback. More importantly, the very low initial virial parameters of the MCs in that study may not be realistic, as those values are inherited from the larger-scale simulation by \citet{Ibanez+MacLow2016ApJ...824...41I}, where the velocity dispersions were greatly underestimated due to numerical dissipation, as discussed above.

\cite{Seifried+2018ApJ} performed high-resolution ($\sim 0.1 \,\rm pc$), MHD simulations, as zoom-in reruns of a galactic-fountain simulation by \cite{Walch+15}, to investigate whether the observed level of MC turbulence could be generated and maintained by external SNe. They exploded single SNe at different distances of $d_{\rm SN}$ = 25 pc, 50 pc, 62.5 pc, or 75 pc from the center-of-mass of their cloud and concluded that SNe are not able to sustain the observed level of MC turbulence. However, our more realistic implementation of SN feedback shows that SNe at distances of 25 pc or less are frequent (see Figure~\ref{fig_f_sn_alpha_vir}), and able to drive MC turbulence and even destroy the clouds. Furthermore, many of the MCs are never bound, so they do not even require internal SNe to be dispersed. It is possible that the initial virial parameter of the MCs in \cite{Seifried+2018ApJ} is underestimated \citep[as in][]{Ibanez+MacLow+2017ApJ}, because the clouds are extracted from the galactic-fountain simulations by \cite{Walch+15} that have a spatial resolution of only 3.9~pc.

\subsection{MC Lifetime}

The observational evidence that MCs may be short lived, that is $t_{\rm life}\sim t_{\rm dyn}$, based on stellar ages was first proposed by \citet{Larson81}, and later demonstrated systematically by \citet{Elmegreen2000}. By analyzing the age difference of stellar clusters in relation to their separation, \citet{Elmegreen2000} showed that the duration of star formation is always of the order of the dynamical time, which implies that the lifetime of MCs is also of that order of magnitude \citep[see also][]{Hartmann+2001, Ballesteros-Paredes+Hartmann07}. A much larger MC lifetime, $t_{\rm life}\gg t_{\rm dyn}$, would require an unrealistically large population of MCs without any sign of star-formation activity. More recent estimates, based on the spatial correlation between MCs and H$\alpha$ in the LMC \citep{Kawamura+09}, free-free emission in the Milky Way \citep{Murray11} or young stellar clusters in M33 \citep{Corbelli+17}, have confirmed that MC lifetimes are typically a few dynamical times. While MC ages in these studies are constrained by stellar ages, short MC timescales were derived in M51 with a completely different method, based on the inter-arm travel time \citep{Meidt+15}. 

To constrain the importance of galactic dynamics in the formation and disruption of MCs, the same method to determine MC ages should be applied to a number of different galaxies. This was recently achieved by \citet{Chevance+Kruijssen2019MNRAS.tmp.3155C}, who analyzed 9 nearby disc galaxies from the PHANGS-ALMA survey. This work has also the advantage that, contrary to previous studies, it does not rely on the definition and identification of individual MCs, but on a well-tested statistical method to study the spatial correlation between CO and H$\alpha$ emission peaks \citep{Kruijssen+Longmore14,Kruijssen+18}. As in previous studies, the absolute timescale comes from the characteristic duration of the H$\alpha$ emission of a stellar population, but in a way that does not require any assumption about the duration of star formation, which is instead an output of the method. \citet{Chevance+Kruijssen2019MNRAS.tmp.3155C} found that the inferred cloud lifetime in those 9 galaxies is short, between approximately 10 and 30~Myr. 

Our result that $t_{\rm life} \sim 3 \, t_{\rm dyn,\alpha}$ is consistent with the observational estimates mentioned above. On the other hand, we have also found that the dispersal time is shorter than the formation time, $t_{\rm disp} \sim t_{\rm dyn,\alpha}$, particularly for gravitationally-bound clouds, where $t_{\rm disp}/t_{\rm life} \sim 1/4$, on the average. This asymmetry in the evolution of MCs that are active regions of star formation (whose virial parameter decreases to a value $\alpha_{\rm vir,min}\le 2$) requires a self-consistent treatment of the SN feedback, and is not correctly captured in simulations with randomly generated SNe. The short dispersal time in the MCs may be related to the short feedback times inferred by \citet{Chevance+Kruijssen2019MNRAS.tmp.3155C}, although they interpreted that result as evidence against SNe being the dominant feedback mechanism. More work is required for a proper comparison of MC lifetimes with the observational estimates, which should be based on the analysis of synthetic observations of the simulation.

\section{Conclusions} \label{conclusions}

In this work we have studied the effect of individual supernovae on molecular clouds, based on a high-resolution 250 pc simulation. The position and time of SNe are computed self-consistently for the first time, by resolving the formation of individual massive stars. We refer to this implementation of the SN feedback as \emph{real SNe}. The large sample size includes 83 MCs and 273 SNe during approximately 20~Myr of a simulation that includes self-gravity for 30~Myr. 

In general, our results illustrate that supernova feedback externally is a main large-scale driver of the turbulence and hence ultimately responsible for making molecular clouds transient, while the overabundance of internal SNe make their dispersal faster.

Our specific conclusions are as follows:

\begin{enumerate}

\item Despite the complex structure of MCs emerging from SN-driven turbulence, their ratio of kinetic and gravitational energy is captured well by the simple expression of the virial parameter for a uniform sphere of size equal to the mass-weighted rms cloud radius. This result applies to MCs across a broad range of values of virial parameter and mass, suggesting that MCs have a universal fractal structure.

\item As a result of SN feedback, all MCs are transient, with a lifetime of order three dynamical times, or 5-20~Myr, even if a fraction of the clouds (32 out of 83) experience a phase during which they are gravitationally bound. 

\item Bound clouds are active sites of star formation, thus they tend to generate internal SNe and, as a result, their dispersal time is short. We find that on average $t_{\rm disp}/t_{\rm life}\sim 1/4$ for these clouds. For clouds that are never bound, the dispersal time is longer, $t_{\rm disp}/t_{\rm life}\sim 1/2$.

\item In our simulation, internal SNe occur frequently, in approximately half of the clouds, while they should occur in only 5\% of the clouds if SNe were uniformly distributed. In the clouds with internal SNe, the number of such SNe is typically a factor $\sim 50$ above the expected value (for the specific cloud size and lifetime) from a uniform distribution. The probability of such overdensities would be many orders of magnitude lower if the SNe had a uniform spatial distribution. These overdensities are due to SNe from locally-born massive stars that we can track thanks to our self-consistent implementation of SN feedback.

\item The quantitative impact of SNe depends crucially on their location relative to MCs, which has been described self-consistently for the first time in this work. Thanks to this realistic implementation of the feedback, we have been able to quantify the efficiency of SN energy injection as a function of SN distance to the clouds and the scatter in such relation. The energy injection from the most significant SN for each MC is on average an order of magnitude lower than that available from the SN remnant shock in the Sedov-Taylor phase. Considering all SNe during the cloud lifetime, we find that the total energy injected greatly exceeds the binding energy of the cloud.

\item The decay of kinetic energy of MCs, after a SN event, is initially much shorter than the dynamical time, inconsistent with turbulence decay, because at this time the dissipation is dominated by the SN remnant shocks. At a later phase a fraction of the injected kinetic energy is distributed over the whole cloud, contributing to the cloud turbulence. 

\item SN explosions can indeed drive and maintain MC turbulence, as shown by the rms velocity of the clouds in the simulation. Using our new results on the efficiency, frequency and decay time of SN feedback, we show that estimating the cloud rms velocity from a global balance of energy injection and dissipation, as previously attempted in the literature, is highly uncertain. 

\end{enumerate}

A reliable assessment of the effect of SNe on MC turbulence requires a self-consistent simulation yielding realistic populations of SNe and MCs formed {\it ab initio}. Having established the important role of SNe in this work, future studies should extend such self-consistent modeling with the inclusion of radiation and winds from massive stars. Although SNe will most likely remain the dominant source of turbulent energy at large scales, photoionizing radiation and stellar winds may contribute significantly to the disruption of MCs.

This work focuses on the turbulence and dispersal process of MCs, under the sole effect of SNe. As MCs are formed {\it ab initio} in the simulation, with properties matching the observations, the most important implication of this work is that SN-driven turbulence explains the origin of MCs as well. Due to the continuous crisscrossing of SN remnants, as some regions expand, leading to the dispersion of MCs, others are compressed, giving birth to new MCs that will lead to new SNe. SNe and MCs are thus arguably the two most essential components of the star-gas life cycle, as the ISM turbulence is self-regulated thanks to their combined effect.

\section*{Acknowledgements}

ZJL acknowledges financial support from China Scholarship Council (CSC) under grant No. 201606660003.
PP and VMP acknowledge support by the Spanish MINECO under project AYA2017-88754-P. 
LP acknowledges financial support from NSFC under grant No. 11973098.
Computing resources for this work were provided by the NASA High-End Computing (HEC) Program through the NASA Advanced Supercomputing (NAS) Division at Ames Research Center. 
We acknowledge PRACE for awarding us access to Curie at GENCI@CEA, France.
Storage and computing resources at the University of Copenhagen HPC centre, funded in part by Villum Fonden (VKR023406), were used to carry out part of the data analysis.

\bibliography{SN_feedback}

\appendix

\section{Physical Parameters of MCs} \label{app_table}

\begin{figure}
\centering
\includegraphics[width=0.49\textwidth]{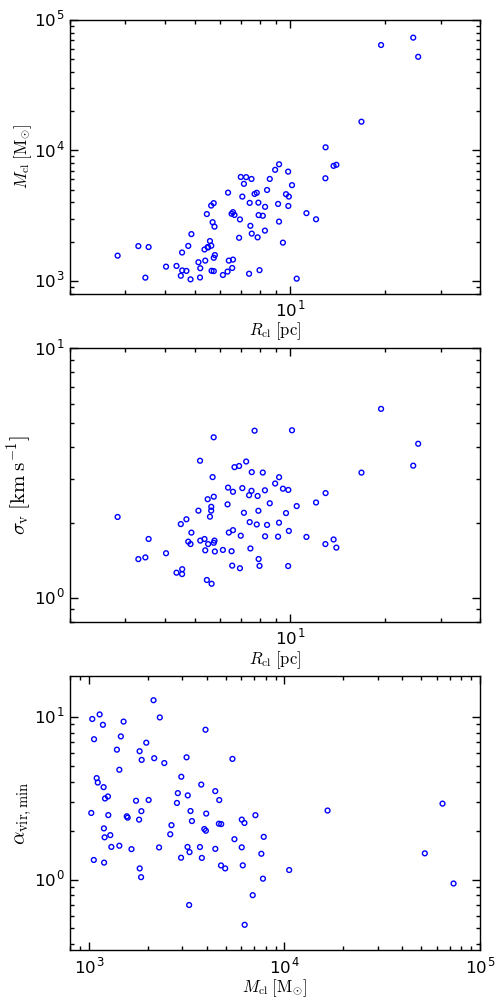}
\caption{Upper panel: Cloud mass versus cloud radius. Middle panel: One-dimensional velocity dispersion versus cloud radius. Lower panel: Virial parameter, $\alpha_{\rm vir,min}$, versus cloud mass. }
\label{fig_mass_r_v}
\end{figure}

\begin{figure}
\centering
\includegraphics[width=0.49\textwidth]{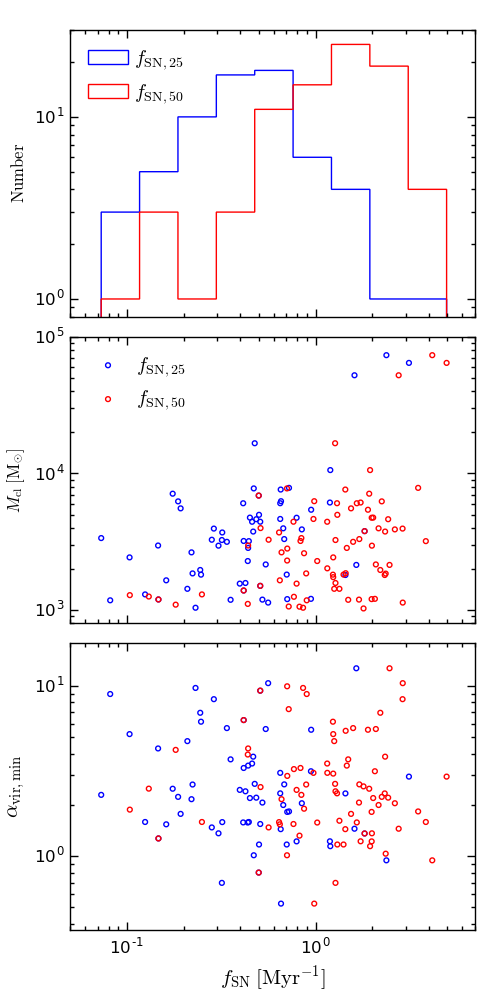}
\caption{Upper panel: The distribution of SN explosion rate in 25 pc (blue line) and 50 pc (red line) to the cloud mass center. Middle panel: Cloud mass versus SN explosion rate $f_{\rm SN}$. Lower panel: $\alpha_{\rm vir,min}$ versus $f_{\rm SN}$. $f_{\rm SN,25}$ (blue opened circles) and $f_{\rm SN,50}$ ( red opened circles) are the SNe rate which explode within $25\,\rm pc$ and $50\,\rm pc$ to the mass center of MC, respectively. The number of locally actual SN rate of $f_{\rm SN,25}$ and $f_{\rm SN,50}$ without the zeros are 65 and 82 out of 83, respectively.}
\label{fig_f_sn_alpha_vir}
\end{figure}

In this Appendix we provide the full versions of Tables~\ref{table_mc_properties}, \ref{table_energy_injection} and \ref{tab_lifetime}, where we list the properties of all the clouds in the sample. Some of these properties are also plotted in Figures~\ref{fig_mass_r_v} and \ref{fig_f_sn_alpha_vir}.

\clearpage

\begin{table*}
\centering
\caption{Basic cloud properties measured at the time when the clouds are identified. From left to right: cloud number, mass, observational virial parameter, star formation rate, radius and one-dimensional velocity dispersion.}
\begin{tabular}{lccccc  cc  cccccc}
\hline
\hline
$\rm Cloud$ & $M_{\rm cl}$ & $\alpha_{\rm vir,obs}$ & $SFR$ & $R_{\rm cl}$ & $\sigma_{\rm v}$ & \, & \, & $\rm Cloud$ & $M_{\rm cl}$ & $\alpha_{\rm vir,obs}$ & $SFR$ & $R_{\rm cl}$ & $\sigma_{\rm v}$ \\
\, & $\rm [M_{\odot}]$ & \, & $\rm [Myr^{-1}]$ & $\rm[pc]$ & $\rm [km\,\rm s^{-1}]$  & \, & \, &  \, & $\rm [M_{\odot}]$ & \, & $\rm [Myr^{-1}]$ & $\rm[pc]$ & $\rm [km\,\rm s^{-1}]$\\
\hline
C1 & 4.62 $ \times 10^3 $ & 3.87 & 0 & 9.69 & 2.18 & \, & \, & C43 & 2.97 $ \times 10^3 $ & 9.13 & 0 & 12.07 & 2.41 \\
C2 & 1.2 $ \times 10^3 $ & 2.47 & 0 & 4.54 & 1.3 & \, & \, & C44 & 4.43 $ \times 10^3 $ & 2.97 & 0 & 9.9 & 1.85 \\
C3 & 3.2 $ \times 10^3 $ & 8.95 & 0.06 & 6.65 & 3.34 & \, & \, & C45 & 4.43 $ \times 10^3 $ & 4.66 & 0.01 & 7.04 & 2.75 \\
C4 & 4.64 $ \times 10^3 $ & 14.03 & 0.02 & 7.7 & 4.67 & \, & \, & C46 & 1.66 $ \times 10^4 $ & 3.95 & 0.04 & 16.82 & 3.17 \\
C5 & 1.74 $ \times 10^3 $ & 3.51 & 0 & 5.34 & 1.72 & \, & \, & C47 & 1.82 $ \times 10^3 $ & 2.24 & 0.18 & 3.55 & 1.72 \\
C6 & 6.05 $ \times 10^3 $ & 3.15 & 0.03 & 8.61 & 2.39 & \, & \, & C48 & 2.3 $ \times 10^3 $ & 12.88 & 0 & 7.55 & 3.19 \\
C7 & 3.89 $ \times 10^3 $ & 2.81 & 0 & 9.15 & 1.76 & \, & \, & C49 & 7.32 $ \times 10^4 $ & 1.49 & 0.03 & 24.56 & 3.38 \\
C8 & 2.96 $ \times 10^3 $ & 1.56 & 0.01 & 6.92 & 1.31 & \, & \, & C50 & 1.46 $ \times 10^3 $ & 12.36 & 0.02 & 6.58 & 2.66 \\
C9 & 2.16 $ \times 10^3 $ & 9.25 & 0 & 7.88 & 2.56 & \, & \, & C51 & 1.39 $ \times 10^3 $ & 7.09 & 0 & 5.11 & 2.23 \\
C10 & 3.25 $ \times 10^3 $ & 0.9 & 0.02 & 5.44 & 1.18 & \, & \, & C52 & 1.13 $ \times 10^3 $ & 16.75 & 0 & 7.4 & 2.57 \\
C11 & 6.12 $ \times 10^3 $ & 2.2 & 0.02 & 12.93 & 1.64 & \, & \, & C53 & 1.11 $ \times 10^3 $ & 5.15 & 0 & 6.11 & 1.55 \\
C12 & 6.26 $ \times 10^3 $ & 1.35 & 0.11 & 6.97 & 1.77 & \, & \, & C54 & 3.75 $ \times 10^3 $ & 7.43 & 0 & 9.86 & 2.7 \\
C13 & 7.76 $ \times 10^3 $ & 1.76 & 0.02 & 14.01 & 1.59 & \, & \, & C55 & 6.24 $ \times 10^3 $ & 5.55 & 0.01 & 7.24 & 3.51 \\
C14 & 1.58 $ \times 10^3 $ & 3.32 & 0 & 5.76 & 1.53 & \, & \, & C56 & 1.65 $ \times 10^3 $ & 1.64 & 0 & 4.54 & 1.24 \\
C15 & 1.26 $ \times 10^3 $ & 3.64 & 0 & 6.54 & 1.34 & \, & \, & C57 & 2.02 $ \times 10^3 $ & 4.75 & 0.01 & 5.56 & 2.11 \\
C16 & 1.25 $ \times 10^3 $ & 4.6 & 0 & 5.18 & 1.69 & \, & \, & C58 & 4.74 $ \times 10^3 $ & 2.47 & 0.07 & 7.83 & 1.96 \\
C17 & 1.21 $ \times 10^3 $ & 4.6 & 0 & 7.99 & 1.34 & \, & \, & C59 & 1.85 $ \times 10^3 $ & 1.41 & 0.08 & 3.3 & 1.43 \\
C18 & 1.96 $ \times 10^3 $ & 13.97 & 0 & 9.48 & 2.73 & \, & \, & C60 & 1.03 $ \times 10^3 $ & 4.89 & 0.08 & 4.83 & 1.64 \\
C19 & 1.29 $ \times 10^3 $ & 2.76 & 0 & 4.03 & 1.51 & \, & \, & C61 & 2.64 $ \times 10^3 $ & 2.71 & 0 & 7.47 & 1.57 \\
C20 & 5.42 $ \times 10^3 $ & 15.92 & 0.1 & 10.13 & 4.69 & \, & \, & C62 & 6.88 $ \times 10^3 $ & 0.99 & 0 & 9.85 & 1.34 \\
C21 & 1.19 $ \times 10^3 $ & 6.46 & 0 & 4.68 & 2.06 & \, & \, & C63 & 1.56 $ \times 10^3 $ & 3.12 & 0 & 2.83 & 2.11 \\
C22 & 2.85 $ \times 10^3 $ & 5 & 0.04 & 9.21 & 2 & \, & \, & C64 & 4.98 $ \times 10^3 $ & 2.52 & 0 & 8.44 & 1.96 \\
C23 & 3.31 $ \times 10^3 $ & 4.04 & 0.05 & 11.25 & 1.75 & \, & \, & C65 & 1.04 $ \times 10^3 $ & 21.16 & 0 & 10.48 & 2.33 \\
C24 & 1.2 $ \times 10^3 $ & 2.36 & 0.03 & 5.63 & 1.14 & \, & \, & C66 & 2.28 $ \times 10^3 $ & 2.74 & 0 & 4.86 & 1.82 \\
C25 & 2.14 $ \times 10^3 $ & 14.16 & 0 & 6.88 & 3.37 & \, & \, & C67 & 1.85 $ \times 10^3 $ & 2.79 & 0 & 4.75 & 1.68 \\
C26 & 2.6 $ \times 10^3 $ & 2.45 & 0.03 & 5.75 & 1.69 & \, & \, & C68 & 2.82 $ \times 10^3 $ & 7.22 & 0.01 & 5.67 & 3.04 \\
C27 & 1.06 $ \times 10^3 $ & 2.67 & 0.18 & 3.47 & 1.45 & \, & \, & C69 & 6.43 $ \times 10^4 $ & 3.82 & 0.05 & 19.42 & 5.71 \\
C28 & 3.95 $ \times 10^3 $ & 10.84 & 0 & 5.71 & 4.39 & \, & \, & C70 & 3.19 $ \times 10^3 $ & 1.96 & 0 & 7.94 & 1.43 \\
C29 & 4.75 $ \times 10^3 $ & 3.95 & 0.06 & 6.35 & 2.76 & \, & \, & C71 & 7.83 $ \times 10^3 $ & 4.21 & 0.02 & 9.22 & 3.04 \\
C30 & 1.18 $ \times 10^3 $ & 11.63 & 0 & 6.32 & 2.37 & \, & \, & C72 & 3.27 $ \times 10^3 $ & 1.82 & 0.01 & 6.51 & 1.53 \\
C31 & 5.22 $ \times 10^4 $ & 3.24 & 0.02 & 25.46 & 4.14 & \, & \, & C73 & 3.36 $ \times 10^3 $ & 2.63 & 0.01 & 6.58 & 1.86 \\
C32 & 7.1 $ \times 10^3 $ & 4.02 & 0.01 & 8.95 & 2.87 & \, & \, & C74 & 1.86 $ \times 10^3 $ & 6.25 & 0.01 & 5.61 & 2.31 \\
C33 & 3.78 $ \times 10^3 $ & 2.87 & 0.06 & 5.61 & 2.23 & \, & \, & C75 & 1.5 $ \times 10^3 $ & 9.5 & 0 & 5.72 & 2.54 \\
C34 & 1.1 $ \times 10^3 $ & 6.18 & 0.03 & 4.49 & 1.97 & \, & \, & C76 & 3.7 $ \times 10^3 $ & 2.71 & 0.01 & 8.32 & 1.76 \\
C35 & 1.43 $ \times 10^3 $ & 5.75 & 0 & 6.38 & 1.82 & \, & \, & C77 & 7.62 $ \times 10^3 $ & 2.04 & 0.01 & 13.72 & 1.71 \\
C36 & 1.82 $ \times 10^3 $ & 7.18 & 0 & 5.47 & 2.48 & \, & \, & C78 & 1.06 $ \times 10^4 $ & 3.27 & 0.09 & 12.94 & 2.62 \\
C37 & 3.98 $ \times 10^3 $ & 3.83 & 0.02 & 7.92 & 2.23 & \, & \, & C79 & 3.16 $ \times 10^3 $ & 10.11 & 0.03 & 8.18 & 3.17 \\
C38 & 2.43 $ \times 10^3 $ & 9.62 & 0 & 8.31 & 2.69 & \, & \, & C80 & 1.06 $ \times 10^3 $ & 23.63 & 0 & 5.17 & 3.54 \\
C39 & 1.19 $ \times 10^3 $ & 5.13 & 0 & 5.72 & 1.66 & \, & \, & C81 & 3.96 $ \times 10^3 $ & 2.93 & 0.01 & 7.43 & 2.01 \\
C40 & 1.3 $ \times 10^3 $ & 2.05 & 0.06 & 4.35 & 1.26 & \, & \, & C82 & 1.81 $ \times 10^3 $ & 3.17 & 0.03 & 5.49 & 1.64 \\
C41 & 6.05 $ \times 10^3 $ & 3.47 & 0.04 & 7.53 & 2.68 & \, & \, & C83 & 1.43 $ \times 10^3 $ & 3.49 & 0 & 5.37 & 1.55 \\
C42 & 5.55 $ \times 10^3 $ & 2.38 & 0.04 & 7.13 & 2.19 \\
\hline
\hline
\end{tabular}
\label{tab_1_all}
\end{table*}

\begin{table*}
\centering
\caption{List of the properties and the distance of the most significant impact SN for all MCs in the sample. From left to right: cloud number, cloud mass, virial parameter, SN rate within $25\,\rm pc$, SN rate within $50\,\rm pc$, total energy injection, the ratio of total energy injection to initial potential energy, distance of the most significant SN, and the energy injection by the most significant SN.}
\begin{tabular}{lccccccccc}
\hline
\hline
$\rm Cloud$ & $M_{\rm cl} \,\rm [M_{\odot}]$ & $\alpha_{\rm vir,min}$ & $f_{\rm SN,25} \,\rm [Myr^{-1}]$ & $f_{\rm SN,50} \,\rm [Myr^{-1}]$ & $E_{\rm inj,tot} \,\rm [erg]$ & $E_{\rm inj,tot}/E_{\rm g}$ & $d_{\rm SN,max} \,\rm [pc]$ &  $E_{\rm inj,max} \,\rm [erg]$ \\
\hline
C1 & 4.62 $ \times 10^3 $ & 2.20 & 0.49 & 2.43 & 1.30 $ \times 10^{50} $ & 9.02 $ \times 10^2 $ & 6.05 & 9.71 $ \times 10^{49} $ \\
C2 & 1.20 $ \times 10^3 $ & 1.82 & 0.71 & 1.98 & 1.70 $ \times 10^{47} $ & 1.03 $ \times 10^1 $ & 49.17 & 5.50 $ \times 10^{46} $ \\
C3 & 3.20 $ \times 10^3 $ & 3.29 & 0.42 & 0.83 & 2.01 $ \times 10^{48} $ & 2.29 $ \times 10^1 $ & 25.25 & 1.05 $ \times 10^{48} $ \\
C4 & 4.64 $ \times 10^3 $ & 3.08 & 0.65 & 0.98 & 1.79 $ \times 10^{49} $ & 1.09 $ \times 10^2 $ & 10.19 & 1.29 $ \times 10^{49} $ \\
C5 & 1.74 $ \times 10^3 $ & 3.05 & 0.00 & 1.24 & 2.36 $ \times 10^{47} $ & 7.22 $ \times 10^0 $ & 49.13 & 1.20 $ \times 10^{47} $ \\
C6 & 6.05 $ \times 10^3 $ & 2.34 & 0.65 & 1.30 & 2.69 $ \times 10^{50} $ & 1.13 $ \times 10^3 $ & 6.40 & 2.16 $ \times 10^{50} $ \\
C7 & 3.89 $ \times 10^3 $ & 2.04 & 0.85 & 2.63 & 1.40 $ \times 10^{49} $ & 1.43 $ \times 10^2 $ & 32.00 & 5.52 $ \times 10^{48} $ \\
C8 & 2.96 $ \times 10^3 $ & 1.36 & 0.31 & 1.99 & 2.15 $ \times 10^{48} $ & 3.11 $ \times 10^1 $ & 24.28 & 1.18 $ \times 10^{48} $ \\
C9 & 2.16 $ \times 10^3 $ & 5.57 & 0.54 & 2.09 & 1.23 $ \times 10^{49} $ & 3.57 $ \times 10^2 $ & 11.20 & 4.33 $ \times 10^{48} $ \\
C10 & 3.25 $ \times 10^3 $ & 0.70 & 0.32 & 1.28 & 9.87 $ \times 10^{49} $ & 9.67 $ \times 10^2 $ & 2.06 & 9.87 $ \times 10^{49} $ \\
C11 & 6.12 $ \times 10^3 $ & 1.22 & 1.19 & 1.73 & 1.28 $ \times 10^{50} $ & 5.87 $ \times 10^2 $ & 11.49 & 3.25 $ \times 10^{49} $ \\
C12 & 6.26 $ \times 10^3 $ & 0.53 & 0.66 & 0.98 & 7.69 $ \times 10^{49} $ & 1.79 $ \times 10^2 $ & 6.98 & 7.69 $ \times 10^{49} $ \\
C13 & 7.76 $ \times 10^3 $ & 1.01 & 0.47 & 0.71 & 7.74 $ \times 10^{49} $ & 2.35 $ \times 10^2 $ & 13.66 & 5.56 $ \times 10^{49} $ \\
C14 & 1.58 $ \times 10^3 $ & 2.40 & 0.43 & 1.28 & 1.45 $ \times 10^{49} $ & 6.68 $ \times 10^2 $ & 6.44 & 6.35 $ \times 10^{48} $ \\
C15 & 1.26 $ \times 10^3 $ & 2.49 & 0.00 & 0.13 & 6.14 $ \times 10^{46} $ & 4.02 $ \times 10^0 $ & 146.68 & 4.14 $ \times 10^{46} $ \\
C16 & 1.25 $ \times 10^3 $ & 3.24 & 0.00 & 0.77 & 1.50 $ \times 10^{47} $ & 8.87 $ \times 10^0 $ & 34.69 & 4.07 $ \times 10^{46} $ \\
C17 & 1.21 $ \times 10^3 $ & 3.15 & 0.95 & 2.06 & 1.97 $ \times 10^{48} $ & 1.44 $ \times 10^2 $ & 3.54 & 1.20 $ \times 10^{48} $ \\
C18 & 1.96 $ \times 10^3 $ & 6.94 & 0.24 & 2.20 & 2.09 $ \times 10^{49} $ & 8.38 $ \times 10^2 $ & 14.86 & 1.39 $ \times 10^{49} $ \\
C19 & 1.29 $ \times 10^3 $ & 1.87 & 0.00 & 0.10 & 5.75 $ \times 10^{46} $ & 2.71 $ \times 10^0 $ & 133.21 & 4.89 $ \times 10^{46} $ \\
C20 & 5.42 $ \times 10^3 $ & 5.52 & 0.95 & 1.90 & 3.13 $ \times 10^{50} $ & 1.15 $ \times 10^3 $ & 9.76 & 1.97 $ \times 10^{50} $ \\
C21 & 1.19 $ \times 10^3 $ & 2.06 & 0.52 & 1.70 & 8.53 $ \times 10^{49} $ & 5.27 $ \times 10^3 $ & 3.62 & 8.50 $ \times 10^{49} $ \\
C22 & 2.85 $ \times 10^3 $ & 3.40 & 0.44 & 1.46 & 9.87 $ \times 10^{48} $ & 1.64 $ \times 10^2 $ & 49.91 & 4.72 $ \times 10^{48} $ \\
C23 & 3.31 $ \times 10^3 $ & 2.64 & 0.68 & 1.71 & 9.73 $ \times 10^{49} $ & 1.27 $ \times 10^3 $ & 4.93 & 7.39 $ \times 10^{49} $ \\
C24 & 1.20 $ \times 10^3 $ & 1.27 & 0.15 & 0.15 & 2.60 $ \times 10^{48} $ & 1.39 $ \times 10^2 $ & 7.84 & 2.60 $ \times 10^{48} $ \\
C25 & 2.14 $ \times 10^3 $ & 12.67 & 1.64 & 2.47 & 6.37 $ \times 10^{49} $ & 1.73 $ \times 10^3 $ & 8.57 & 3.76 $ \times 10^{49} $ \\
C26 & 2.60 $ \times 10^3 $ & 1.90 & 0.00 & 0.87 & 1.21 $ \times 10^{48} $ & 1.64 $ \times 10^1 $ & 57.78 & 4.33 $ \times 10^{47} $ \\
C27 & 1.06 $ \times 10^3 $ & 1.32 & 0.00 & 0.82 & 3.21 $ \times 10^{47} $ & 1.06 $ \times 10^1 $ & 129.62 & 2.07 $ \times 10^{47} $ \\
C28 & 3.95 $ \times 10^3 $ & 8.34 & 0.29 & 2.89 & 4.67 $ \times 10^{48} $ & 2.76 $ \times 10^1 $ & 23.60 & 2.13 $ \times 10^{48} $ \\
C29 & 4.75 $ \times 10^3 $ & 2.19 & 0.45 & 2.02 & 2.10 $ \times 10^{49} $ & 9.81 $ \times 10^1 $ & 4.33 & 2.02 $ \times 10^{49} $ \\
C30 & 1.18 $ \times 10^3 $ & 8.94 & 0.08 & 0.90 & 1.09 $ \times 10^{48} $ & 8.90 $ \times 10^1 $ & 14.63 & 9.30 $ \times 10^{47} $ \\
C31 & 5.22 $ \times 10^4 $ & 1.45 & 1.61 & 2.76 & 9.66 $ \times 10^{50} $ & 1.63 $ \times 10^2 $ & 9.79 & 1.31 $ \times 10^{50} $ \\
C32 & 7.10 $ \times 10^3 $ & 2.49 & 0.18 & 1.93 & 4.58 $ \times 10^{49} $ & 1.58 $ \times 10^2 $ & 8.77 & 4.17 $ \times 10^{49} $ \\
C33 & 3.78 $ \times 10^3 $ & 1.36 & 1.82 & 1.82 & 8.34 $ \times 10^{49} $ & 3.29 $ \times 10^2 $ & 8.71 & 4.38 $ \times 10^{49} $ \\
C34 & 1.10 $ \times 10^3 $ & 4.20 & 0.00 & 0.18 & 2.43 $ \times 10^{47} $ & 1.50 $ \times 10^1 $ & 139.38 & 2.42 $ \times 10^{47} $ \\
C35 & 1.43 $ \times 10^3 $ & 4.73 & 0.21 & 1.26 & 2.19 $ \times 10^{48} $ & 1.14 $ \times 10^2 $ & 14.45 & 1.98 $ \times 10^{48} $ \\
C36 & 1.82 $ \times 10^3 $ & 6.15 & 0.25 & 1.24 & 4.74 $ \times 10^{49} $ & 1.31 $ \times 10^3 $ & 7.32 & 4.73 $ \times 10^{49} $ \\
C37 & 3.98 $ \times 10^3 $ & 2.54 & 0.00 & 0.51 & 8.69 $ \times 10^{47} $ & 7.49 $ \times 10^0 $ & 54.89 & 2.35 $ \times 10^{47} $ \\
C38 & 2.43 $ \times 10^3 $ & 5.20 & 0.10 & 1.24 & 1.76 $ \times 10^{49} $ & 4.03 $ \times 10^2 $ & 45.74 & 4.33 $ \times 10^{48} $ \\
C39 & 1.19 $ \times 10^3 $ & 3.70 & 0.35 & 1.49 & 2.92 $ \times 10^{48} $ & 2.04 $ \times 10^2 $ & 9.52 & 1.66 $ \times 10^{48} $ \\
C40 & 1.30 $ \times 10^3 $ & 1.59 & 0.13 & 0.25 & 1.52 $ \times 10^{48} $ & 6.36 $ \times 10^1 $ & 12.97 & 1.45 $ \times 10^{48} $ \\
C41 & 6.05 $ \times 10^3 $ & 1.58 & 0.41 & 1.65 & 1.87 $ \times 10^{50} $ & 6.17 $ \times 10^2 $ & 11.62 & 1.47 $ \times 10^{50} $ \\
C42 & 5.55 $ \times 10^3 $ & 1.77 & 0.19 & 1.54 & 7.03 $ \times 10^{48} $ & 2.54 $ \times 10^1 $ & 27.52 & 4.23 $ \times 10^{48} $ \\
\hline
\hline
\end{tabular}
\label{tab_2_all}
\end{table*}

\begin{table*}
\center{Table \ref{tab_2_all} --- Continued}
\begin{tabular}{lccccccccc}
\hline
\hline
$\rm Cloud$ & $M_{\rm cl} \,\rm [M_{\odot}]$ & $\alpha_{\rm vir,min}$ & $f_{\rm SN,25} \,\rm [Myr^{-1}]$ & $f_{\rm SN,50} \,\rm [Myr^{-1}]$ & $E_{\rm inj,tot} \,\rm [erg]$ & $E_{\rm inj,tot}/E_{\rm g}$ & $d_{\rm SN,max} \,\rm [pc]$ \tablenotemark{a} &  $E_{\rm inj,max} \,\rm [erg]$ \\
\hline
C43 & 2.97 $ \times 10^3 $ & 4.29 & 0.15 & 0.44 & 4.13 $ \times 10^{49} $ & 7.70 $ \times 10^2 $ & 12.04 & 3.60 $ \times 10^{49} $ \\
C44 & 4.43 $ \times 10^3 $ & 1.54 & 0.51 & 0.76 & 4.34 $ \times 10^{49} $ & 4.09 $ \times 10^2 $ & 10.76 & 1.55 $ \times 10^{49} $ \\
C45 & 4.43 $ \times 10^3 $ & 3.50 & 0.46 & 1.15 & 1.94 $ \times 10^{49} $ & 1.15 $ \times 10^2 $ & 23.47 & 1.52 $ \times 10^{49} $ \\
C46 & 1.66 $ \times 10^4 $ & 2.66 & 0.48 & 1.27 & 6.01 $ \times 10^{49} $ & 5.07 $ \times 10^1 $ & 36.01 & 2.90 $ \times 10^{49} $ \\
C47 & 1.82 $ \times 10^3 $ & 1.17 & 0.70 & 1.41 & 7.61 $ \times 10^{48} $ & 9.12 $ \times 10^1 $ & 14.87 & 4.46 $ \times 10^{48} $ \\
C48 & 2.30 $ \times 10^3 $ & 9.93 & 0.00 & 0.71 & 1.83 $ \times 10^{46} $ & 4.79 $ \times 10^{-1} $ & 70.83 & 9.12 $ \times 10^{45} $ \\
C49 & 7.32 $ \times 10^4 $ & 0.94 & 2.37 & 4.15 & 1.60 $ \times 10^{51} $ & 1.19 $ \times 10^2 $ & 13.40 & 2.14 $ \times 10^{50} $ \\
C50 & 1.46 $ \times 10^3 $ & 7.59 & 0.00 & 0.00 & 0.00 $ \times 10^{0} $ & 0.00 $ \times 10^0 $ & ... & 0.00 $ \times 10^{0} $ \\
C51 & 1.39 $ \times 10^3 $ & 6.29 & 0.42 & 0.42 & 2.16 $ \times 10^{47} $ & 1.02 $ \times 10^1 $ & ... & 2.00 $ \times 10^{47} $ \\
C52 & 1.13 $ \times 10^3 $ & 10.36 & 0.56 & 2.90 & 6.54 $ \times 10^{48} $ & 5.80 $ \times 10^2 $ & 17.53 & 2.31 $ \times 10^{48} $ \\
C53 & 1.11 $ \times 10^3 $ & 3.95 & 0.00 & 0.44 & 5.35 $ \times 10^{46} $ & 4.20 $ \times 10^0 $ & 113.12 & 3.46 $ \times 10^{46} $ \\
C54 & 3.75 $ \times 10^3 $ & 3.84 & 0.47 & 2.34 & 1.49 $ \times 10^{49} $ & 1.73 $ \times 10^2 $ & 10.21 & 7.52 $ \times 10^{48} $ \\
C55 & 6.24 $ \times 10^3 $ & 2.23 & 0.19 & 2.24 & 6.64 $ \times 10^{49} $ & 2.36 $ \times 10^2 $ & 13.34 & 5.22 $ \times 10^{49} $ \\
C56 & 1.65 $ \times 10^3 $ & 1.54 & 0.16 & 0.65 & 9.69 $ \times 10^{49} $ & 3.00 $ \times 10^3 $ & 3.95 & 9.68 $ \times 10^{49} $ \\
C57 & 2.02 $ \times 10^3 $ & 3.08 & 0.00 & 1.15 & 1.12 $ \times 10^{48} $ & 2.61 $ \times 10^1 $ & 47.91 & 4.21 $ \times 10^{47} $ \\
C58 & 4.74 $ \times 10^3 $ & 1.22 & 0.79 & 1.98 & 9.88 $ \times 10^{49} $ & 3.85 $ \times 10^2 $ & 6.31 & 9.86 $ \times 10^{49} $ \\
C59 & 1.85 $ \times 10^3 $ & 1.03 & 0.00 & 2.35 & 1.66 $ \times 10^{46} $ & 2.29 $ \times 10^{-1} $ & 42.32 & 1.46 $ \times 10^{46} $ \\
C60 & 1.03 $ \times 10^3 $ & 2.57 & 0.00 & 1.80 & 4.15 $ \times 10^{46} $ & 1.96 $ \times 10^0 $ & 53.26 & 2.48 $ \times 10^{46} $ \\
C61 & 2.64 $ \times 10^3 $ & 2.16 & 0.22 & 0.66 & 4.85 $ \times 10^{49} $ & 8.65 $ \times 10^2 $ & 7.09 & 4.78 $ \times 10^{49} $ \\
C62 & 6.88 $ \times 10^3 $ & 0.80 & 0.50 & 0.50 & 4.50 $ \times 10^{49} $ & 1.66 $ \times 10^2 $ & 6.66 & 2.71 $ \times 10^{49} $ \\
C63 & 1.56 $ \times 10^3 $ & 2.45 & 0.40 & 0.79 & 6.96 $ \times 10^{49} $ & 1.48 $ \times 10^3 $ & 0.58 & 6.96 $ \times 10^{49} $ \\
C64 & 4.98 $ \times 10^3 $ & 1.17 & 0.50 & 1.30 & 8.75 $ \times 10^{49} $ & 4.94 $ \times 10^2 $ & 2.05 & 3.42 $ \times 10^{49} $ \\
C65 & 1.04 $ \times 10^3 $ & 9.71 & 0.23 & 0.86 & 3.71 $ \times 10^{48} $ & 4.43 $ \times 10^2 $ & 9.53 & 1.56 $ \times 10^{48} $ \\
C66 & 2.28 $ \times 10^3 $ & 1.57 & 0.44 & 1.02 & 1.95 $ \times 10^{48} $ & 3.62 $ \times 10^1 $ & 19.34 & 1.25 $ \times 10^{48} $ \\
C67 & 1.85 $ \times 10^3 $ & 2.63 & 0.22 & 0.89 & 3.67 $ \times 10^{48} $ & 9.33 $ \times 10^1 $ & 38.90 & 2.53 $ \times 10^{48} $ \\
C68 & 2.82 $ \times 10^3 $ & 2.96 & 0.00 & 0.71 & 1.42 $ \times 10^{48} $ & 2.02 $ \times 10^1 $ & 74.60 & 2.76 $ \times 10^{47} $ \\
C69 & 6.43 $ \times 10^4 $ & 2.93 & 3.13 & 4.95 & 1.18 $ \times 10^{51} $ & 8.25 $ \times 10^1 $ & 13.54 & 2.68 $ \times 10^{50} $ \\
C70 & 3.19 $ \times 10^3 $ & 1.59 & 0.44 & 3.84 & 2.98 $ \times 10^{48} $ & 4.06 $ \times 10^1 $ & 8.21 & 6.31 $ \times 10^{47} $ \\
C71 & 7.83 $ \times 10^3 $ & 1.83 & 0.72 & 3.50 & 5.63 $ \times 10^{49} $ & 1.49 $ \times 10^2 $ & 6.96 & 4.12 $ \times 10^{49} $ \\
C72 & 3.27 $ \times 10^3 $ & 1.47 & 0.28 & 0.56 & 7.82 $ \times 10^{49} $ & 7.54 $ \times 10^2 $ & 5.53 & 6.25 $ \times 10^{49} $ \\
C73 & 3.36 $ \times 10^3 $ & 2.29 & 0.07 & 0.84 & 1.18 $ \times 10^{48} $ & 1.19 $ \times 10^1 $ & 48.74 & 5.20 $ \times 10^{47} $ \\
C74 & 1.86 $ \times 10^3 $ & 5.44 & 0.00 & 1.44 & 7.11 $ \times 10^{47} $ & 1.97 $ \times 10^1 $ & 28.01 & 1.07 $ \times 10^{47} $ \\
C75 & 1.50 $ \times 10^3 $ & 9.36 & 0.51 & 0.51 & 5.75 $ \times 10^{48} $ & 2.79 $ \times 10^2 $ & 8.91 & 5.73 $ \times 10^{48} $ \\
C76 & 3.70 $ \times 10^3 $ & 1.58 & 0.32 & 0.64 & 5.53 $ \times 10^{49} $ & 5.68 $ \times 10^2 $ & 4.43 & 5.50 $ \times 10^{49} $ \\
C77 & 7.62 $ \times 10^3 $ & 1.44 & 0.65 & 1.44 & 1.33 $ \times 10^{50} $ & 4.43 $ \times 10^2 $ & 8.78 & 5.59 $ \times 10^{49} $ \\
C78 & 1.06 $ \times 10^4 $ & 1.14 & 1.20 & 1.95 & 1.13 $ \times 10^{50} $ & 1.77 $ \times 10^2 $ & 11.95 & 4.42 $ \times 10^{49} $ \\
C79 & 3.16 $ \times 10^3 $ & 5.64 & 0.34 & 1.58 & 9.46 $ \times 10^{48} $ & 1.29 $ \times 10^2 $ & 22.83 & 2.93 $ \times 10^{48} $ \\
C80 & 1.06 $ \times 10^3 $ & 7.29 & 0.00 & 0.72 & 1.65 $ \times 10^{47} $ & 1.46 $ \times 10^1 $ & 91.02 & 6.23 $ \times 10^{46} $ \\
C81 & 3.96 $ \times 10^3 $ & 2.00 & 0.67 & 2.16 & 6.57 $ \times 10^{49} $ & 5.51 $ \times 10^2 $ & 7.16 & 6.29 $ \times 10^{49} $ \\
C82 & 1.81 $ \times 10^3 $ & 2.33 & 1.44 & 2.33 & 5.05 $ \times 10^{48} $ & 1.47 $ \times 10^2 $ & 15.27 & 2.47 $ \times 10^{48} $ \\
C83 & 1.43 $ \times 10^3 $ & 1.61 & 0.00 & 1.34 & 0.00 $ \times 10^{0} $ & 0.00 $ \times 10^0 $ & ... & 0.00 $ \times 10^{0} $ \\
\hline
\hline
\end{tabular}
\tablenotetext{a}{For the clouds C50, C51, C83 we could not identify the SNe corresponding to the most significant kinetic energy peak.}
\label{tab_2_all_continued}
\end{table*}

\begin{table*}
\centering
\caption{List of timescales for all the clouds in the sample. From left to right: cloud number, cloud formation time, cloud dispersal time, cloud lifetime, cloud delay time and the cloud dynamical time.}
\begin{tabular}{lccccc  ccc  lccccc}
\hline
\hline
$\rm Cloud$ & $t_{\rm form}$ & $t_{\rm disp}$ & $t_{\rm life}$ & $t_{\rm delay}$ & $t_{\rm dyn,\alpha}$   & \, & \, & \,   & $\rm Cloud$ & $t_{\rm form}$ & $t_{\rm disp}$ & $t_{\rm life}$ & $t_{\rm delay}$ & $t_{\rm dyn,\alpha}$ \\
\, & $\rm [Myr]$ & $\rm [Myr]$ & $\rm [Myr]$ & $\rm [Myr]$ & $\rm [Myr]$    & \, & \, & \,   & \, & $\rm [Myr]$ & $\rm [Myr]$ & $\rm [Myr]$ & $\rm [Myr]$ & $\rm [Myr]$ \\
\hline
C1 & $>$ 9.28 & 5.33 & $>$ 14.61 & 0.35 & 5 & \, & \, & \, & C43 & $>$ 12.33 & 12.69 & $>$ 25.02 & 2.26 & 6.23 \\
C2 & 8.91 & 3.19 & 12.1 & 3.14 & 3.89 & \, & \, & \, & C44 & 6.24 & 12.75 & 18.99 & 2.96 & 7.31 \\
C3 & 7.79 & 4.78 & 12.57 & 1.41 & 4.29 & \, & \, & \, & C45 & 5.01 & 4.51 & 9.52 & 0.91 & 2.78 \\
C4 & $>$ 8.35 & 3.16 & $>$ 11.51 & 1.64 & 4.23 & \, & \, & \, & C46 & $>$ 12.68 & 7.91 & $>$ 20.59 & 2.87 & 5.4 \\
C5 & 5.1 & 6.18 & 11.28 & 1.35 & 3.29 & \, & \, & \, & C47 & 8.12 & 2.78 & 10.9 & 0.67 & 2.04 \\
C6 & $>$ 10.54 & 0.97 & $>$ 11.51 & 0.94 & 3.26 & \, & \, & \, & C48 & 11.98 & 5.8 & 17.78 & 3.72 & 3.28 \\
C7 & $>$ 15.35 & 3.72 & $>$ 19.07 & 0.91 & 5.18 & \, & \, & \, & C49 & $>$ 16.75 & 4.28 & $>$ 21.03 & 0.12 & 7.02 \\
C8 & $>$ 6.82 & 8.15 & $>$ 14.97 & 0.15 & 5.65 & \, & \, & \, & C50 & 6.65 & 2.55 & 9.2 & ... & 2.42 \\
C9 & $>$ 8.32 & 11.13 & $>$ 19.45 & 1.93 & 5.25 & \, & \, & \, & C51 & 3.08 & 3.08 & 6.16 & ... & 2.68 \\
C10 & 7.41 & 3.02 & 10.43 & 0.85 & 4.88 & \, & \, & \, & C52 & 6.65 & 10.2 & 16.85 & 6.59 & 4.29 \\
C11 & $>$ 11.69 & 5.98 & $>$ 17.67 & 0.88 & 8.01 & \, & \, & \, & C53 & 6.65 & 6.68 & 13.33 & 1.26 & 4.17 \\
C12 & $>$9.08 & 2.4 & $>$11.48 & 1.14 & 4.83 & \, & \, & \, & C54 & 15.91 & 3.49 & 19.4 & 2.11 & 5.22 \\
C13 & $>$ 8.49 & 4.19 & $>$ 12.68 & 3.52 & 9.29 & \, & \, & \, & C55 & 3.81 & 10.4 & 14.21 & 4.78 & 5.62 \\
C14 & 8.79 & 5.3 & 14.09 & 0.59 & 4.52 & \, & \, & \, & C56 & 8.17 & 4.69 & 12.86 & 0.06 & 3.59 \\
C15 & $>$ 5.83 & 10.25 & $>$ 16.08 & 1.99 & 5.87 & \, & \, & \, & C57 & 4.51 & 8.85 & 13.36 & 2.81 & 2.9 \\
C16 & 4.4 & 10.23 & 14.63 & 3.37 & 3.92 & \, & \, & \, & C58 & $>$ 13.21 & 3.6 & $>$ 16.81 & 0.76 & 4.3 \\
C17 & $>$ 11.77 & 8.29 & $>$ 20.06 & 3.34 & 5.82 & \, & \, & \, & C59 & 3.57 & 2.78 & 6.35 & 0.62 & 2.26 \\
C18 & 4.57 & 12.54 & 17.11 & 3.96 & 5.12 & \, & \, & \, & C60 & 5.39 & 4.19 & 9.58 & 1.14 & 3.07 \\
C19 & 4.78 & 8.53 & 13.31 & 2.87 & 3.57 & \, & \, & \, & C61 & $>$ 18.75 & 4.63 & $>$ 23.38 & 0.62 & 4.95 \\
C20 & $>$ 5.09 & 5.45 & $>$ 10.54 & 0.47 & 3.72 & \, & \, & \, & C62 & 13.04 & 4.01 & 17.05 & 2.11 & 7.47 \\
C21 & $>$ 9.61 & 6.48 & $>$ 16.09 & 0.03 & 3.8 & \, & \, & \, & C63 & 6.62 & 2.14 & 8.76 & 0.03 & 1.5 \\
C22 & $>$ 10.87 & 4.39 & $>$ 15.26 & 1.11 & 4.05 & \, & \, & \, & C64 & 18.34 & 1.26 & 19.6 & 1.23 & 5.99 \\
C23 & $>$ 4.54 & 6.83 & $>$ 11.37 & 0.03 & 6.29 & \, & \, & \, & C65 & 10.14 & $>$ 16.11 & $>$ 26.25 & $>$ 20.63 & 5.47 \\
C24 & $>$ 9.81 & 8.35 & $>$ 18.16 & 3.46 & 5.79 & \, & \, & \, & C66 & 9.14 & 6.3 & 15.44 & 1.52 & 4 \\
C25 & 2.43 & 2.55 & 4.98 & 0.47 & 2.08 & \, & \, & \, & C67 & 6.5 & 5.24 & 11.74 & 2.08 & 2.77 \\
C26 & 5.1 & 6.56 & 11.66 & 1.32 & 3.9 & \, & \, & \, & C68 & 2.75 & 8.29 & 11.04 & 0.21 & 3.89 \\
C27 & $>$ 11.01 & 5.22 & $>$ 16.23 & 1.49 & 2.38 & \, & \, & \, & C69 & 14.42 & 3.55 & 17.97 & 0.15 & 3.33 \\
C28 & 5.71 & 3.6 & 9.31 & 1.46 & 1.56 & \, & \, & \, & C70 & 8.88 & 1.52 & 10.4 & 1.79 & 5.76 \\
C29 & 8.03 & 4.22 & 12.25 & 0.73 & 2.72 & \, & \, & \, & C71 & 9.23 & 2.96 & 12.19 & 0.5 & 4.55 \\
C30 & $>$ 14.06 & 9.55 & $>$ 23.61 & 3.16 & 2.77 & \, & \, & \, & C72 & 9.46 & 4.86 & 14.32 & 0.35 & 4.15 \\
C31 & $>$ 16.55 & 3.52 & $>$ 20.07 & 1.26 & 6.79 & \, & \, & \, & C73 & 4.19 & $>$ 11.78 & $>$ 15.97 & $>$ 23.41 & 4.43 \\
C32 & 9.32 & 4.07 & 13.39 & 2.29 & 3.73 & \, & \, & \, & C74 & 7.03 & 4.89 & 11.92 & 0.44 & 2.38 \\
C33 & 9.23 & 2.43 & 11.66 & 1.32 & 2.75 & \, & \, & \, & C75 & 4.28 & 2.93 & 7.21 & 0.26 & 2.2 \\
C34 & 7.91 & 5.65 & 13.56 & 1.38 & 2.88 & \, & \, & \, & C76 & $>$ 19.25 & 4.22 & $>$ 23.47 & 0.7 & 4.98 \\
C35 & $>$ 13.03 & 3.11 & $>$ 16.14 & 0.76 & 3.22 & \, & \, & \, & C77 & 18.17 & 1.85 & 20.02 & 1.52 & 7.91 \\
C36 & 6.97 & 4.25 & 11.22 & 0.09 & 2.16 & \, & \, & \, & C78 & 15.76 & 3.78 & 19.54 & 3.72 & 5.29 \\
C37 & 7.65 & 8.7 & 16.35 & 0.18 & 5.15 & \, & \, & \, & C79 & 9.38 & 7.94 & 17.32 & 1.03 & 3.21 \\
C38 & 11.46 & 9.58 & 21.04 & 3.6 & 3.97 & \, & \, & \, & C80 & 8.03 & 7.94 & 15.97 & 4.75 & 2.94 \\
C39 & $>$ 11.28 & 14.18 & $>$ 25.46 & 0.44 & 3.79 & \, & \, & \, & C81 & 9.7 & 8.38 & 18.08 & 1.35 & 4.93 \\
C40 & 5.1 & 8.82 & 13.92 & 0.97 & 3.63 & \, & \, & \, & C82 & 8.73 & 9.29 & 18.02 & 0.21 & 4.16 \\
C41 & $>$ 12.45 & 3.75 & $>$ 16.2 & 0.15 & 3.85 & \, & \, & \, & C83 & 4.95 & 3.46 & 8.41 & ... & 4.7 \\
C42 & 10.93 & 5.27 & 16.2 & 0.18 & 3.31 \\
\hline
\hline
\end{tabular}
\label{tab_lifetime_all}
\end{table*}

\end{document}